\documentclass[aps,prl,reprint,twocolumn,superscriptaddress,floatfix,nofootinbib]{revtex4-1}
\usepackage{amsmath,amssymb,color,comment,physics}
\usepackage[makeroom]{cancel}
\usepackage[caption=false]{subfig}
\usepackage{mathrsfs}
\usepackage{graphicx}
\usepackage{subfig}
\usepackage[countmax]{subfloat}
\usepackage[english]{babel}
\usepackage{dsfont}
\definecolor{jadclr}{rgb}{0,0.5,0}
\definecolor{jadcolor}{rgb}{0.5,0,0}
\usepackage[bookmarks=true,colorlinks,linkcolor=jadclr,urlcolor=jadcolor,citecolor=jadcolor]{hyperref}

\definecolor{mypurple}{rgb}{0.49,0.18,0.56}
\definecolor{mygold}{rgb}{0.93,0.69,0.13}
\definecolor{mygreen}{rgb}{0,0.5,0}
\definecolor{myblue}{rgb}{0,0,0.75}
\definecolor{mymagenta}{cmyk}{0,1,0,0.12}
\definecolor{mygray}{rgb}{0.5,0.5,0.5}

\usepackage{umoline}

\def\d{\mathrm d}

\definecolor{mypink1}{rgb}{0.858, 0.188, 0.478}

\def\d{\mathrm d}
\begin{document}
	
	\title{Reliability of lattice gauge theories}
	\author{Jad C.~Halimeh}
	\affiliation{Kirchhoff Institute for Physics, Ruprecht-Karls-Universit\"{a}t Heidelberg, Im Neuenheimer Feld 227, 69120 Heidelberg, Germany}
	\affiliation{Institute for Theoretical Physics, Ruprecht-Karls-Universit\"{a}t Heidelberg, Philosophenweg 16, 69120 Heidelberg, Germany} 
	\affiliation{INO-CNR BEC Center and Department of Physics, University of Trento, Via Sommarive 14, I-38123 Trento, Italy}
	\affiliation{Max Planck Institute for the Physics of Complex Systems, N\"othnitzer Stra{\ss}e 38, 01187 Dresden, Germany}
	
	\author{Philipp Hauke}
	\affiliation{Kirchhoff Institute for Physics, Ruprecht-Karls-Universit\"{a}t Heidelberg, Im Neuenheimer Feld 227, 69120 Heidelberg, Germany}
	\affiliation{Institute for Theoretical Physics, Ruprecht-Karls-Universit\"{a}t Heidelberg, Philosophenweg 16, 69120 Heidelberg, Germany}
	\affiliation{INO-CNR BEC Center and Department of Physics, University of Trento, Via Sommarive 14, I-38123 Trento, Italy}
	
	\begin{abstract}
		Currently, there are intense experimental efforts to realize lattice gauge theories in quantum simulators. Except for specific models, however, practical quantum simulators can never be fine-tuned to perfect local gauge invariance. There is thus a strong need for a rigorous understanding of gauge-invariance violation and how to reliably protect against it. As we show through analytic and numerical evidence, in the presence of a gauge invariance-breaking term the gauge violation accumulates only perturbatively at short times before proliferating only at very long times. This proliferation can be suppressed up to infinite times by energetically penalizing processes that drive the dynamics away from the initial gauge-invariant sector. Our results provide a theoretical basis that highlights a surprising robustness of gauge-theory quantum simulators.
	\end{abstract}
	
	\date{\today}
	\maketitle
	\textbf{\emph{Introduction.---}}In modern physics, gauge theories assume a central role, ranging from the Standard Model of Particle Physics \cite{Cheng_book} to emergent exotic solid-state phases \cite{Balents_NatureReview,Savary2016}. 
	The defining feature of any gauge theory is its local conservation laws, such as Gauss's law for a $\mathrm{U}(1)$ gauge symmetry. 
	Despite their elegance, the computation of gauge theories on classical computers is a daunting task \cite{Gattringer_book,Calzetta_book}, particularly for out-of-equilibrium phenomena.	Currently, a complementary tool to probe gauge theories is emerging in the form of low-energy tabletop devices, so-called quantum simulators \cite{Wiese2013,Zohar2015,Dalmonte2016,MariCarmen2019}.
	Although still in the developmental phase, these experiments are rapidly advancing \cite{Martinez2016,Bernien2017,Dai2017,Klco2018,Goerg2019,Kokail2019,Schweizer2019,Mil2019,Yang2019}, 
	making one open issue all the more pressing: how can we ensure the reliability of quantum simulators once they scale beyond problems that can be benchmarked by classical computers \cite{Hauke2012,Berges2019}? 
	For gauge theories, this issue is particularly subtle, since a faithful quantum simulation necessitates not only the correct engineering of the Hamiltonian dynamics, but crucially also of the defining local gauge symmetry. 
	It becomes thus an outstanding challenge to understand how gauge-invariant dynamics may be faithfully simulated~--~without unrealistically fine-tuned interactions between the constituents of the quantum simulator and in the consequently unavoidable presence of gauge-violating errors.
	
	\begin{figure}[htp]
		\centering
		\hspace{-.25 cm}
		\includegraphics[width=.445\textwidth]{{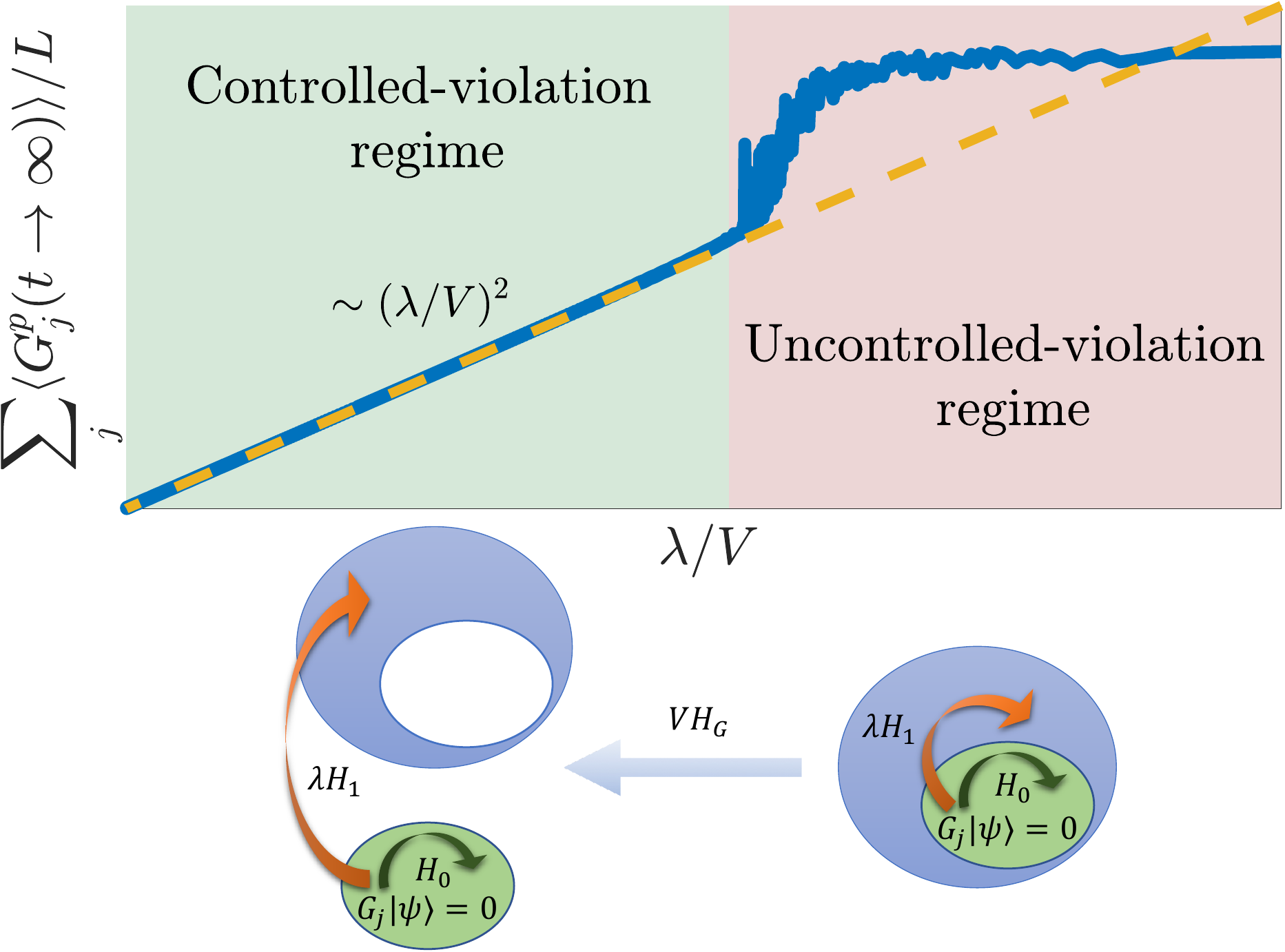}}\quad
		\hspace{-.15 cm}
		\caption{(Color online). 
			Right:~Given a gauge theory with Hamiltonian $H_0$ on $L$ matter sites, undesired processes $\propto\lambda H_1$ that break gauge symmetry drive the dynamics away from the initial gauge-invariant sector (green bubble) given by the analog of Gauss's law, $G_j\ket{\psi}=0$ $\forall j$ to other sectors in the total Hilbert space (blue bubble). At late times, the average gauge violation assumes a random value. 
			Left:~Through the introduction of a protection term $\propto V H_G$ that energetically penalizes gauge-violating processes, gauge invariance is retained up to infinite times. Two sharply distinct regimes emerge: an \textit{uncontrolled-violation regime} where $V$ is too small to energetically isolate the initial gauge-invariant sector, and a \textit{controlled-violation regime} for sufficiently large $V$ where the infinite-time violation scales as $\sim(\lambda/V)^2$. The axes are both log-scale. See Fig.~\ref{fig:scaling} for a detailed quantitative presentation on $\mathrm{Z}_2$ ($p=1$) and $\mathrm{U}(1)$ ($p=2$) gauge theories.}
		\label{fig:illustration} 
	\end{figure}
	
	As we show in this Letter, quantum simulators can reliably reproduce the out-of-equilibrium dynamics of gauge theories even if the prohibitive restriction of perfect gauge invariance is relaxed. 
	Our results are based on numerical studies of $\mathrm{Z}_2$ and $\mathrm{U}(1)$ gauge theories as well as analytic proofs. 
	In the presence of gauge invariance-violating terms, the short-time dynamics deviates only perturbatively slowly from the idealized scenario, and observables reproduce the ideal dynamics during a time frame controlled by the strength of the gauge-violating errors. 
	Upon adding a protection term that energetically penalizes such errors, gauge invariance becomes bounded from above even up to infinite times; cf.~Fig.~\ref{fig:illustration}.
	As we explain below and in the Supplemental Material (SM) \cite{SM5}, this protection can be mathematically understood as an emergent, deformed gauge symmetry \cite{Chubb2017}. 
	In surprising contrast to expectations from perturbation theory, our numerics suggests that the strength of the protection term does not need to scale with system size.

	Our results thus form a theoretical basis for some previous works, which have found evidence that quantum simulators may approximately retain gauge invariance \cite{Banerjee2012,Hauke2013,Kuehn2014,Negretti2017,Barros2019}. Moreover, our work complements existing results on equilibrium theories: 
	Gauge-invariant equilibrium phases can emerge in a low-energy effective theory, even if the microscopic description breaks gauge invariance, e.g., in topological phases of matter \cite{Hastings2005}, when the gauge degree of freedom decouples because of a large mass \cite{Wetterich2017}, or when gauge-noninvariant terms at a small scale renormalize away at large distances (``light from chaos") \cite{Foerster1980,Poppitz2008,Golterman2001}. 
	
	\textbf{\emph{Model.---}}Here, we focus on out-of-equilibrium dynamics, which is highly pertinent for current quantum simulators in ultracold atomic gases \cite{Schweizer2019,Goerg2019,Mil2019,Yang2019} and for which emergent gauge symmetry has received considerably less attention. Although our discussion is general, for concreteness, we consider a $\mathrm{Z}_2$ gauge theory \cite{Zohar2017,Barbiero2019,Frank2019,Borla2019} coupled to matter inspired by a recent experiment \cite{Schweizer2019}~--~we mostly consign similar results for a $\mathrm{U}(1)$ gauge theory to the SM \cite{SM0}. 
	
	The $\mathrm{Z}_2$ gauge theory lives on a lattice of $L$ matter sites in one spatial dimension \cite{footnote}, described by the Hamiltonian
	\begin{align}\label{eq:H0}
	H_0=&\,\sum_{j=1}^L\big[J_a\big(a^\dagger_j\tau^z_{j,j+1}a_{j+1}+\text{H.c.}\big)-J_f\tau^x_{j,j+1}\big],
	\end{align}
	with periodic boundary conditions. Here, $\tau_{j,j+1}^{\{x,z\}}$ are spin-$1/2$ Pauli matrices, where the $z$ ($x$) component stands for the $\mathrm{Z}_2$ gauge (electric) field on the link between matter sites $j$ and $j+1$. The matter fields are represented by hardcore bosons with the annihilation operator $a_j$ on site $j$. The dynamics of the matter field couples to the $\mathrm{Z}_2$ gauge field with strength $J_a$. The electric field has energy $J_f$. 
	
	Gauge invariance is encoded in a set of local symmetry generators
	\begin{align}\label{eq:Gauss}
	G_j=1-(-1)^j\tau^x_{j-1,j}Q_j\tau^x_{j,j+1},
	\end{align}
	with eigenvalues $g_j$ and where $Q_j=1-2a_j^\dagger a_j$ is the charge of the matter field. 
	As a $\mathrm{Z}_2$ lattice equivalent of Gauss's law, the $G_j$ commute with the model Hamiltonian $[H_0,G_j]=0$,~$\forall j$. 
	For a perfectly gauge-invariant system, the Hilbert space thus decouples into different symmetry sectors with fixed local charge $g_j=0,2$.   
	
	In the Standard Model of Particle Physics, such local gauge invariance is postulated at a fundamental level. 
	In a quantum simulator, however, gauge invariance needs to be engineered. 
	Some experiments have used Gauss's law to integrate out either matter or gauge fields, which yields an effective spin theory, suitable for implementation in a quantum computer and with encoded gauge invariance \cite{Martinez2016,Bernien2017,Kokail2019}. 
	However, this approach is viable only in one spatial dimension. 
	If, in contrast, both matter and gauge fields are retained as active degrees of freedom, such as in recent ultracold-atom  experiments~\cite{Schweizer2019,Mil2019,Yang2019}, 
	exact gauge invariance would require fine-tuning at unrealistic levels of accuracy. 
	Consequently, such quantum simulators will always have some inherent gauge invariance-breaking processes. 
	
	To study the severity of such terms, we add for concreteness the error Hamiltonian
	\begin{align}\nonumber
	\lambda H_1=\,\lambda \sum_{j=1}^L\Big[&\big(c_1a_j^\dagger\tau^+_{j,j+1} a_{j+1}+c_2a_j^\dagger \tau^-_{j,j+1} a_{j+1}+\text{H.c.}\big)\\\label{eq:H1}
	&\,+a_j^\dagger a_j\big(c_3\tau^z_{j,j+1}-c_4\tau^z_{j-1,j}\big)\Big].  
	\end{align}
	These terms are inspired by Ref.~\cite{Schweizer2019}, though our conclusions do not depend on their precise form. 
	Here, $\lambda$ is a parameter representing an adjustable gauge-noninvariant error strength and the constants $c_l$ are modeled after experimental parameters \cite{SM3}. Generically, the error term of Eq.~\eqref{eq:H1} will drive the dynamics out of the gauge-invariant subspace, such that after a certain time the expectation value of the Gauss-law operator will be the same as in a random state from the entire Hilbert space (i.e., the space that contains gauge-invariant states as well as all states with $G_j\ket{\psi}\neq 0$).
	
	\begin{figure}[t!]
		\centering
		\hspace{-.2 cm}
		\includegraphics[width=.49\textwidth]{{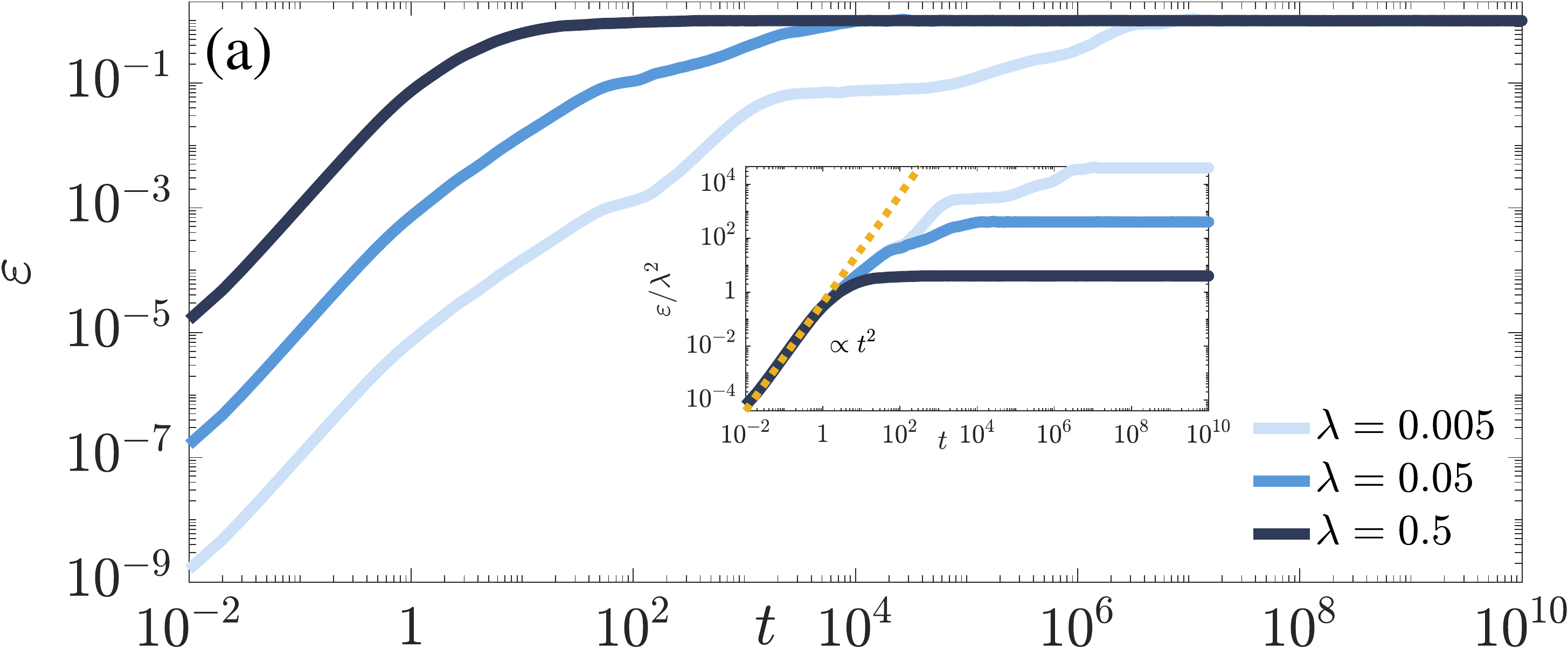}}\quad
		\hspace{-.2 cm}
		\includegraphics[width=.49\textwidth]{{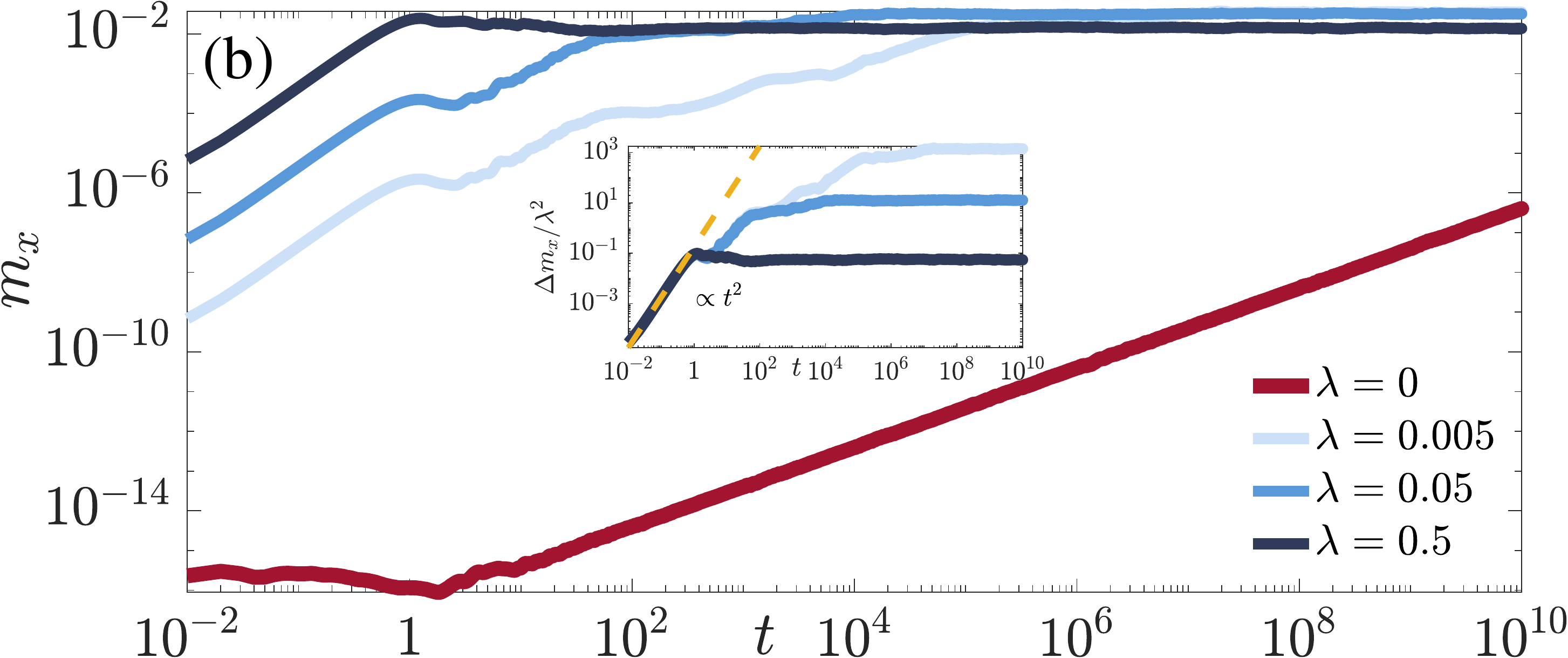}}\quad
		\hspace{-.2 cm}
		\includegraphics[width=.49\textwidth]{{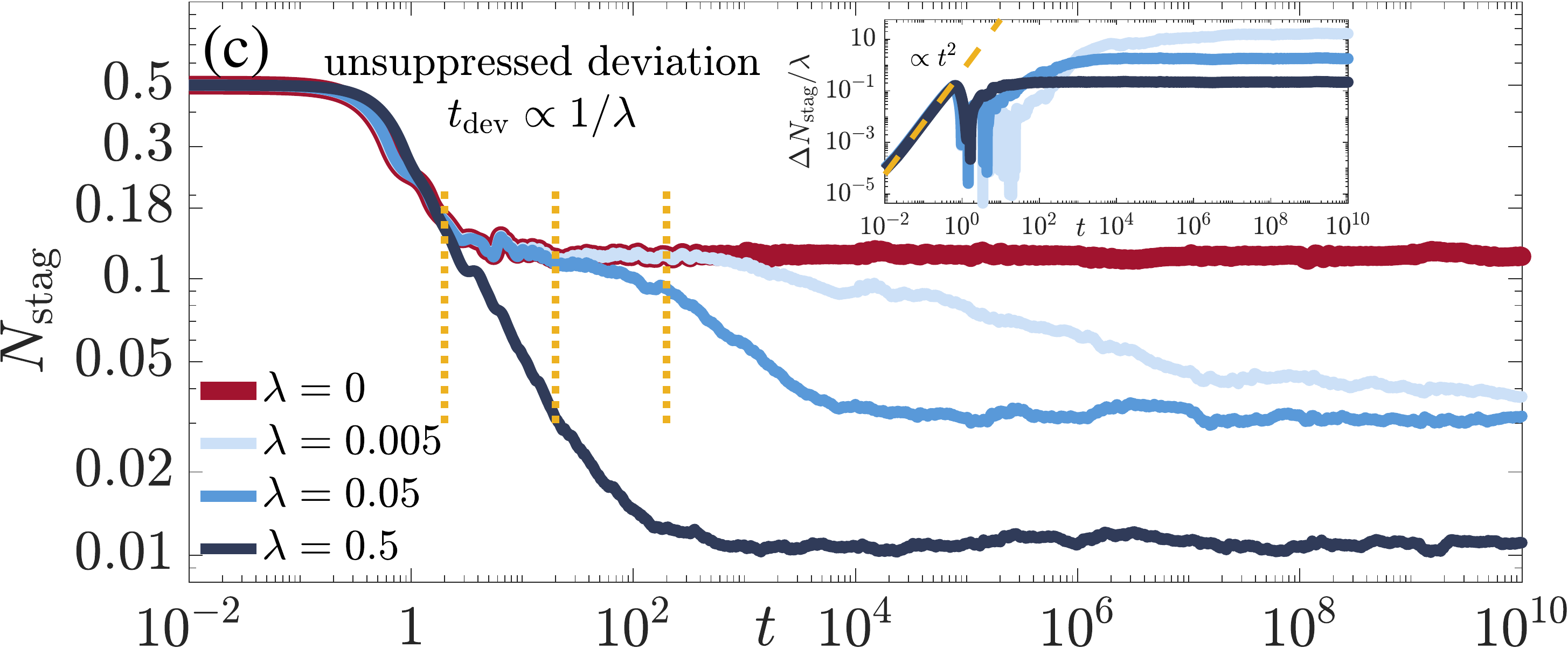}}\quad
		\hspace{-.2 cm}
		\caption{(Color online). 
			Effect of gauge invariance-breaking terms on the dynamics of a $\mathrm{Z}_2$ gauge theory. Spatiotemporal averages of (a) the
			gauge-invariance violation in Eq.~\eqref{eq:error}, (b) the electric field in Eq.~\eqref{eq:mag}, and (c) the staggered boson number in Eq.~\eqref{eq:nstag}, deviate only gradually from the ideal dynamics, before gauge-noninvariant effects begin to dominate after a timescale of $t_\mathrm{dev}\propto 1/\lambda$. Insets: Rescaled deviations from ideal dynamics, showing the perturbative growth with $\lambda$. Data for $L=6$ matter sites. See SM \cite{SM4} for further information.
		} 
		\label{fig:Fig2} 
	\end{figure}
	
	\begin{figure}[t!]
		\centering
		\hspace{-.25 cm}
		\includegraphics[width=.49\textwidth]{{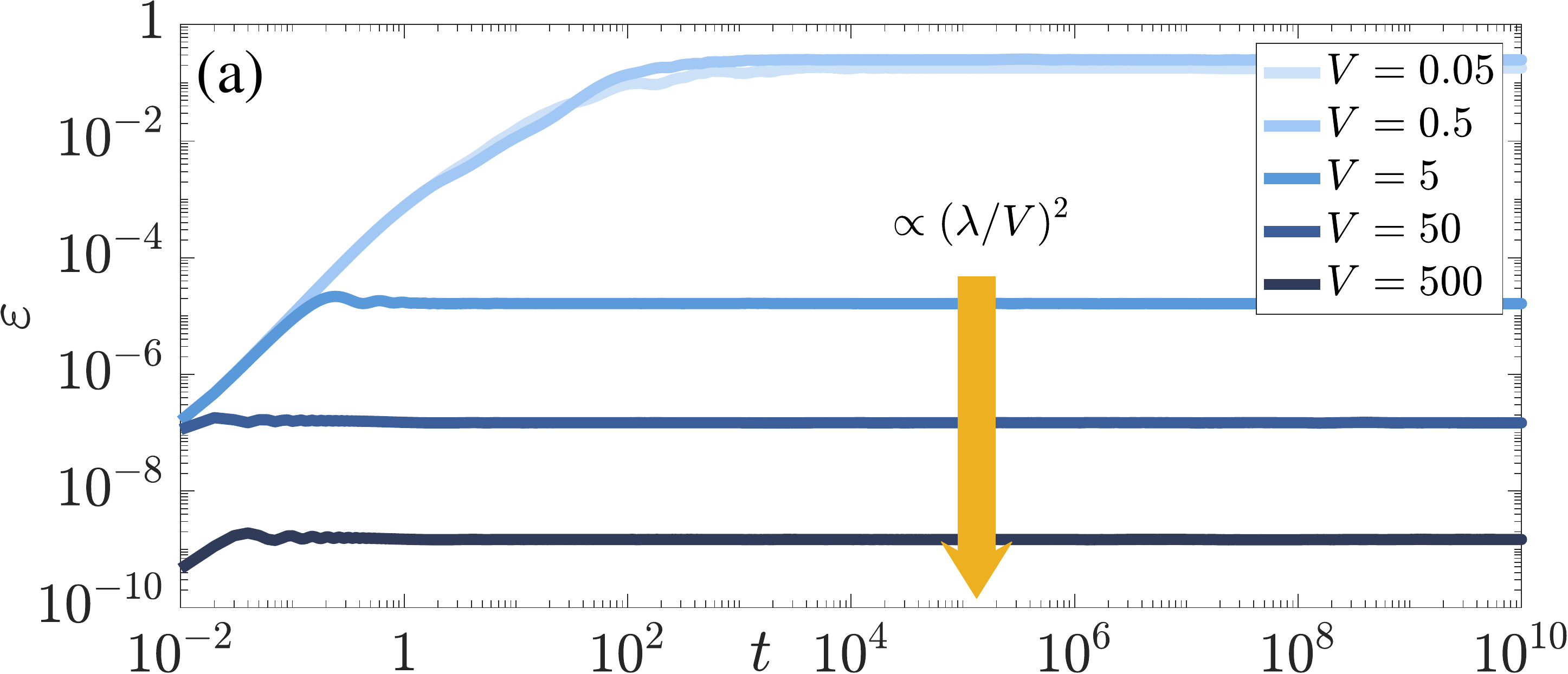}}\\
		\hspace{-.25 cm}
		\includegraphics[width=.49\textwidth]{{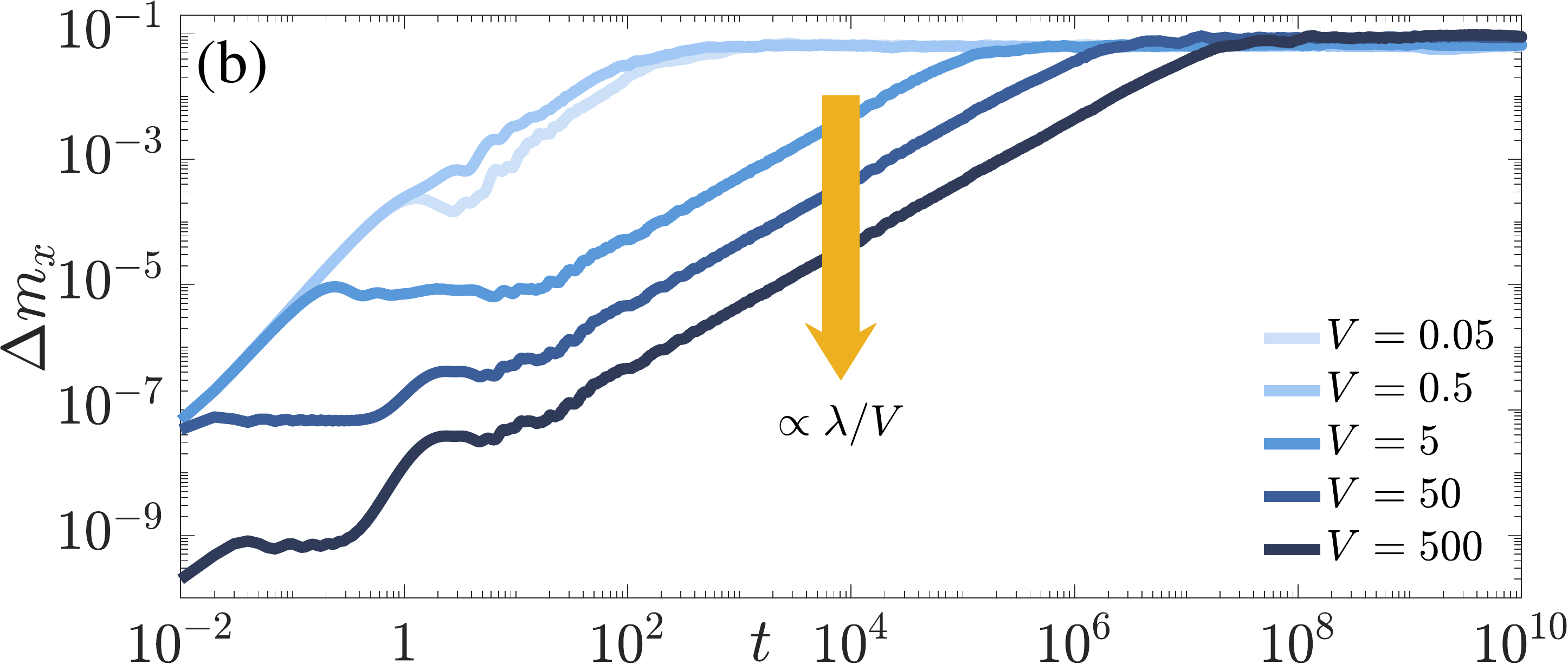}}\\
		\hspace{-.25 cm}
		\includegraphics[width=.49\textwidth]{{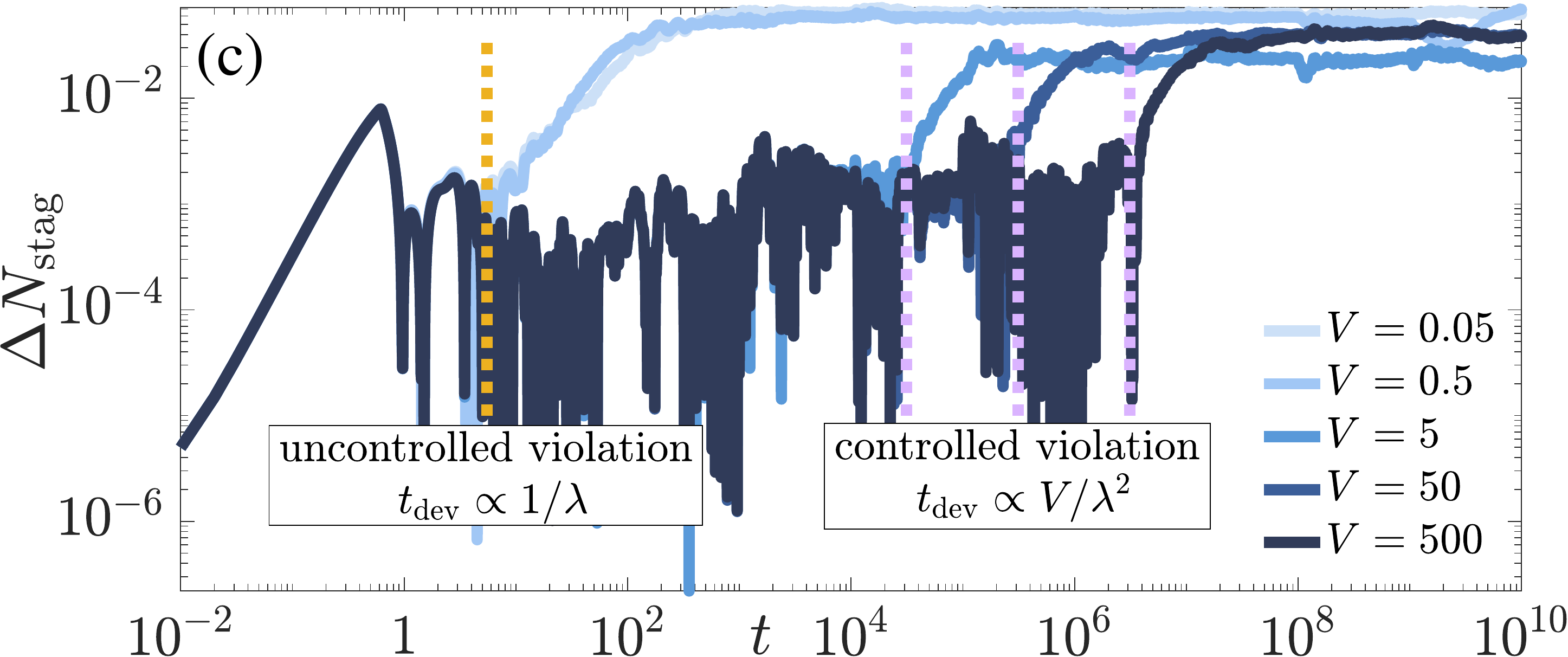}}\\
		\hspace{-.25 cm}
		\caption{(Color online). Dynamics of the spatiotemporal averages of (a) the
			gauge-invariance violation in Eq.~\eqref{eq:error}, (b) the electric field in Eq.~\eqref{eq:mag}, and (c) the staggered boson number in Eq.~\eqref{eq:nstag} at error strength $\lambda=0.05$ and at various values of the protection strength $V$ for $L=6$ matter sites. We see two clear regimes: at small $V$ the gauge-noninvariant behavior dominates after a timescale $\propto 1/\lambda$. When $V$ is sufficiently large (controlled-violation regime), the gauge-invariance violation in (a) is indefinitely suppressed by $(\lambda/V)^2$, while deviations in the electric field in (c) are suppressed by $\lambda/V$ up to a timescale $\propto V/\lambda^2$. The deviation in the staggered boson number becomes uncontrolled only at times beyond $t\propto V/\lambda^2$. See SM \cite{SM4} for corresponding results at $\lambda=0.005$ and $0.5$.
		} 
		\label{fig:Fig3} 
	\end{figure}
	
	\textbf{\emph{Quench dynamics with gauge violation.---}}We mimic a typical quantum simulator experiment, where the system is initialized in a simple product state $\ket{\psi_0}$ in the gauge-invariant sector $G_j\ket{\psi_0}=0$, $\forall j$, which is then quenched with the Hamiltonian $H_0+\lambda H_1$. 
	We choose $\ket{\psi_0}$ such that $\langle a^\dagger_ja_j\rangle=[1+(-1)^j]/2$ and $\langle \tau^x_{j,j+1}\rangle=(-1)^{j+1}$, and, following Ref.~\cite{Schweizer2019}, we set $J_a=1$ and $J_f=0.54$ (our conclusions do not depend on these precise choices). 
	Our numerical results are obtained using the QuTiP \cite{Johansson2012,Johansson2013} and QuSpin \cite{Weinberg2017,Weinberg2019} exact diagonalization toolkits. For time evolution, we have opted to use our own exact exponentation routine in lieu of these toolkits' time-evolution functions, given the extremely long times we simulate, for which solvers based on integrating ordinary differential equations are not optimal.
	
	To evaluate the effect of $H_1$ on the gauge-invariant dynamics, we consider the dynamics of the spatiotemporal averages of the violation of Gauss's law, 
	\begin{equation}\label{eq:error}
	\varepsilon(t)=\frac{1}{Lt}\int_0^t\d s\, \sum_{j=1}^L\bra{\psi(s)}G_j\ket{\psi(s)},
	\end{equation}
	the magnetization in the $x$ direction (the `electric field')
	\begin{align}\label{eq:mag}
	m_{x}(t)=\frac{1}{Lt}\int_0^t\d s\,\Big|\sum_{j=1}^L\bra{\psi(s)}\tau^{x}_{j,j+1}\ket{\psi(s)}\Big|,
	\end{align} 
	as well as the staggered boson number
	\begin{equation}\label{eq:nstag}
	N_\text{stag}(t)=\frac{1}{Lt}\int_0^t\d s\,\Big| \sum_{j=1}^L(-1)^j\bra{\psi(s)}a_j^\dagger a_j\ket{\psi(s)}\Big|.
	\end{equation}
	In the above, $\ket{\psi(s)}=\exp[-\mathrm{i}(H_0+\lambda H_1)s]\ket{\psi_0}$. 
	The respective deviations from the ideal gauge-invariant case are denoted by $\Delta N_\text{stag}(t)$ and $\Delta m_x(t)$. 
	Further observables are discussed in the SM \cite{SM4}. Note that the gauge violation in Eq.~\eqref{eq:error} suffices for the $\mathrm{Z}_2$ gauge theory, but would require an even power in $G_j$ or its absolute value for the $\mathrm{U}(1)$ gauge theory \cite{SM0}.

	Our numerical results are presented in Fig.~\ref{fig:Fig2}. 
	The gauge violation grows only gradually, $\varepsilon(t)\sim(\lambda t)^2$ at short times, before it saturates at long times. This subleading increase of the gauge violation directly stems from the fact that $\sum_j G_j$ is gauge invariant and also commutes with $H_0$, which necessarily leads to a vanishing first-order contribution in perturbation theory; cf.~SM \cite{SM1}. 
	In contrast, the short-time scaling behavior can change for gauge-invariant observables that do not commute with $H_0$, such as $N_\text{stag}$ and $m_x$. Even though $\Delta m_x$ scales as $(\lambda t)^2$ at short times, as seen in the inset of Fig.~\ref{fig:Fig2}(b), $\Delta N_\text{stag}$ shown in the inset of Fig.~\ref{fig:Fig2}(c) scales instead as $\lambda t^2$, with the latter emenating from the first-order term in perturbation theory; cf.~SM \cite{SM1}.  
	Importantly, $N_\text{stag}(t)$ remains nevertheless close to the gauge-invariant dynamics up to $t\approx1/\lambda$. 
	Thus, there is a clear time frame over which gauge-noninvariant terms do not compromise observable properties, and by decreasing $\lambda$ this time frame can be improved in a controlled manner.
	
	\textbf{\emph{Quench dynamics with energy protection.---}}
	Some promising proposals have suggested to perturbatively generate the desired gauge-theory Hamiltonian by adding a term proportional to Gauss's law, which energetically penalizes violations of gauge invariance \cite{Zohar2015,Dalmonte2016,Hauke2013}. 
	A similar term has been discussed for protecting stored quantum information \cite{Jiang2015}.
	We now systematically address the question of how such a term restores the ability to quantum simulate gauge-invariant dynamics.  
	
	\begin{figure}[t!]
		\centering
		\hspace{-.25 cm}
		\includegraphics[width=.49\textwidth]{{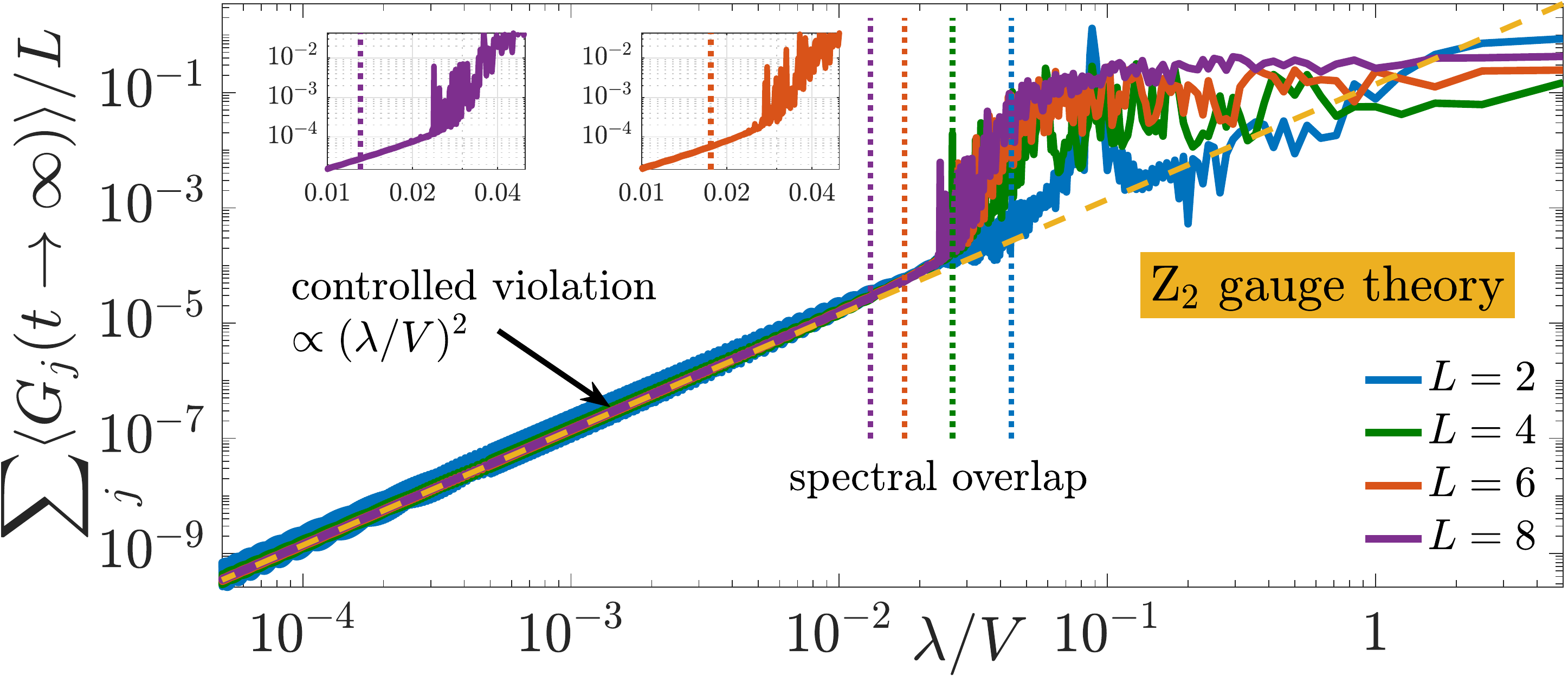}}\\
		\hspace{-.25 cm}
		\includegraphics[width=.49\textwidth]{{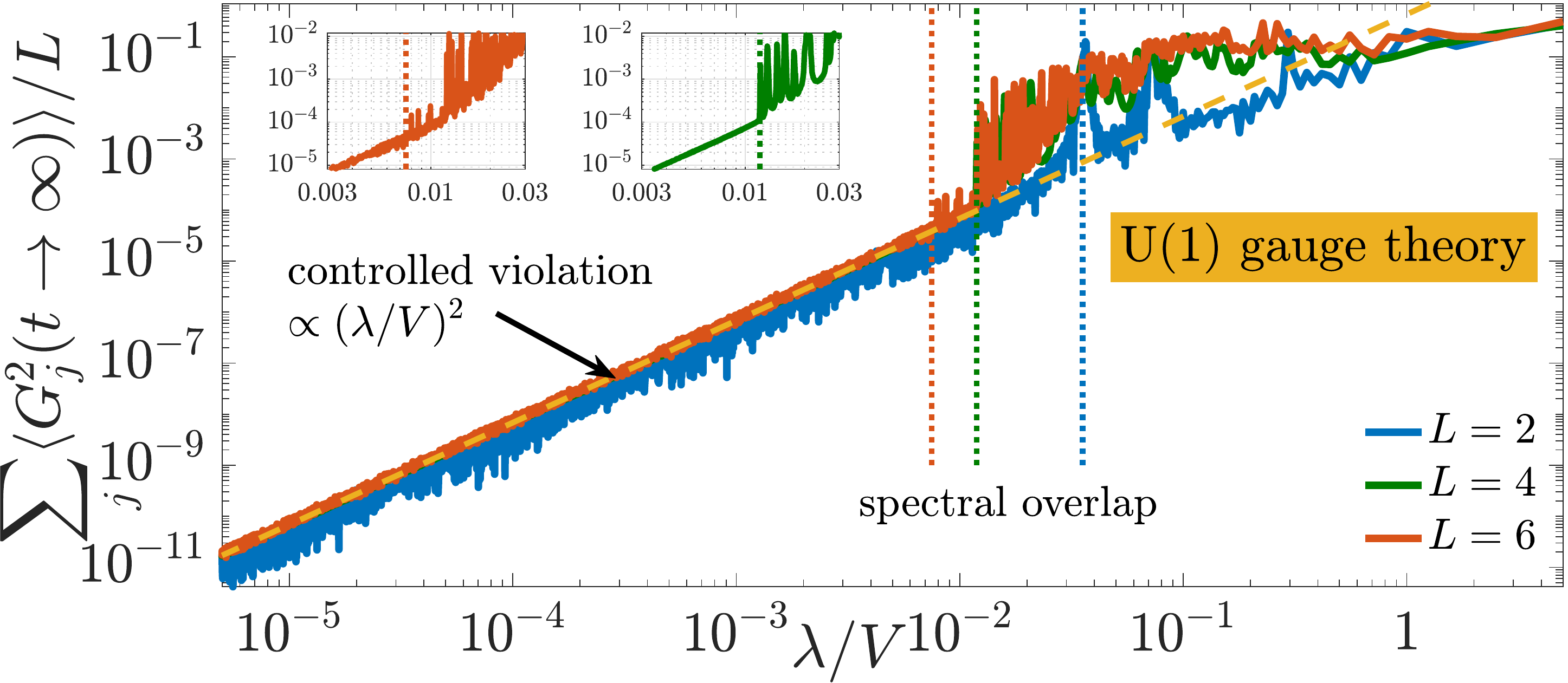}}\quad
		\hspace{-.25 cm}
		\caption{(Color online). Infinite-time violation $\sum_j\langle G_j(t\to\infty)\rangle/L$ in the $\mathrm{Z}_2$ gauge theory (top panel) and $\sum_j\langle G_j^2(t\to\infty)\rangle/L$ in the $\mathrm{U}(1)$ gauge theory (bottom panel) as a function of $\lambda/V$ for $\lambda=0.05$, computed for several system sizes $L$. We see two separate regimes where at large $V$ the violation is controlled and scales as $(\lambda/V)^2$, while at small $V$ the violation is uncontrolled. The dotted lines indicate the largest $V$ at which there is an energy overlap between the initial gauge-invariant sector, and other symmetry sectors, which we protect against. In the controlled-violation regime, the violation is system size-independent. The vertical dotted lines correspond to the largest value of $V$ (for a given system size $L$) at which the initial gauge-invariant sector still has spectral overlap with other sectors.} 
		\label{fig:scaling} 
	\end{figure}
	
	To analyze this scenario, we introduce a protection term 
	\begin{align}\label{eq:protection}
	VH_G=V\sum_jG_j,
	\end{align}
	which energetically penalizes all violations of $G_j\ket{\psi_0}=0$, with adjustable protection strength $V$. Note that Eq.~\eqref{eq:protection} is not sufficient for the $\mathrm{U}(1)$ gauge theory, and should rather be $VH_G=V\sum_jG_j^2$ \cite{SM0}. Our quench Hamiltonian is thus given by $H=H_0+\lambda H_1+VH_G$. Consequently, in Eqs.~\eqref{eq:error},~\eqref{eq:mag}, and~\eqref{eq:nstag} we now have $\ket{\psi(s)}=\exp[-\mathrm{i}(H_0+\lambda H_1+VH_G)s]\ket{\psi_0}$. 
	In Fig.~\ref{fig:Fig3}, we show the effect of $V$ on $\varepsilon(t)$, $\Delta N_\text{stag}(t)$, and $\Delta m_x(t)$. 
	As can be shown in degenerate perturbation theory \cite{SM2}, the violation $\varepsilon(t)$ of Gauss's law is suppressed by $(\lambda/V)^2$ at sufficiently large $V$; cf.~SM \cite{SM4}. One might suspect that such an energy penalty would only suppress gauge-noninvariant processes at short times before the system eventually accesses all possible gauge-invariant sectors at long times. Interestingly, however, once $V$ is on the order of a few $\lambda$, a ``controlled-violation'' regime is reached, where $\varepsilon(t)$ is protected for all simulated times (which are many orders of magnitude larger than what is reachable in current experiments). 
	
	In contrast, for sufficiently large $V$ the deviation of $m_{x}(t)$ is suppressed only by $\lambda/V$, as shown in Fig.~\ref{fig:Fig3}(b), with the 
	suppression ceasing at a timescale $\propto V/\lambda^2$; cf.~SM~\cite{SM4}.
	At the same timescale, $N_\text{stag}$ starts deviating from the gauge-invariant case if the values of $V$ lie in the controlled-violation regime; see Fig.~\ref{fig:Fig3}(c). Furthermore, the terms that drive these deviations can be described as gauge-invariant processes in degenerate perturbation theory, and could thus be absorbed into a renormalized, gauge-invariant $H_0$. See SM \cite{SM2} for more details. 
	The existence of these timescales is a considerable improvement, since outside the controlled-violation regime deviations from the ideal gauge-invariant dynamics proliferate already at an earlier timescale $\propto1/\lambda$.

	Physically, $V$ protects gauge invariance by opening a large energy gap between the sector $G_j\ket{\psi}=0$, $\forall j$, and all other sectors. Since the bandwidth of each sector increases with system size, one could expect the necessity to scale $V$ with $L$, which would invalidate this protection mechanism for large-scale quantum simulators. 
	However, our numerics suggest that this is not the case. We illustrate this in Fig.~\ref{fig:scaling} for the infinite-time per-site violation of Gauss's law as a function of $\lambda/V$ (with fixed $\lambda=0.05$). 
	We see two clear regimes. 
	The first at small values of $V$ shows an uncontrolled violation that heavily depends on how the different gauge-invariant sectors are coupled to one another. 
	The second regime, however, displays a controlled violation that scales as $(\lambda/V)^2$. 
	This regime is expected once the gauge-invariant sector we start in is energetically well separated from other sectors, as can be shown in degenerate perturbation theory and made mathematically rigorous by adapting the results of Ref.~\cite{Chubb2017} (see SM \cite{SM2}): 
	For any unitary symmetry that is broken on a scale $\sim\lambda$, the opening of a gap generates an emergent symmetry that is perturbatively in $(\lambda/V)^2$ close to the original one.  
	Surprisingly, however, the scaling as $(\lambda/V)^2$ sets in much before full separation between sectors is achieved. 
	This onset appears to be largely independent of system size, contrary to the analytic arguments based on perturbation theory and the emergent gauge symmetry (see SM \cite{SM2}). 
	The reason is that the relevant gap to gauge-violating sectors is not to be counted relative to the entire gauge-invariant sector, but only to the energy region that is populated during the quench. 
	A similar effect has been observed for the robustness of the ground-state degeneracy in topological matter \cite{Hastings2005}.
	
	Our results in Figs.~\ref{fig:Fig3} and~\ref{fig:scaling} illustrate how once the controlled-violation regime is reached at a sufficiently large $V$, extrapolations to the ideal case become possible in both the gauge-invariance violation and gauge-invariant observables. Indeed, Fig.~\ref{fig:Fig3} shows a clear timescale before which deviations in a gauge-invariant observable are well-determined as a function of $\lambda$ and $V$. Even better, in this regime the control in the gauge-invariance violation is not limited by any timescale, but rather persists indefinitely $\propto (\lambda/V)^2$.

	\textbf{\emph{Conclusions and outlook.---}}We have carried out a thorough analysis, through exact diagonalization and perturbation theory, of the reliability of lattice gauge theories in out-of-equilibrium dynamics. 
	We have found that small gauge-nonivariant processes (of strength $\lambda$) do not compromise the desired dynamics of observables up to a clear time frame $\sim \lambda^{-1}$. 
	Moreover, when introducing a sufficiently expensive energy penalty of strength $V$ for such processes, the gauge-invariance violation enters a \textit{controlled-violation regime} where it scales as $(\lambda/V)^2$, up to infinite times, and is also robust with respect to system size.
	This is a very encouraging result for experimental efforts on quantum simulators as it indicates that introducing an energy penalty leads to an indefinite protection of gauge invariance. This enables a well-defined extrapolation of a perfect gauge theory from gauge-invariance-violating data.
	
	The suppression in observables' deviations from their gauge-invariant dynamics presents a more varied picture. This is to be expected because when a gauge theory is broken by a small parameter, another gauge theory perturbatively close to it emerges \cite{Chubb2017} that, even though gauge-invariant, is still different from its initial counterpart. The original theory's exact dynamics can, however, be recovered by an appropriate absorption of the new terms into renormalized parameters \cite{Foerster1980,Poppitz2008}.

	\textbf{\emph{Acknowledgements.---}}The authors are grateful to Z.~Jiang and G.~Morigi for useful comments, and to J.~Berges, F.~Jendrzejewski, R.~Ott, B.~Yang, and T.~Zache, for stimulating discussions and collaboration on related work. J.C.H.~thanks J.~C.~Louw for interesting discussions related to exact diagonalization.
	This work is part of and supported by the DFG Collaborative Research Centre SFB 1225 (ISOQUANT), the Provincia Autonoma di Trento, and the ERC Starting Grant StrEnQTh (Project-ID 804305).

\newpage
\bigskip
\pagebreak
\widetext
\begin{center}
	\textbf{\large --- Supplemental Material ---\\ Reliability of lattice gauge theories}\\
	\medskip
	\text{Jad C.~Halimeh and Philipp Hauke}
\end{center}
\setcounter{equation}{0}
\setcounter{figure}{0}
\setcounter{table}{0}
\makeatletter
\renewcommand{\theequation}{S\arabic{equation}}
\renewcommand{\thefigure}{S\arabic{figure}}
\renewcommand{\bibnumfmt}[1]{[S#1]}
\renewcommand{\citenumfont}[1]{S#1}
\newcommand{\rmf}{{\rm f}}

In this Supplemental Material, we discuss our analysis of the $\mathrm{U}(1)$ gauge theory, which supports our conclusions on the $\mathrm{Z}_2$ gauge theory discussed in the main text. Moreover, we add supplemental results on the $\mathrm{Z}_2$ gauge theory that further substantiate our conclusions. Further, we present details of our perturbation theory derivations, in addition to analytic arguments demonstrating the emergent gauge theory in the controlled-violation regime.

\section{Supplemental results for $\mathrm{Z}_2$ gauge theory in support of main conclusions}
\begin{figure}[!ht]
	\centering
	\hspace{-.01 cm}
	\vspace{-.01 cm}
	\includegraphics[width=.45\textwidth]{{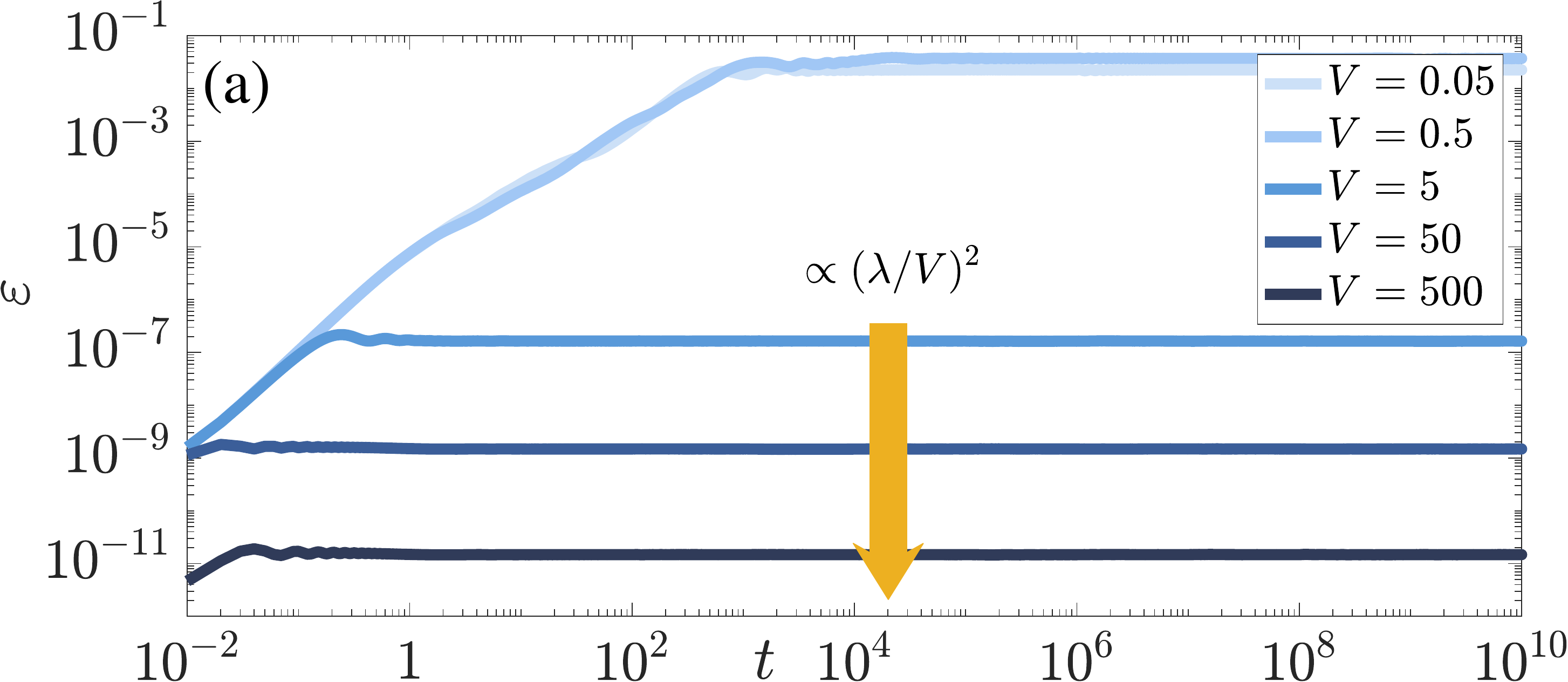}}\quad
	\vspace{-.01 cm}
	\includegraphics[width=.45\textwidth]{{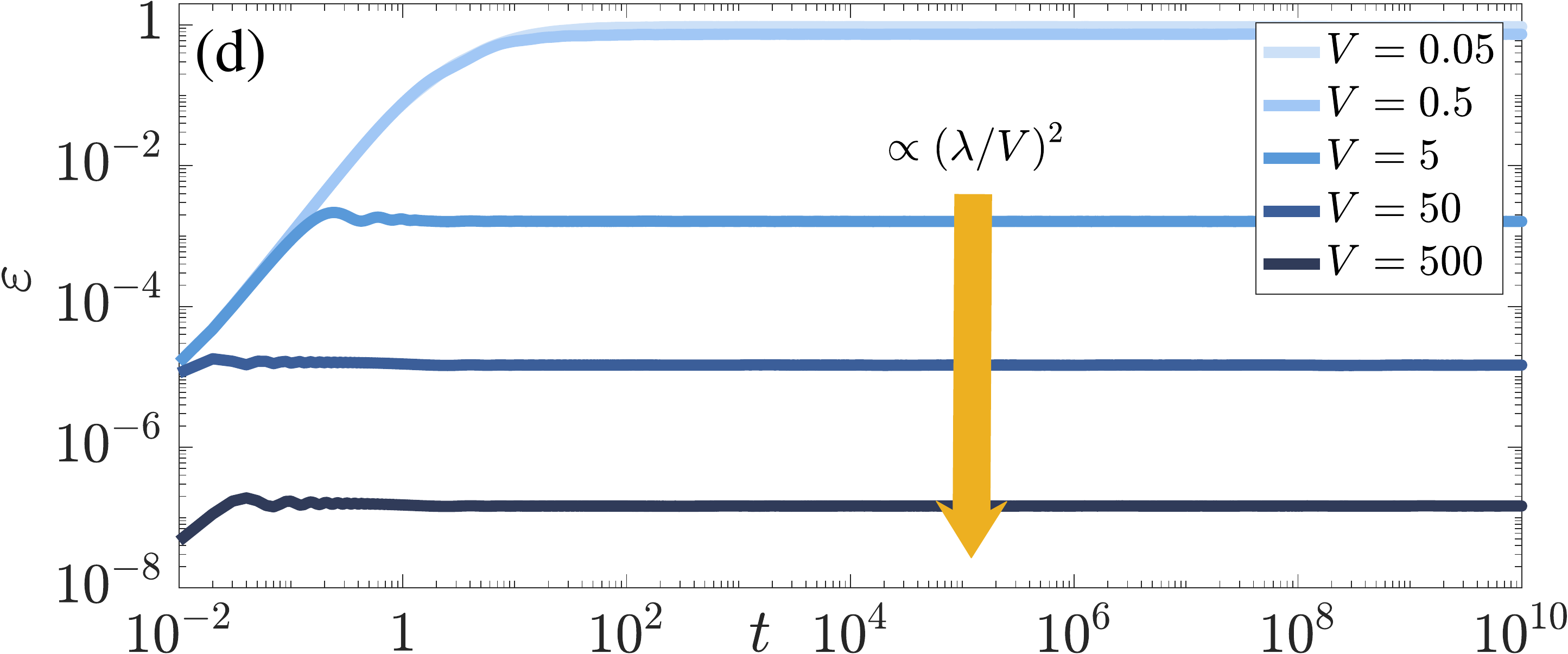}}\quad
	\vspace{-.01 cm}\\
	\hspace{-.01 cm}
	\includegraphics[width=.45\textwidth]{{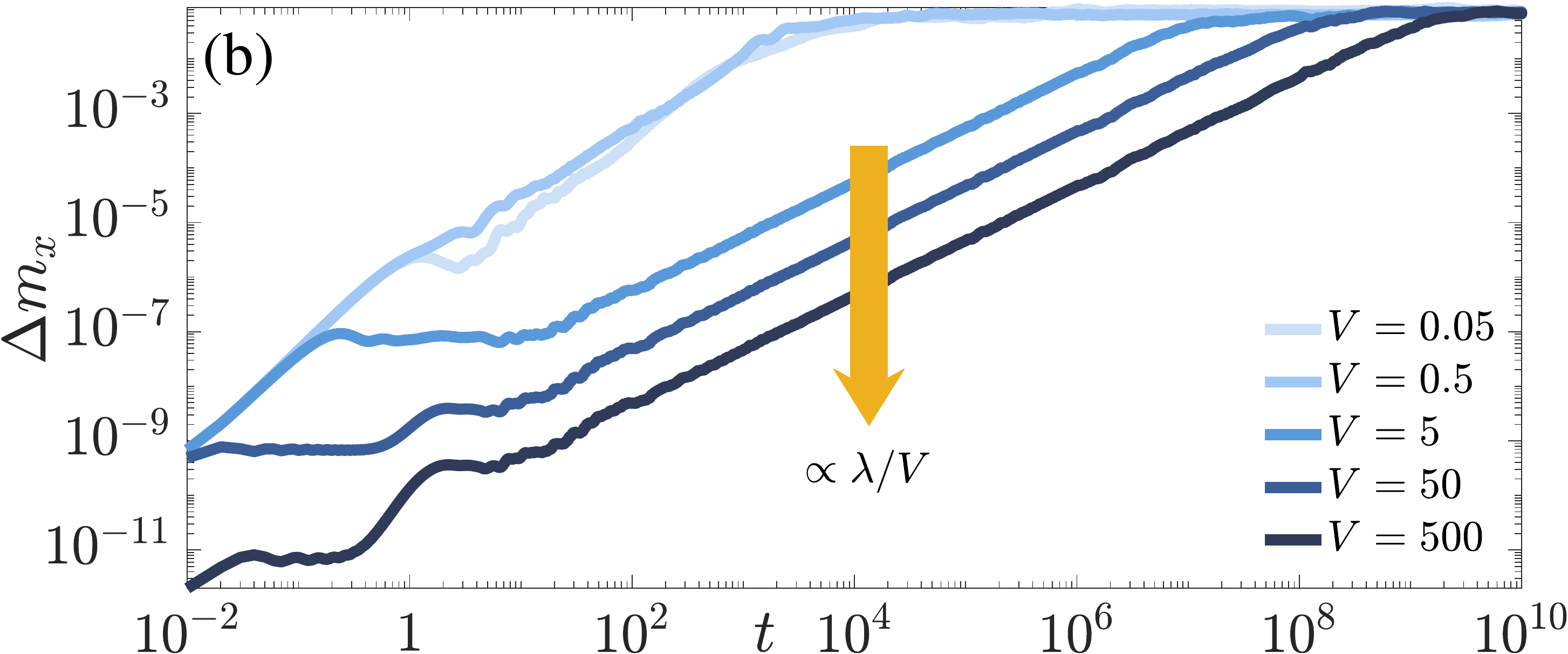}}\quad
	\vspace{-.01 cm}
	\includegraphics[width=.45\textwidth]{{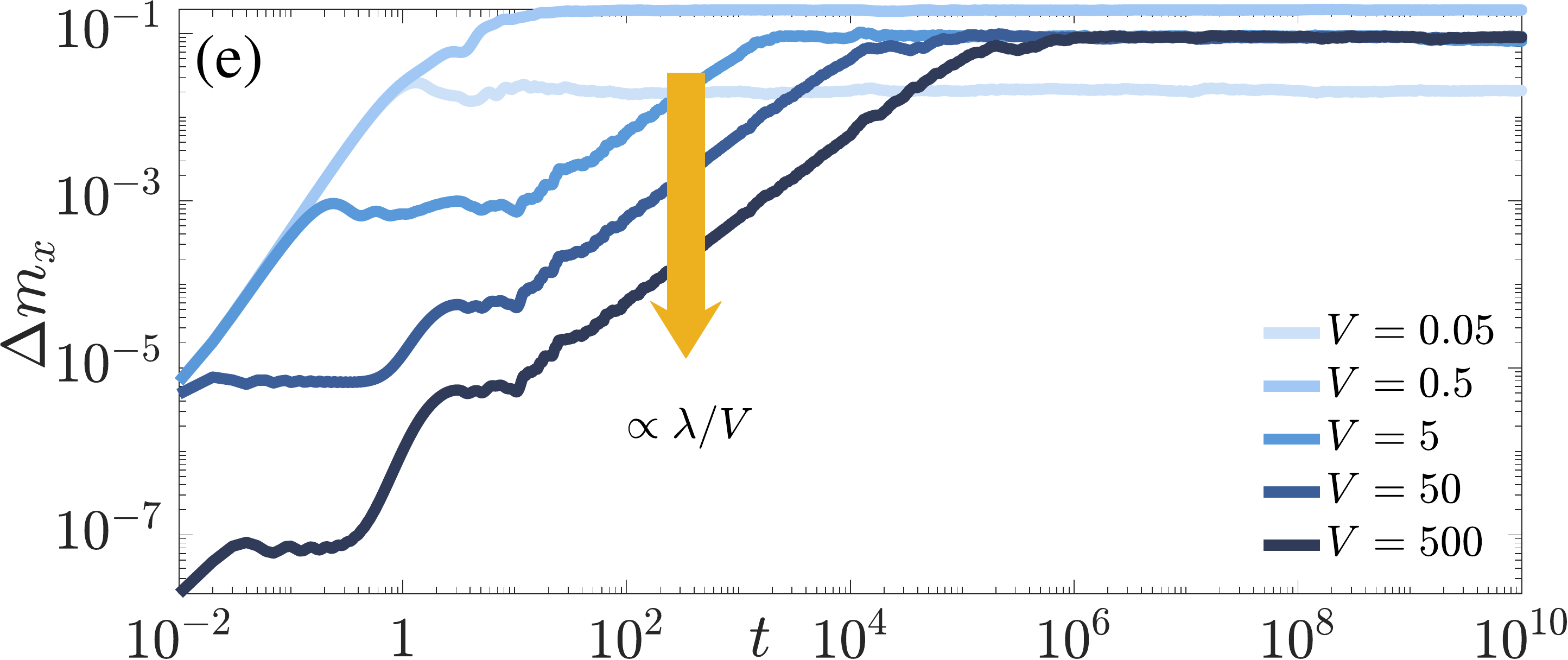}}\quad
	\vspace{-.01 cm}\\
	\hspace{-.01 cm}
	\includegraphics[width=.45\textwidth]{{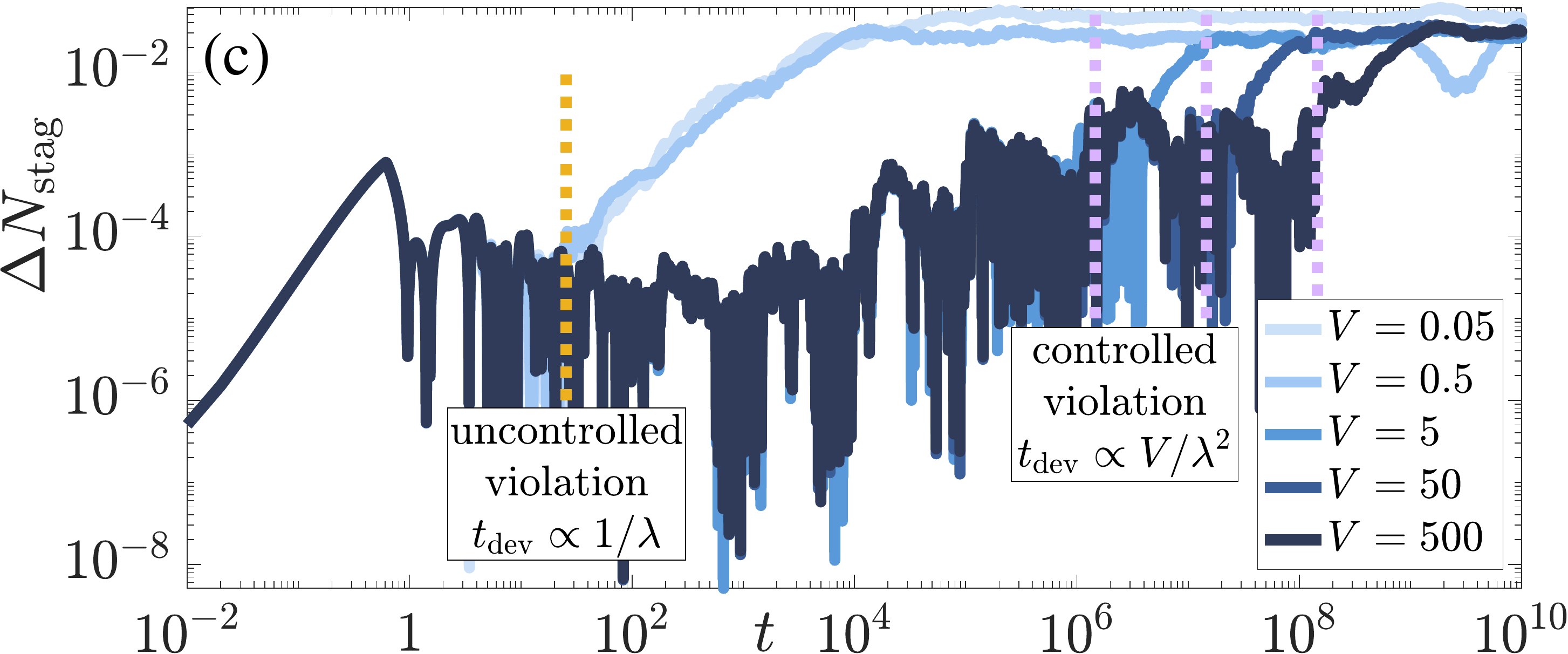}}\quad
	\vspace{-.01 cm}
	\includegraphics[width=.45\textwidth]{{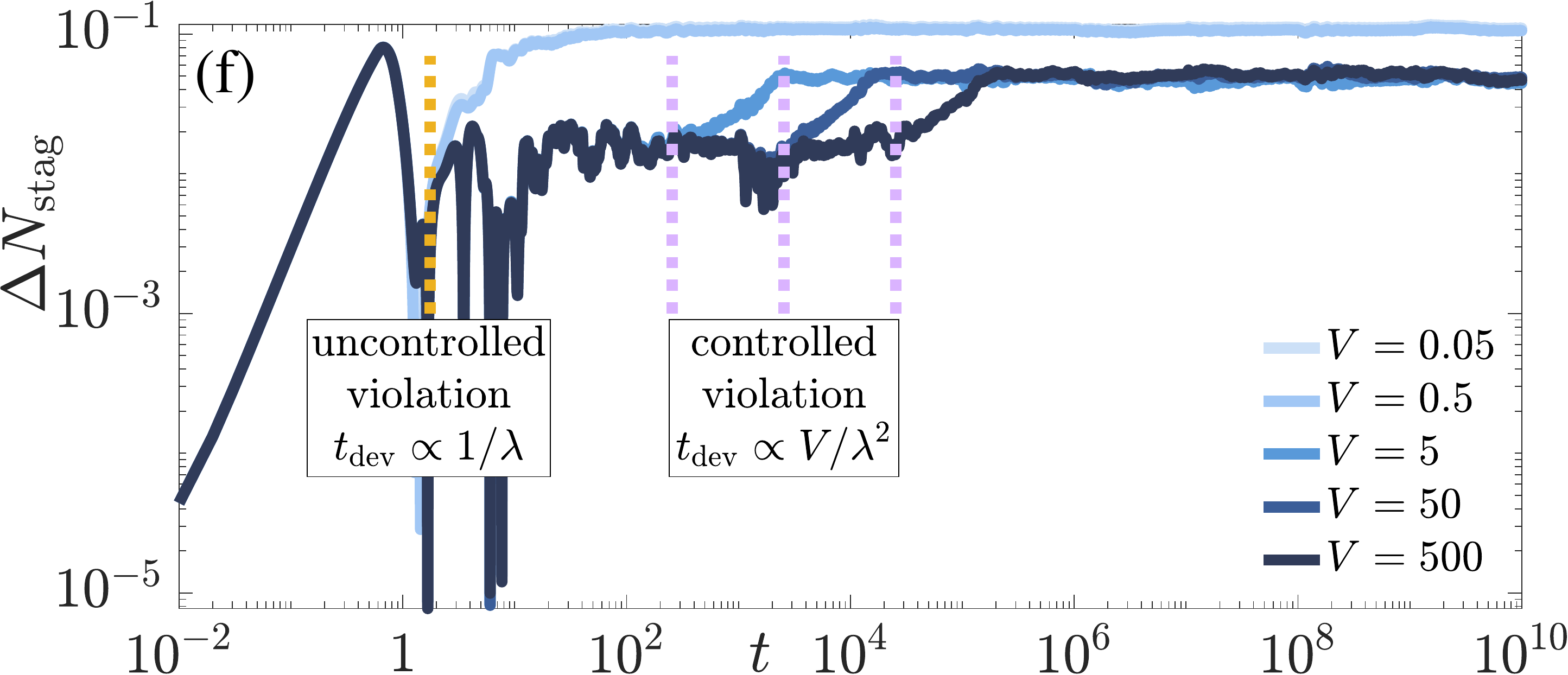}}\quad
	\vspace{-.01 cm}
	\hspace{-.01 cm}
	\caption{(Color online). Same as Fig.~\ref{fig:Fig3} of the main text, but for gauge invariance-breaking strength of $\lambda=0.005$ (left-column panels) and $\lambda=0.5$ (right-column panels). The qualitative picture is unchanged form that of the main text.} 
	\label{fig:Fig3S} 
\end{figure}

In the main text, we present in Fig.~\ref{fig:Fig3} results for the quench dynamics in the $\mathrm{Z}_2$ gauge theory of various observables upon adding a protection term in the Hamiltonian in order to mitigate gauge invariance-breaking processes of strength $\lambda=0.05$. In Fig.~\ref{fig:Fig3S} we repeat these results for two more values of $\lambda=0.005$ and $0.5$. These results affirm the conclusions of the main text, and further corroborate the scaling of the violation and the suppression of the deviation of observables from their ideal gauge-invariant dynamics. 

In the main text, we have resorted to using the spatiotemporal average $\varepsilon(t)$ of the gauge-invariance violation $\sum_j\langle G_j(t)\rangle$ given in Eq.~\eqref{eq:error}. Another, more stringent, way to measure gauge-invariance breaking is through the projector
\begin{align}\label{eq:proj}
P_0&=\bigotimes_{j=1}^LP_{j,0},\\
P_{j,0}&=\sum_q\ket{j,0;q}\bra{j,0;q},
\end{align}
where $\ket{j,0;q}$ is an eigenstate of the local gauge-invariance operator $G_j$ with eigenvalue $0$, and where $q$ denotes all remaining good quantum numbers. Now the violation in gauge invariance can be alternatively measured as
\begin{align}\label{eq:errorproj}
\varepsilon_\text{proj}(t)=1-\frac{1}{t}\int_0^t\d s\,\bra{\psi(s)}P_0\ket{\psi(s)}.
\end{align}
When the initial state $\ket{\psi_0}$ is quenched by $H_0+\lambda H_1$,~i.e., when $\ket{\psi(s)}=\exp[-\mathrm{i}(H_0+\lambda H_1)s]\ket{\psi_0}$, the dynamics of $\varepsilon_\text{proj}(t)$ is qualitatively identical to that of $\varepsilon(t)$. After an initial growth that scales as $(\lambda t)^2$ at short times, $\varepsilon_\text{proj}(t)$ saturates at a timescale of $1/\lambda^2$, as shown in Fig.~\ref{fig:FigPG}. Upon adding an energy protection term $VH_G$,~i.e., $\ket{\psi(s)}=\exp[-\mathrm{i}(H_0+\lambda H_1+VH_G)s]\ket{\psi_0}$ in Eq.~\eqref{eq:errorproj}, we find that for sufficiently large $V$ there is a controlled-violation regime, where the violation is suppressed as $(\lambda/V)^2$ up to infinite times. Indeed, as shown in Fig.~\ref{fig:FigPG}, the scaling behavior of the gauge-invariance violation as measured by $1-\langle P_0(t\to\infty)\rangle$ is qualitatively identical to that measured by $\sum_j\langle G_j(t\to\infty)\rangle/L$ in Fig.~\ref{fig:scaling} of the main text.

\begin{figure}[t!]
	\centering
	\hspace{-.25 cm}
	\includegraphics[width=.49\textwidth]{{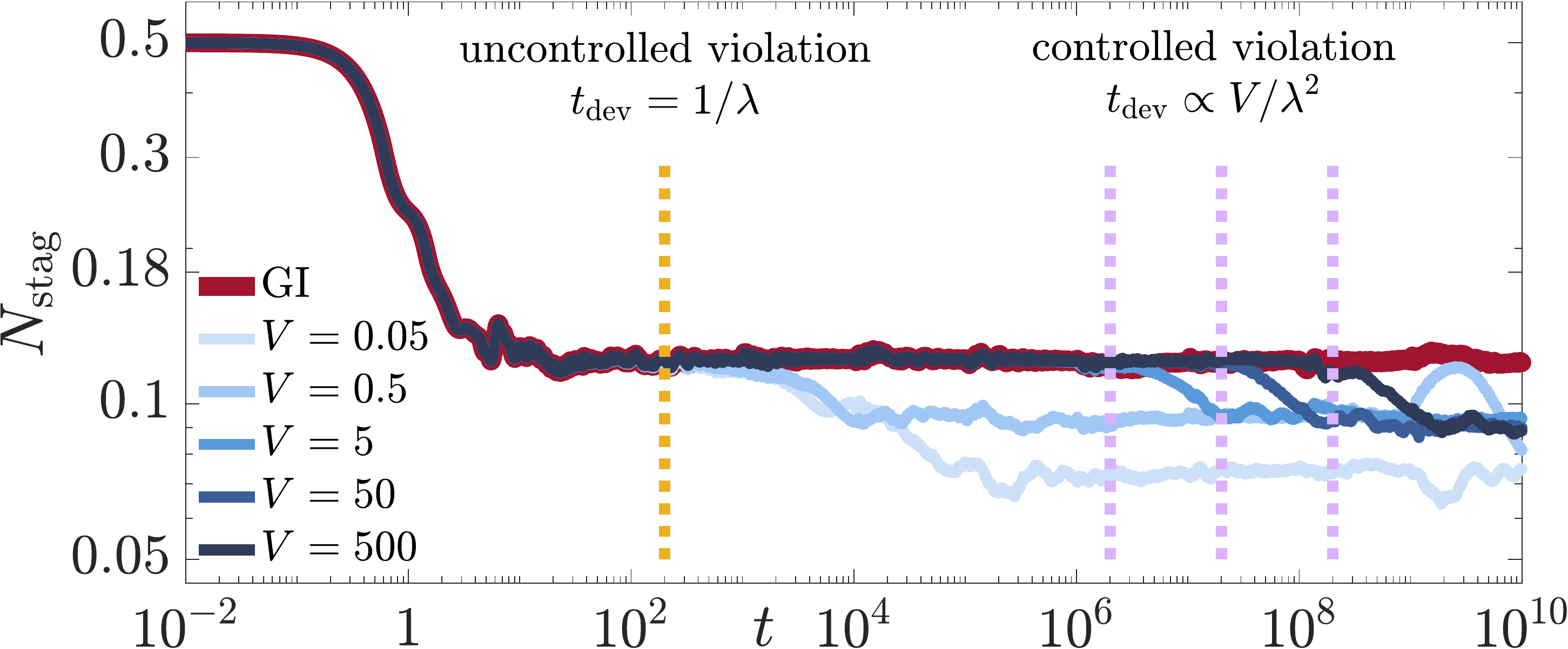}}\\
	\hspace{-.25 cm}
	\includegraphics[width=.49\textwidth]{{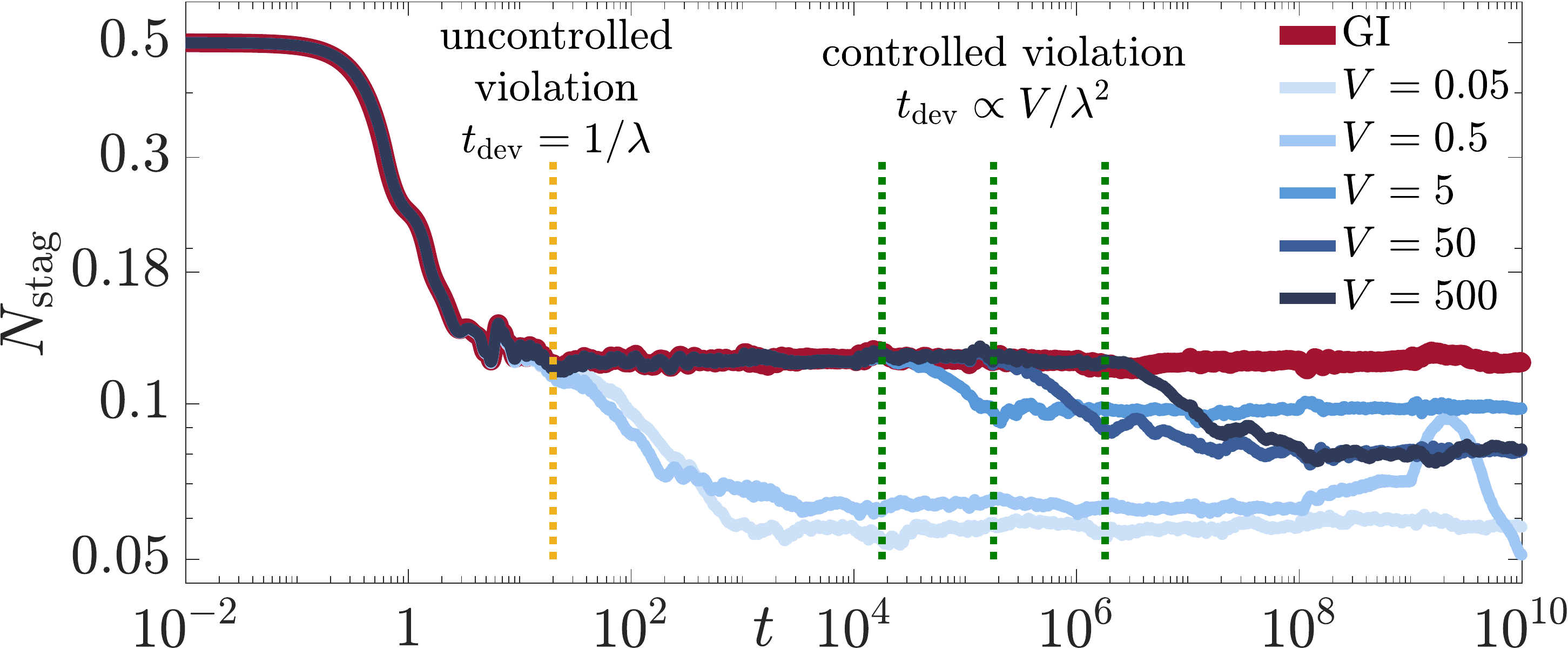}}\\
	\hspace{-.25 cm}
	\includegraphics[width=.49\textwidth]{{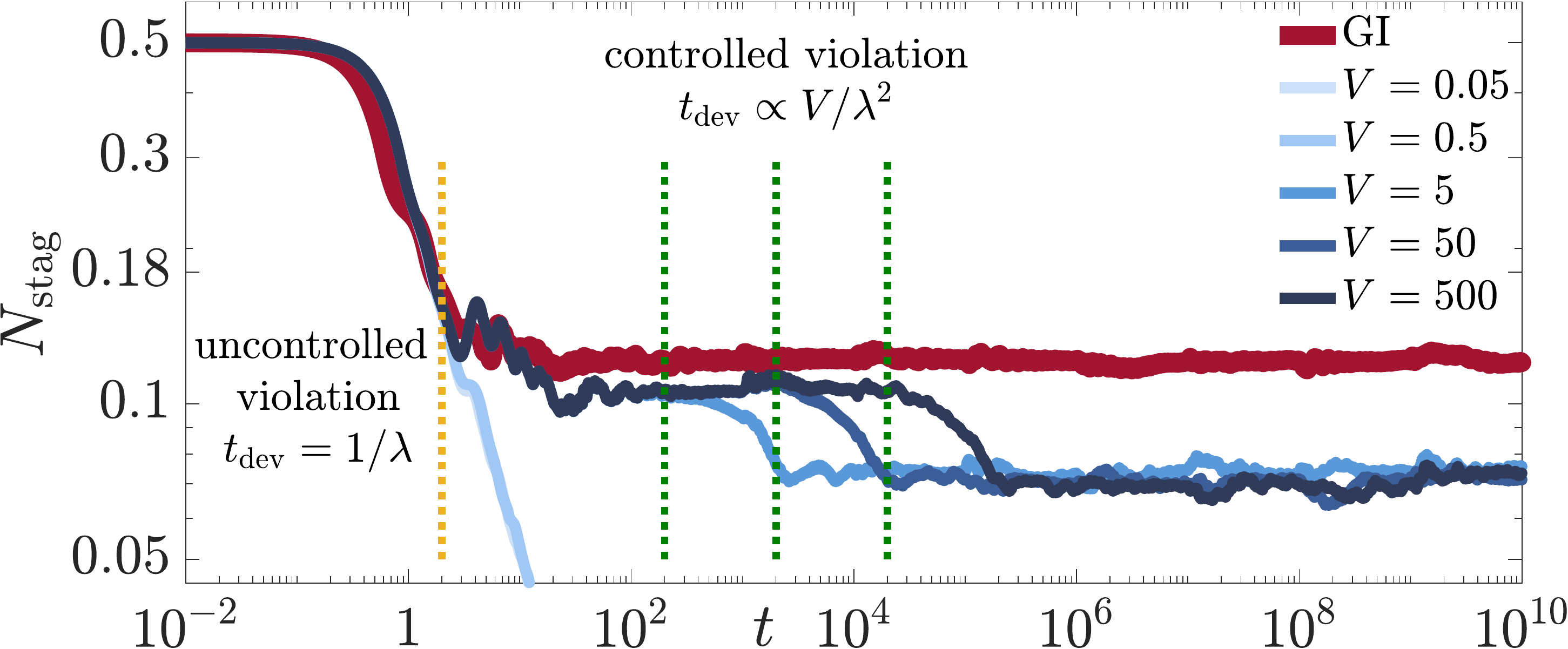}}\\
	\hspace{-.25 cm}
	\caption{(Color online). Dynamics of the spatiotemporally averaged staggered boson number at various values of protection strength $V$ for fixed $\lambda=0.005$ (top panel), $0.05$ (middle panel), and $0.5$ (bottom panel) as compared to the ideal gauge-invariant case (red curves).} 
	\label{fig:FigNstag} 
\end{figure}

\begin{figure}[t!]
	\centering
	\hspace{-.25 cm}
	\includegraphics[width=.49\textwidth]{{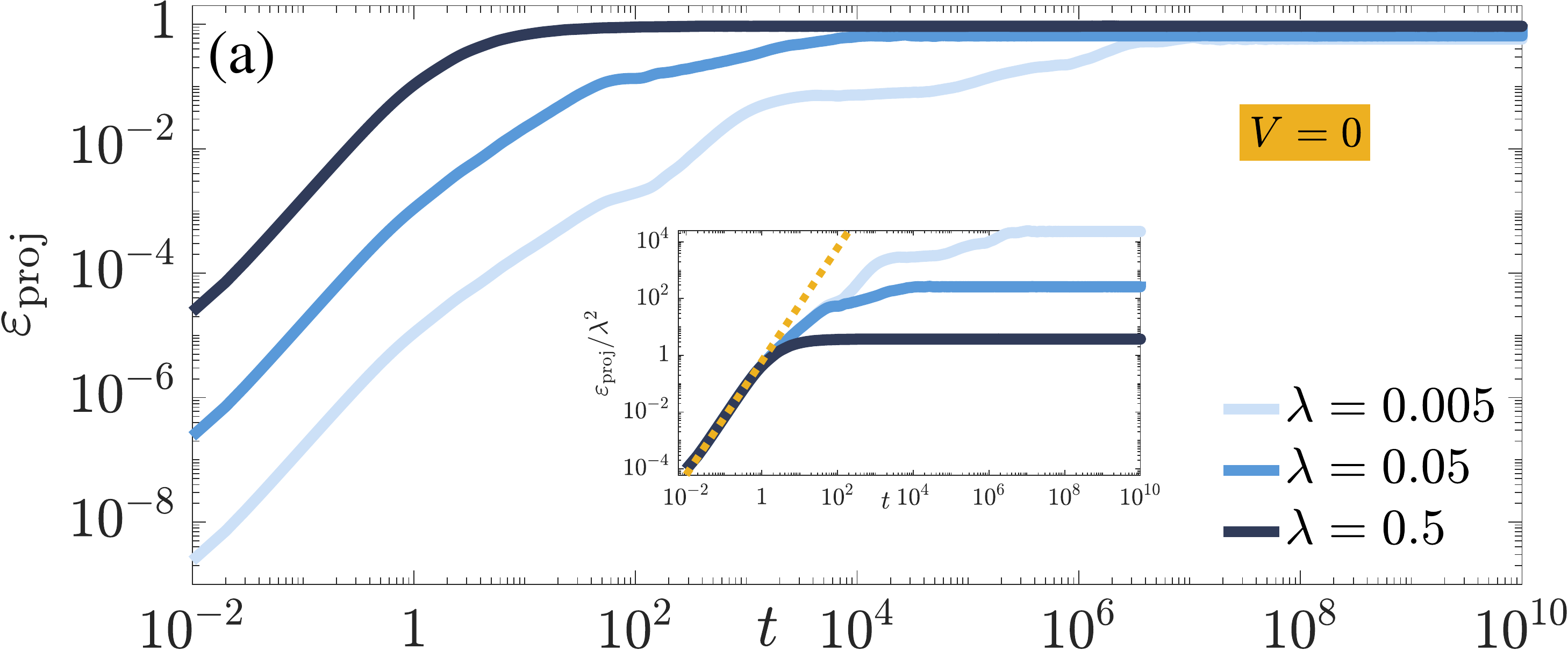}}\\
	\hspace{-.25 cm}
	\includegraphics[width=.49\textwidth]{{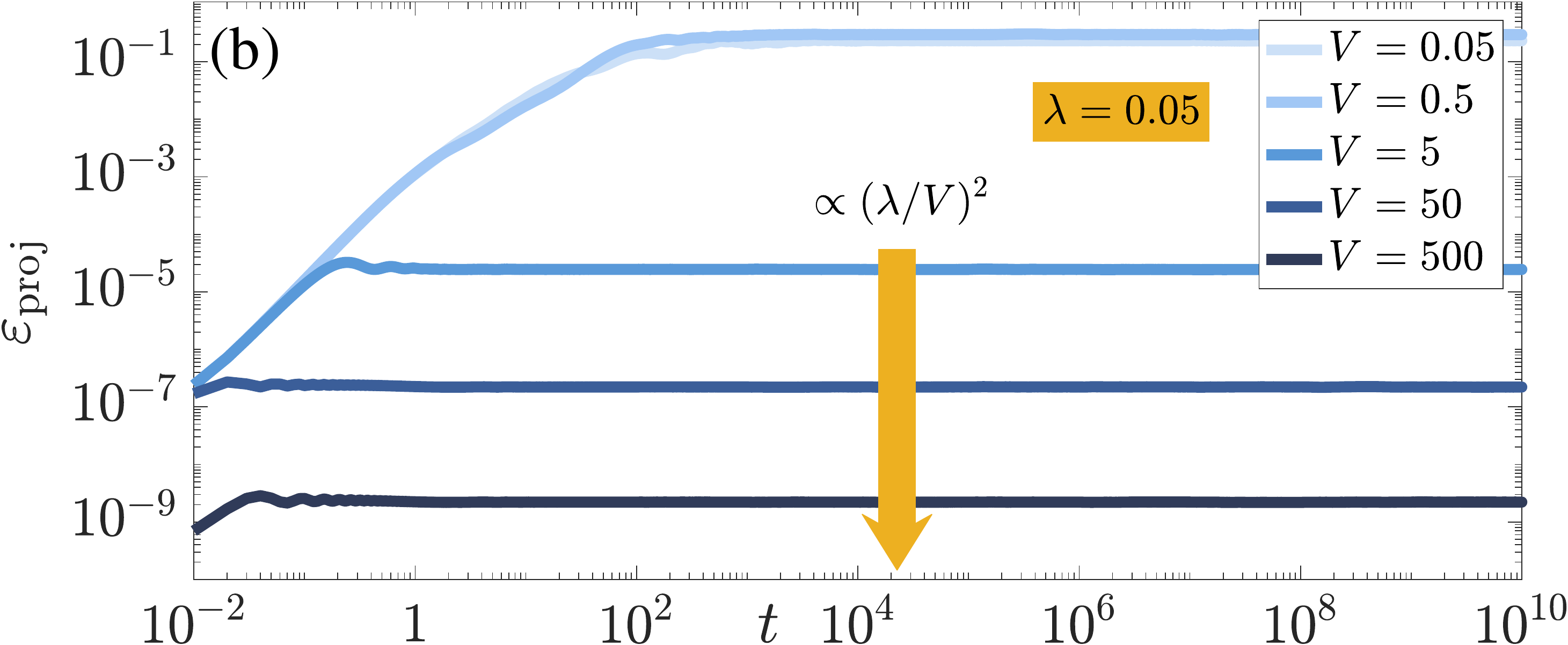}}\\
	\hspace{-.25 cm}
	\includegraphics[width=.49\textwidth]{{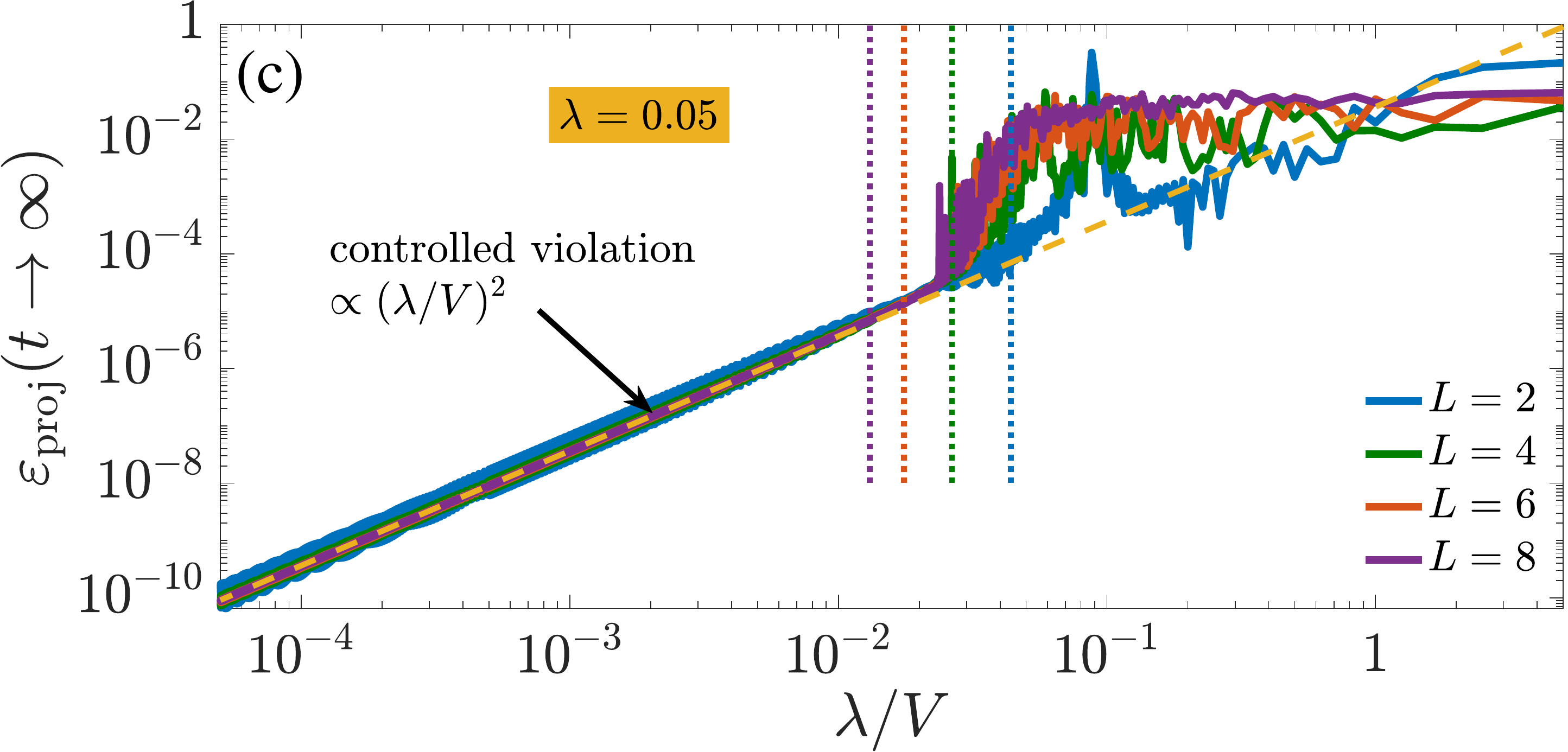}}\\
	\hspace{-.25 cm}
	\caption{(Color online). Gauge violation quantified through projector. In (a), we show the dynamics of $\varepsilon_\mathrm{proj}(t)$ defined in Eq.~\eqref{eq:errorproj} at $V=0$ for various values of $\lambda$. In (b), we fix $\lambda=0.05$ and show the dynamics of $\varepsilon_\mathrm{proj}(t)$ for various values of $V$. In (c), we show the scaling of the infinite-time gauge-invariance violation based on Eq.~\eqref{eq:proj}. The behavior is identical to the gauge-invariance violation based on Eq.~\eqref{eq:error} in the main text.} 
	\label{fig:FigPG} 
\end{figure}

\begin{figure}[!ht]
	\centering
	\vspace{-.01 cm}
	\includegraphics[width=.45\textwidth]{{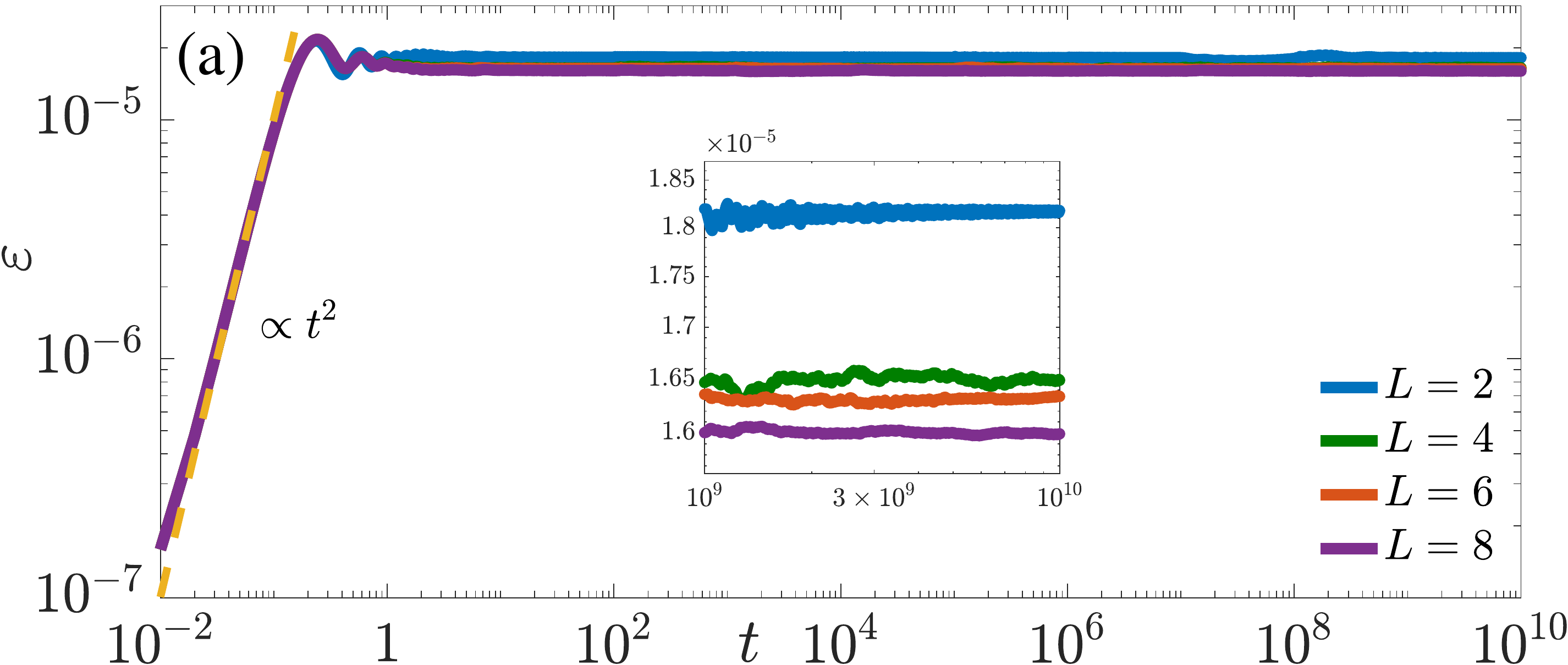}}\\
	\vspace{-.01 cm}
	\includegraphics[width=.45\textwidth]{{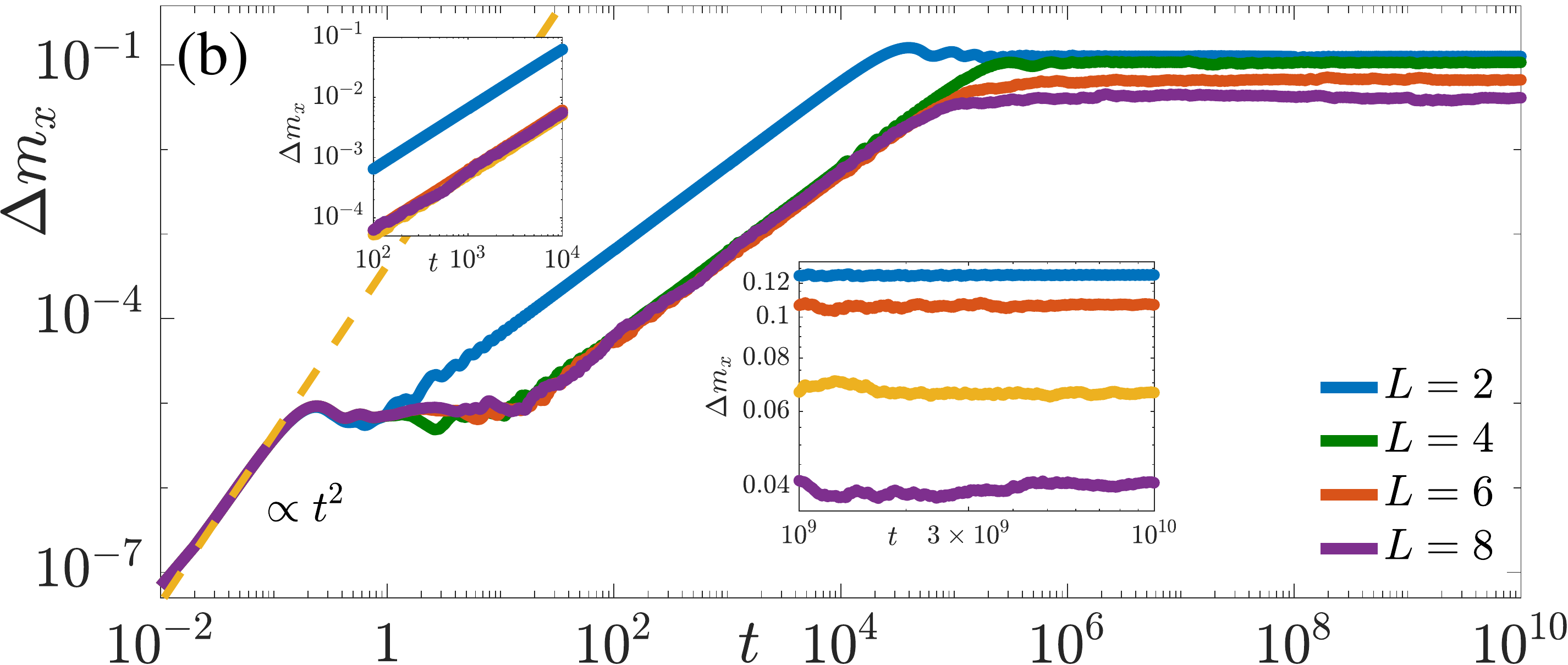}}\\
	\vspace{.01 cm}
	\includegraphics[width=.45\textwidth]{{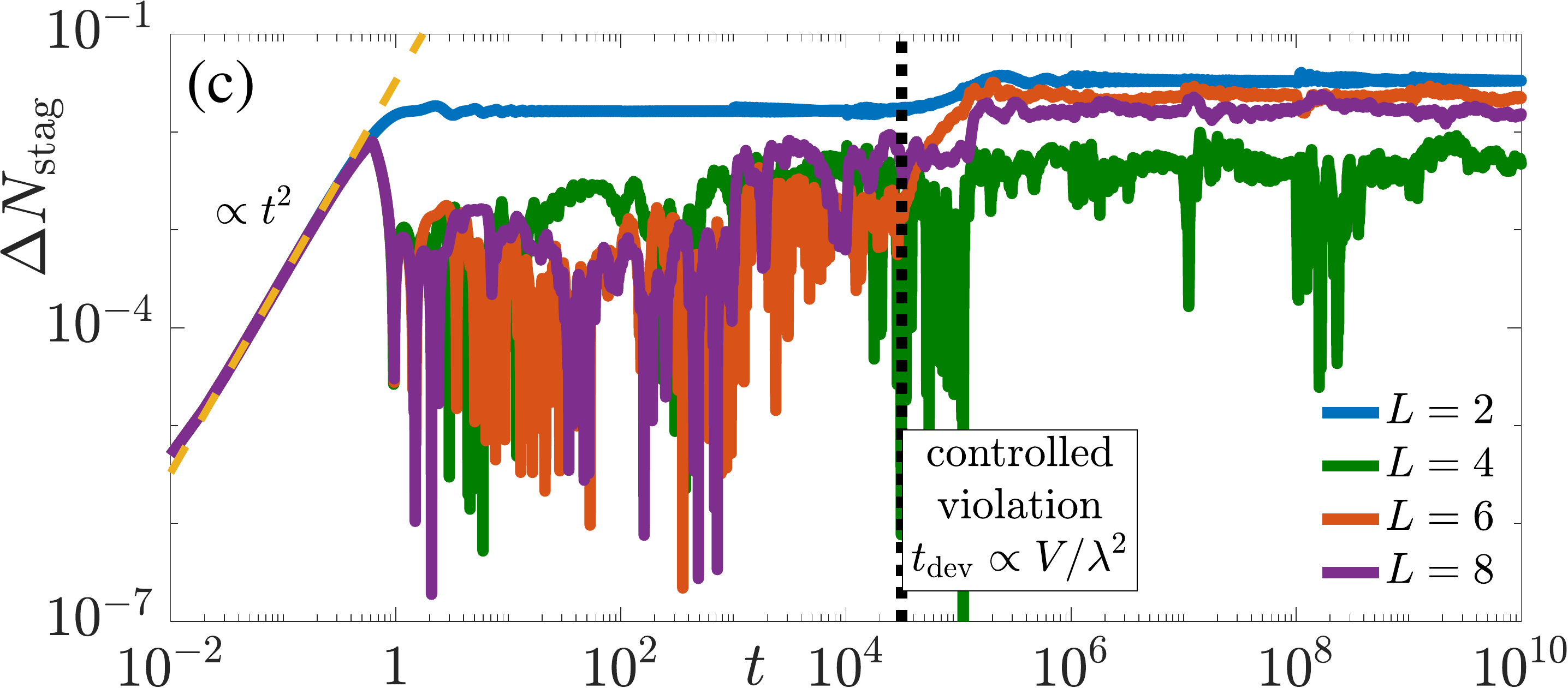}}
	\vspace{-.01 cm}
	\caption{(Color online). Finite-size effects on the dynamics of the spatiotemporal averages of the gauge-invariance violation, electric field, and staggered boson number in the $\mathrm{Z}_2$ gauge theory for $\lambda=0.05$ and $V=5$, which is in the controlled-violation regime as shown in Fig.~\ref{fig:scaling} of the main text. We see that in the controlled-violation regime larger system sizes indicate better protection of gauge invariance.} 
	\label{fig:fss} 
\end{figure}

Let us now look more closely at finite-size effects. In the main text, Fig.~\ref{fig:scaling} indicated that in the controlled-violation regime the infinite-time gauge-invariance violation becomes system size-independent. It is thus interesting to see how the spatiotemporally averaged gauge-invariance violation $\varepsilon(t)$ in Eq.~\eqref{eq:error}, electric-field deviation $\Delta m_x(t)$ in Eq.~\eqref{eq:mag} and staggered boson-number deviation $\Delta N_\text{stag}$ in Eq.~\eqref{eq:nstag} behave dynamically with respect to system size. As we can see in Fig.~\ref{fig:fss}(a) for $\lambda=0.05$ and $V=5$ in the controlled-violation regime, $\varepsilon(t)$ actually goes down with system size at long times, while $\Delta m_x(t)$ in Fig.~\ref{fig:fss}(b) exhibits volume convergence at larger system sizes. The picture is more subtle when it comes to $\Delta N_\text{stag}$ in Fig.~\ref{fig:fss}(c), but nevertheless it seems that the deviation does not significantly increase with system size when comparing the largest system sizes $L=6$ and $L=8$ matter sites. 

Furthermore, we note that the coefficients $c_1,c_2,c_3,c_4$ of Eq.~\eqref{eq:H1} in the main text are inspired by the error terms discussed in Ref.~\cite{Schweizer2019-S}. The error terms are dominated by pair tunneling~--~processes where both species used in the two-component ultracold-atom experiment tunnel~--~and by detuning. We set the dimensionless driving parameter $\chi$ to various values, including ones that are similar to those in the experiment of Ref.~\cite{Schweizer2019-S}, while always ensuring that $c_1+c_2+c_3+c_4=1$. This is done to ensure a systematic control of the error strength through only the parameter $\lambda$ in Eq.~\eqref{eq:H1}.

\begin{figure}[!ht]
	\centering
	\hspace{-.01 cm}
	\vspace{-.01 cm}
	\includegraphics[width=.45\textwidth]{{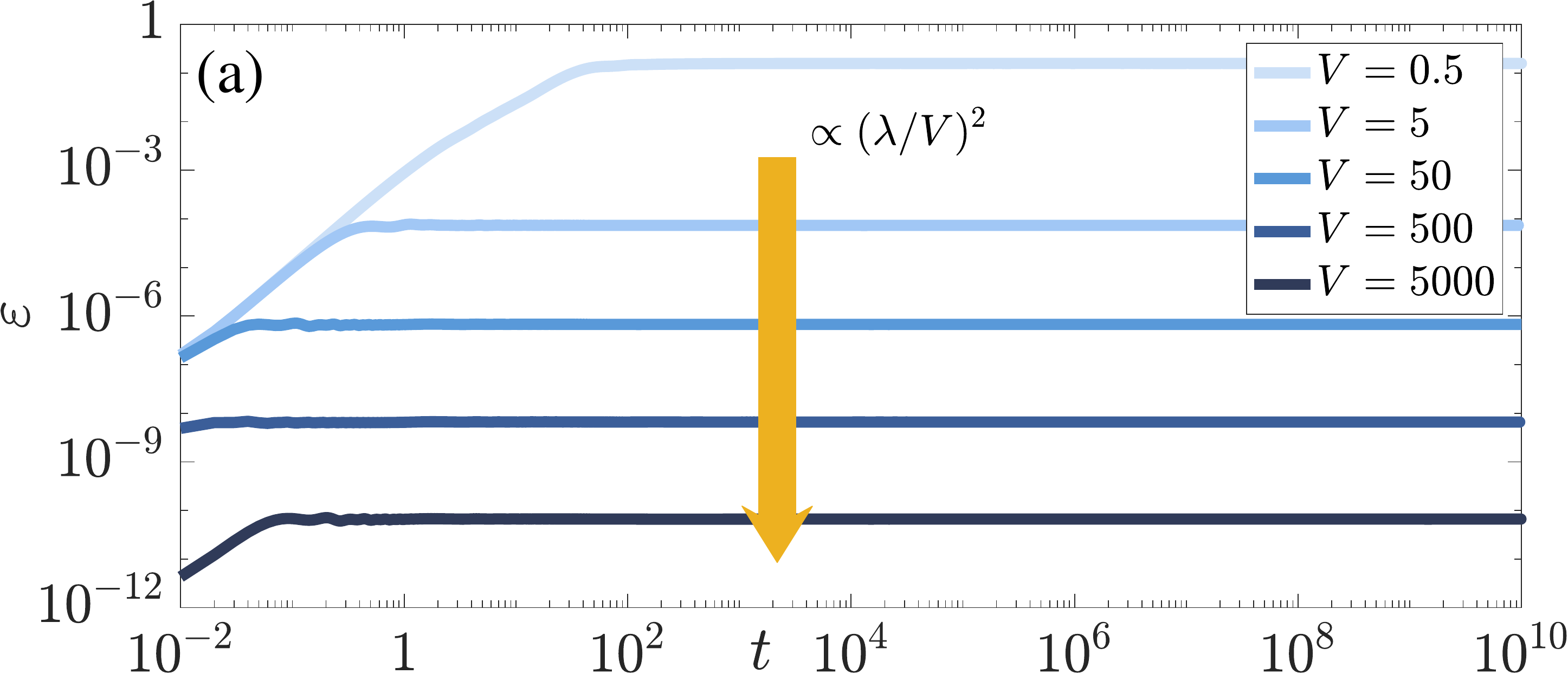}}\quad
	\vspace{-.01 cm}
	\includegraphics[width=.45\textwidth]{{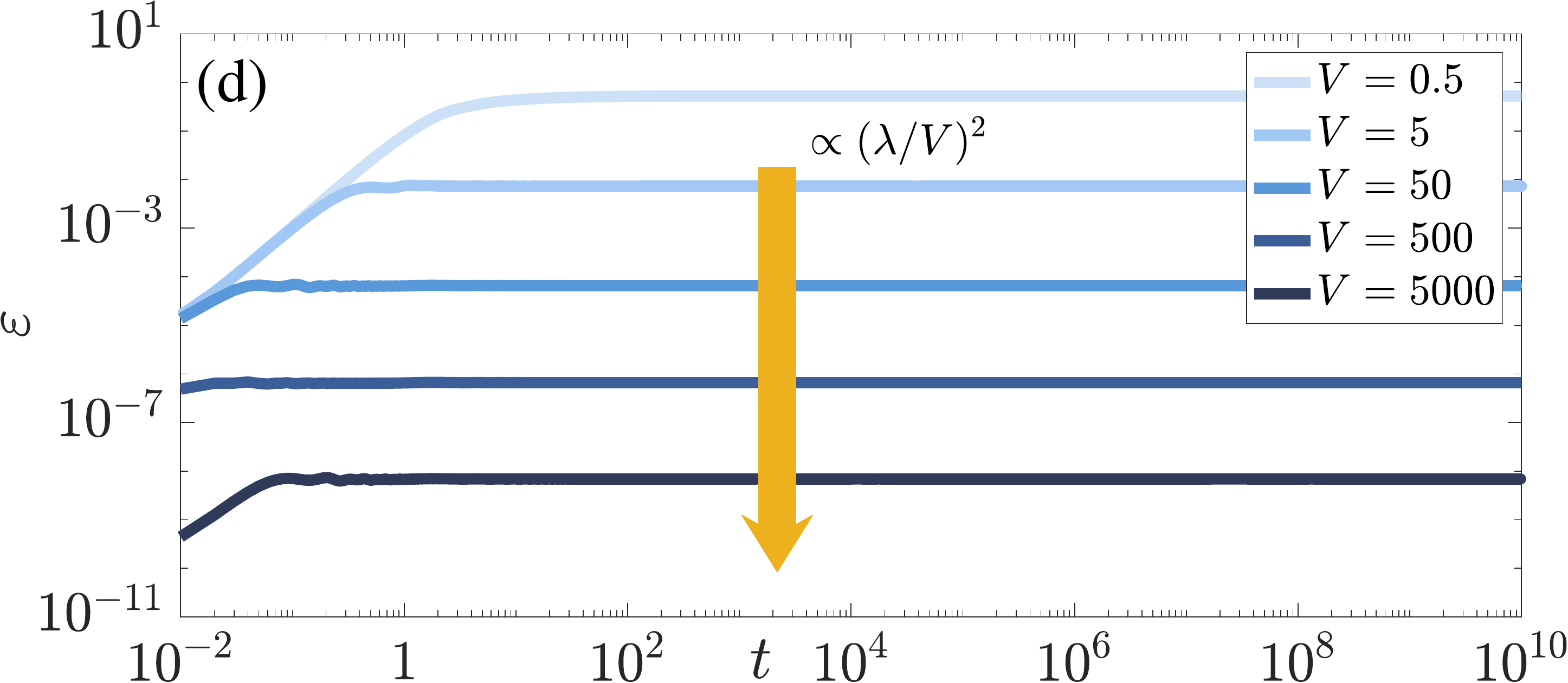}}\quad
	\vspace{-.01 cm}\\
	\hspace{-.01 cm}
	\includegraphics[width=.45\textwidth]{{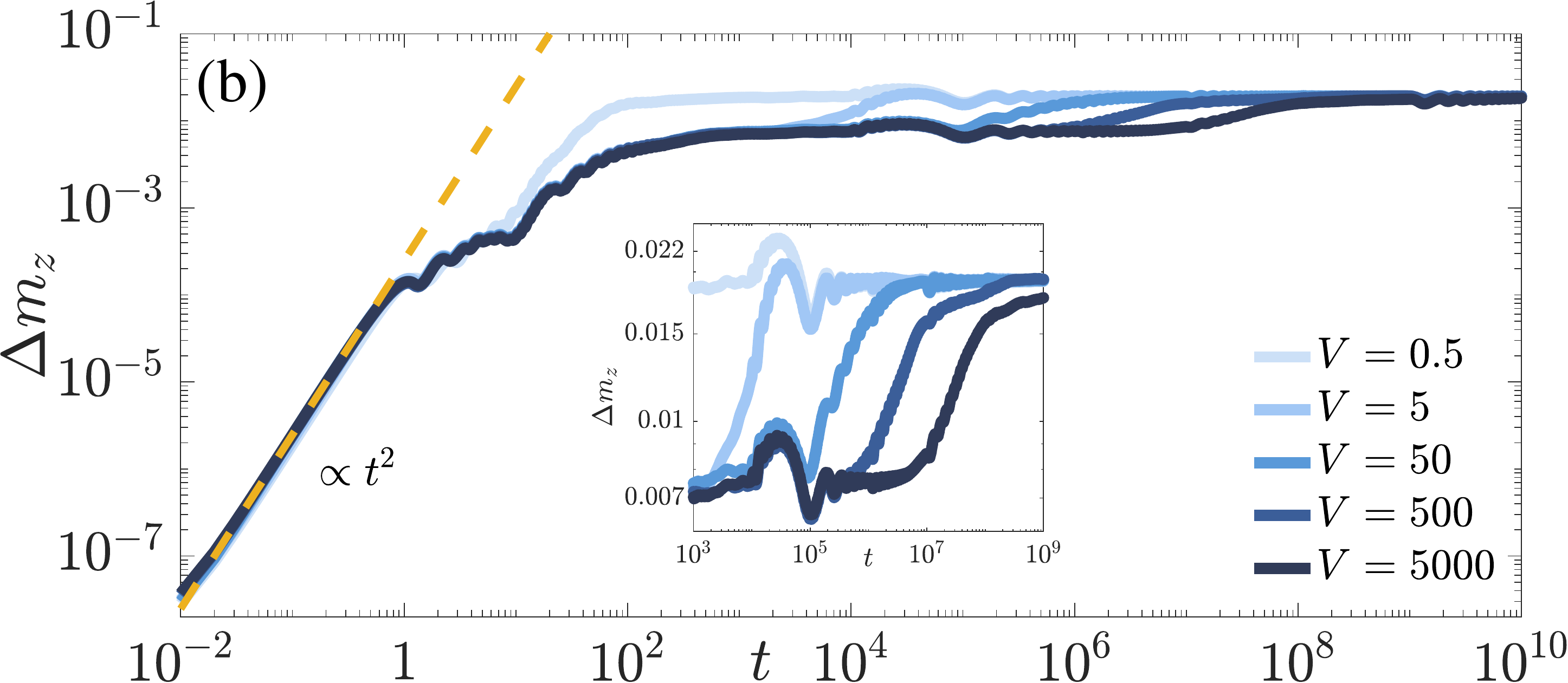}}\quad
	\vspace{-.01 cm}
	\includegraphics[width=.45\textwidth]{{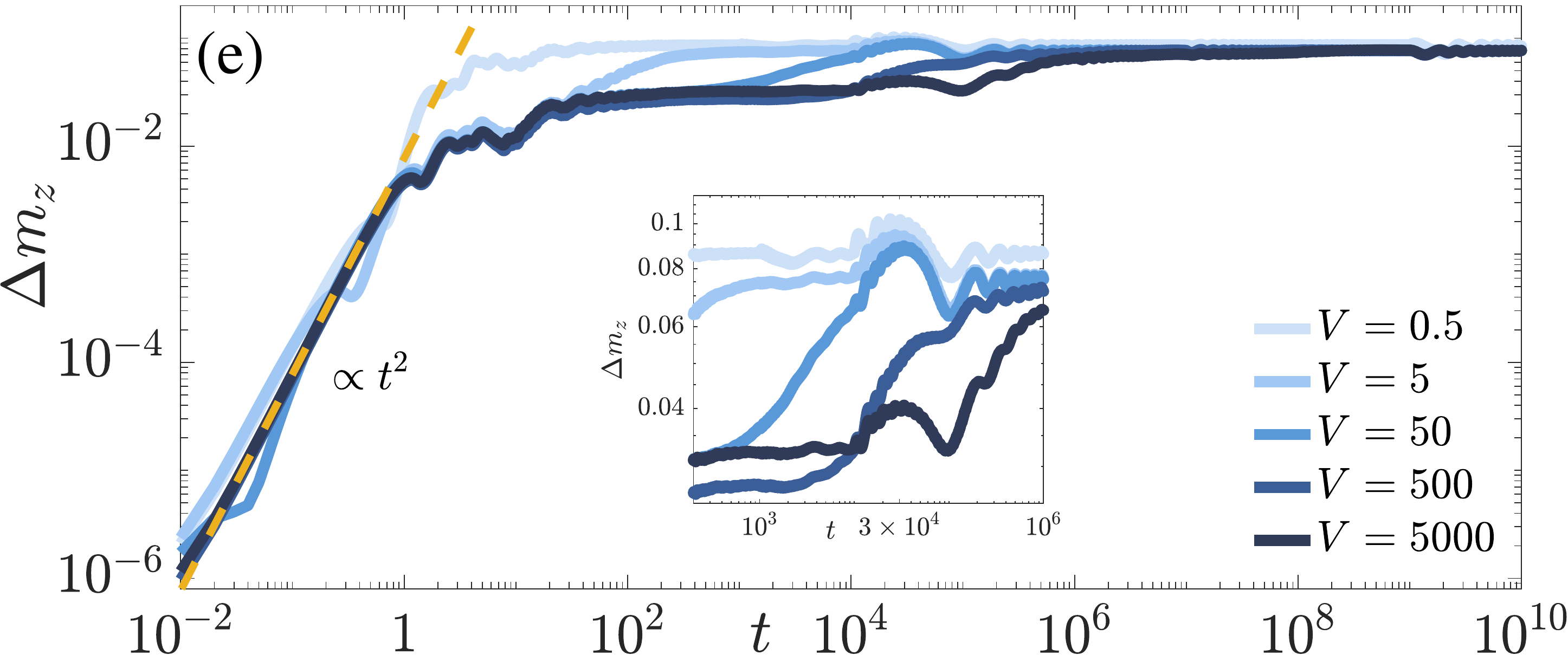}}\quad
	\vspace{-.01 cm}\\
	\hspace{-.01 cm}
	\includegraphics[width=.45\textwidth]{{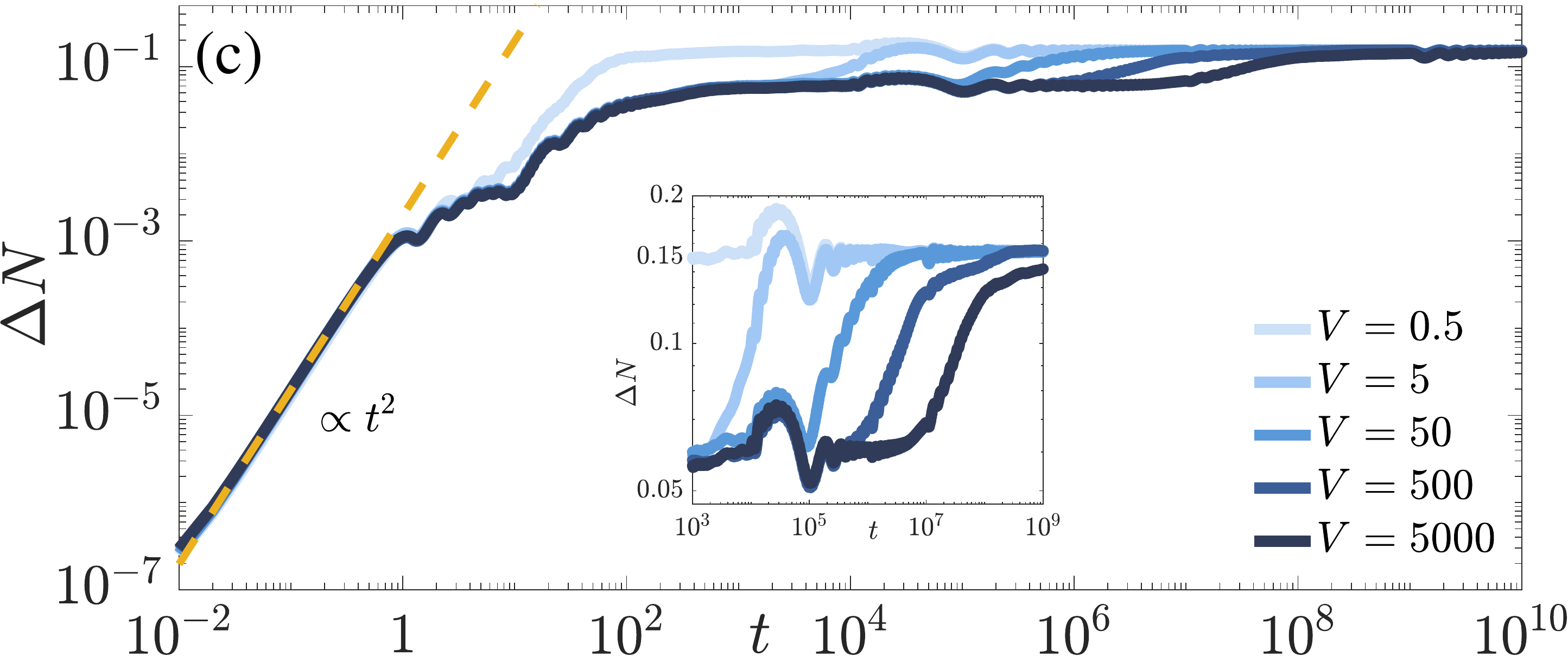}}\quad
	\vspace{-.01 cm}
	\includegraphics[width=.45\textwidth]{{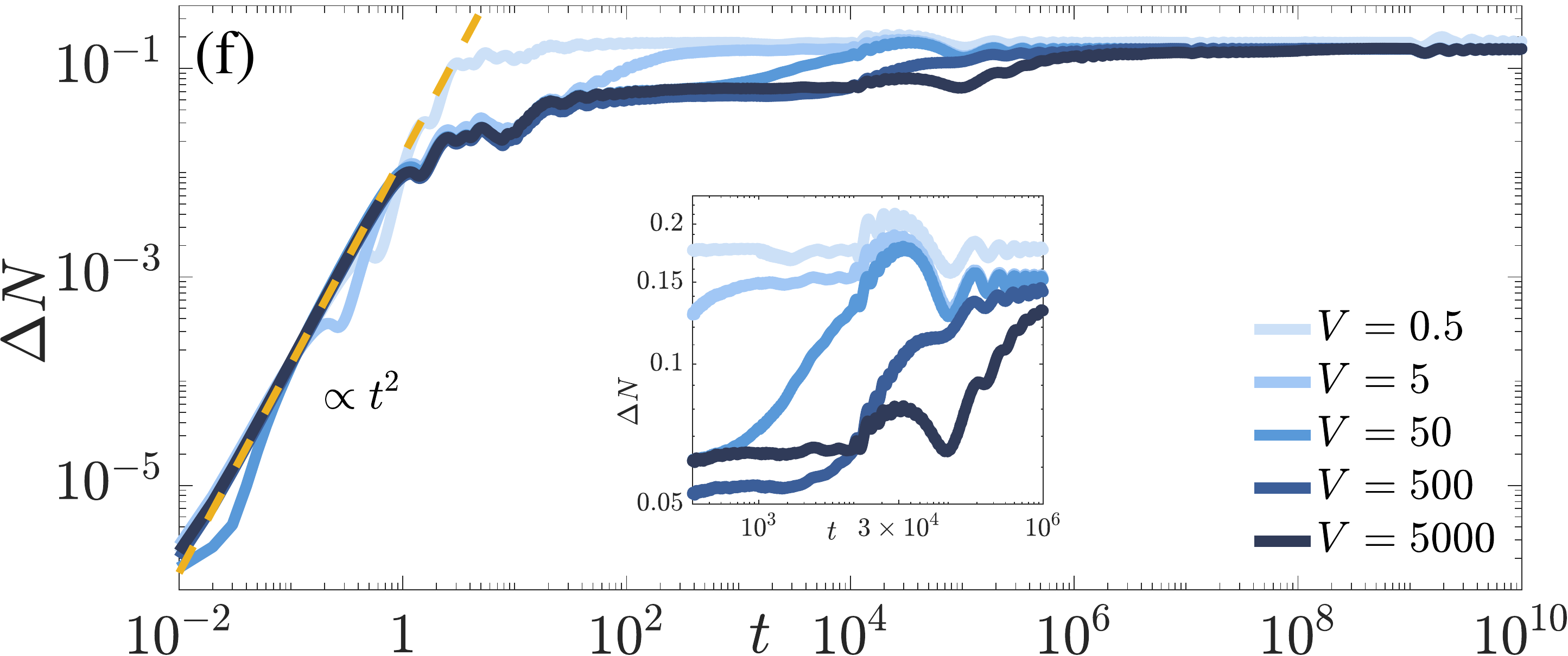}}\quad
	\vspace{-.01 cm}
	\hspace{-.01 cm}
	\caption{(Color online). Quench dynamics of the $\mathrm{U}(1)$ gauge theory given in Eq.~\eqref{eq:Ham0} in the presence of gauge invariance-breaking terms defined in Eq.~\eqref{eq:Ham1} of strength $\lambda=0.05$ (left-column panels) and $0.5$ (right-column panels) and the protection term of Eq.~\eqref{eq:HamG} at various values of protection strength $V$. The conclusions are identical to those of the $\mathrm{Z}_2$ gauge theory in that we see two distinct scales brought about by a sufficiently large $V$ (controlled-violation regime). The first scale is that of the suppression of the gauge-invariance violation by $(\lambda/V)^2$, while the second is the timescale of $V/\lambda^2$ after which the electric field and boson number deviate from their ideal gauge-invariant dynamics.} 
	\label{fig:U1_obs} 
\end{figure}

\section{Analysis of the $\mathrm{U}(1)$ quantum link model}\label{sec:U1}
We now show how the conclusions of the main text, which mostly focus on the $\mathrm{Z}_2$ gauge theory, also hold for the $\mathrm{U}(1)$ gauge theory. We consider the $\mathrm{U}(1)$ QLM given by the Hamiltonian
\begin{equation}\label{eq:Ham0}
H_0=-J\sum_j\big(a_j s_{j,j+1}^+a_{j+1}+\text{H.c.}\big)+\mu\sum_ja_j^\dagger a_j,
\end{equation}
where again $a_j,a_j^\dagger$ denote the ladder operators of the hardcore bosons representing the matter field, but now the gauge field is given by the spin-$1/2$ matrix $s^+_{j,j+1}$ linking sites $j$ and $j+1$, and the electric field by $s^z_{j,j+1}$. The simplicity of this model makes it a perfect benchmark for numerical calculations. Yet, it hosts a number of relevant physical phenomena, such as Coleman's phase transition \cite{Coleman1976-S}, a dynamical quantum phase transition \cite{Huang2019-S}, and extremely slow dynamics generated by gauge invariance \cite{Brenes2018-S}. 
The Hamiltonian $H_0$ is invariant under unitary operations generated by the local Gauss's-law operator

\begin{align}\label{eq:GaussU1}
G_j=a_j^\dagger a_j+\big(s_{j,j+1}^z+s_{j-1,j}^z\big),
\end{align}
$[H_0,G_j]=0$. 
The gauge-invariant subspace is defined by states that satisfy $G_j\left|\psi\right\rangle=0$ $\forall j$. 

Proposals for quantum simulating this QLM have been put forward, e.g., in trapped ions \cite{Hauke2013-S,Yang2016-S}, where $H_0$ is realized as an effective theory with gauge invariance only approximately preserved. To account for this situation, we add the following error term, inspired by (but more general than) Ref.~\cite{Yang2016-S}: 
\begin{equation}\label{eq:Ham1}
\lambda H_1=\,\lambda\sum_j\big(\kappa_ja_js_{x_j,x_j+1}^+ a_{j+y_j}  +\eta_js_{j,j+1}^+ +\gamma_ja_j s_{z_j,z_j+1}^++\text{H.c.}\big).
\end{equation}
Here, in order to show that the features are independent of a specific choice of parameters, $x_j,y_j,z_j\in\mathbb{N}^*$ ($z_j>1$) and $\kappa_j,\eta_j,\gamma_j\in[0,1]$ are randomly selected from a uniform distribution. Generically, as in the case of the $\mathrm{Z}_2$ gauge theory discussed in the main text, the error term $\lambda H_1$ will drive the dynamics out of the gauge-invariant subspace. In order to enforce the gauge invariance of the entire $H$, we will later also add $H_G$ to the Hamiltonian, 
\begin{equation}\label{eq:HamG}
V H_G=V\sum_jG_j^2,
\end{equation}
which differs from that of the $\mathrm{Z}_2$ gauge theory in that $G_j$ is squared. The reason is that $G_j$ of Eq.~\eqref{eq:Gauss} of the $\mathrm{Z}_2$ gauge theory has only two eigenvalues $g_j=0,2$. However, $G_j$ of Eq.~\eqref{eq:GaussU1} of the $\mathrm{U}(1)$ gauge theory has four eigenvalues $g_j=-1,0,1,2$, one of which is negative. As such $\sum_j\langle G_j\rangle$ may be zero even though locally $\langle G_j\rangle$ itself may not vanish. More generally, using $\sum_j\langle G_j\rangle$ as a measure of the violation in Gauss's law would underestimate it in the $\mathrm{U}(1)$ gauge theory, and $V\sum_j G_j$ would not energetically penalize all violations of Gauss's law. Consequently, $G_j$ needs to be raised to an even power in Eq.~\eqref{eq:HamG}, where, for simplicity, we choose this power to be $2$. Of course, one can alternatively use $VH_G=\sum_j|G_j|$.

We now prepare the system in the ground state of Eq.~\eqref{eq:Ham0} at $J=0$, $\mu=1$, $\lambda=0$, and $V=1$, and then quench it with $H_0+\lambda H_1+VH_G$ at $J=1$, $\mu=0$, and some $\lambda$ and $V$, and average the results over $1000$ configurations of the integers $x_j,y_j,z_j$ and coefficients $\kappa_j,\eta_j,\gamma_j$. In the following, however, we note that our conclusions remain the same even without this disorder averaging. In Fig.~\ref{fig:U1_obs} we show results for $\lambda=0.05$ (left-column panels) and $\lambda=0.5$ (right-column panels) at various values of $V$. We consider the spatiotemporal averages of the gauge-invariance violation
\begin{equation}\label{eq:errorU1}
\varepsilon(t)=\frac{1}{Lt}\int_0^t\d s\, \sum_{j=1}^L\bra{\psi(s)}G_j^2\ket{\psi(s)},
\end{equation}
the magnetization in the $z$ direction (the `electric field')
\begin{align}\label{eq:magU1}
m_{z}(t)=\frac{1}{Lt}\int_0^t\d s\,\Big|\sum_{j=1}^L\bra{\psi(s)}s^{z}_{j,j+1}\ket{\psi(s)}\Big|,
\end{align} 
as well as the boson number
\begin{equation}\label{eq:NU1}
N(t)=\frac{1}{Lt}\int_0^t\d s\,\sum_{j=1}^L\bra{\psi(s)}a_j^\dagger a_j\ket{\psi(s)},
\end{equation}
where for the latter two we look at their deviation from the ideal gauge-invariant case. Here $\ket{\psi(t)}=\exp[-\mathrm{i}(H_0+\lambda H_1+VH_G)]\ket{\psi_0}$. As in the case of the $\mathrm{Z}_2$ gauge theory, the violation is suppressed by $(\lambda/V)^2$ up to infinite times in the controlled-violation regime as shown in Fig.~\ref{fig:U1_obs}(a) for $\lambda=0.05$ and Fig.~\ref{fig:U1_obs}(d) for $\lambda=0.5$. Also identically to the $\mathrm{Z}_2$ gauge theory, the gauge-invariant observables $m_{z}(t)$ and $N(t)$ deviate from their ideal dynamics at a timescale $\propto V/\lambda^2$ in the controlled-violation regime, as shown in Fig.~\ref{fig:U1_obs}(b,c) for $\lambda=0.05$ and Fig.~\ref{fig:U1_obs}(e,f) for $\lambda=0.5$. In conclusion, we see two distinct regimes: one of uncontrolled violation when $V$ is too small to counter the effects of Eq.~\eqref{eq:Ham1}, and a controlled-violation regime when $V$ is sufficiently large where the gauge-invariance violation is suppressed by $(\lambda/V)^2$ up to infinite times, as shown in the bottom panel of Fig.~\ref{fig:scaling} in the main text. As such, our results are independent of whether the model is a $\mathrm{Z}_2$ or $\mathrm{U}(1)$ gauge theory, and we expect our conclusions to be valid for more general lattice gauge theories.

\begin{figure}[t!]
	\centering
	\hspace{-.25 cm}
	\includegraphics[width=.49\textwidth]{{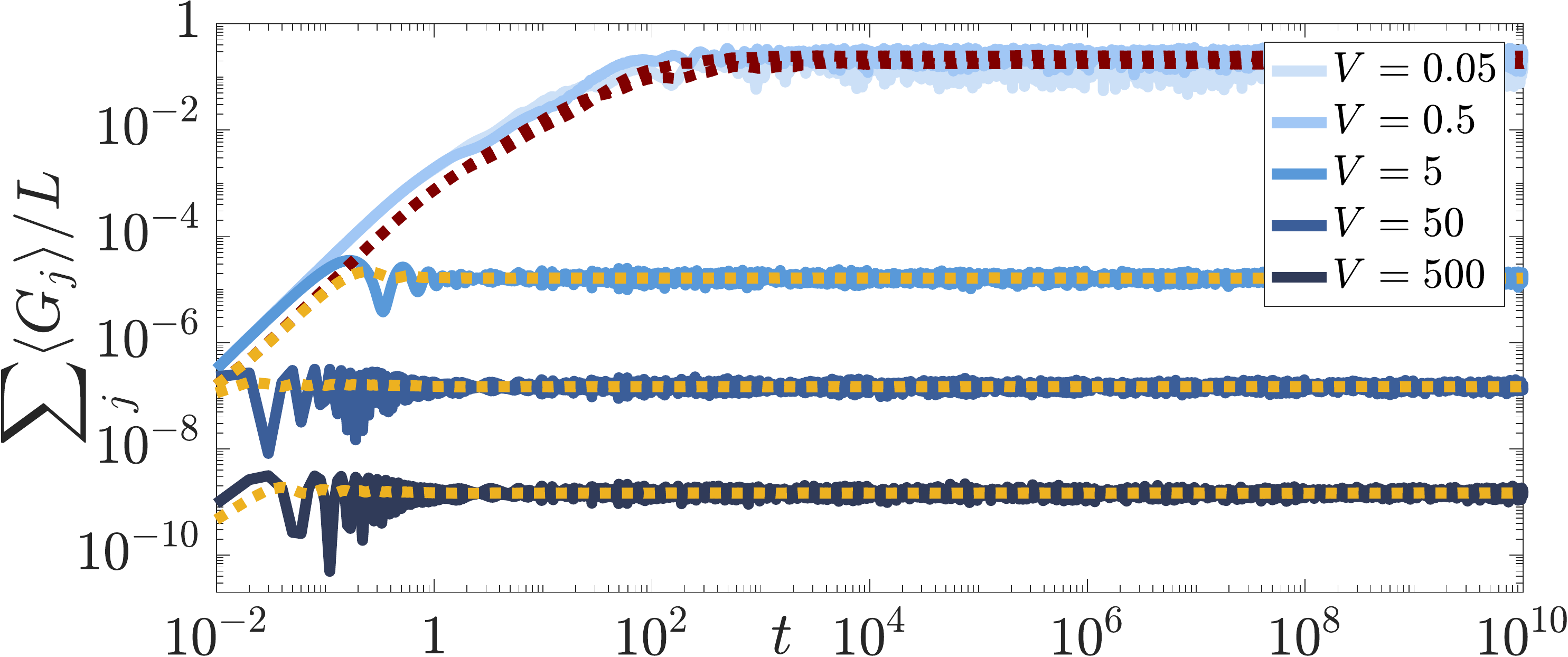}}\\
	\hspace{-.25 cm}
	\caption{(Color online). The spatially averaged gauge-invariance violation (different shades of blue) with its temporal average (dashed lines: red for uncontrolled-violation regime, yellow for controlled-violation regime) overlaid on top of it for $\lambda=0.05$ at various values of protection strength $V$. It is clear from this plot that the conclusions of the main text are unchanged whether looking at the spatially averaged gauge-invariance violation itself or its temporal average. Nevertheless, we opt to present the spatiotemporal average as that mitigates oscillations due to finite system size.}
	\label{fig:UnaveragedViolation} 
\end{figure}

\begin{figure}[t!]
	\centering
	\hspace{-.25 cm}
	\includegraphics[width=.49\textwidth]{{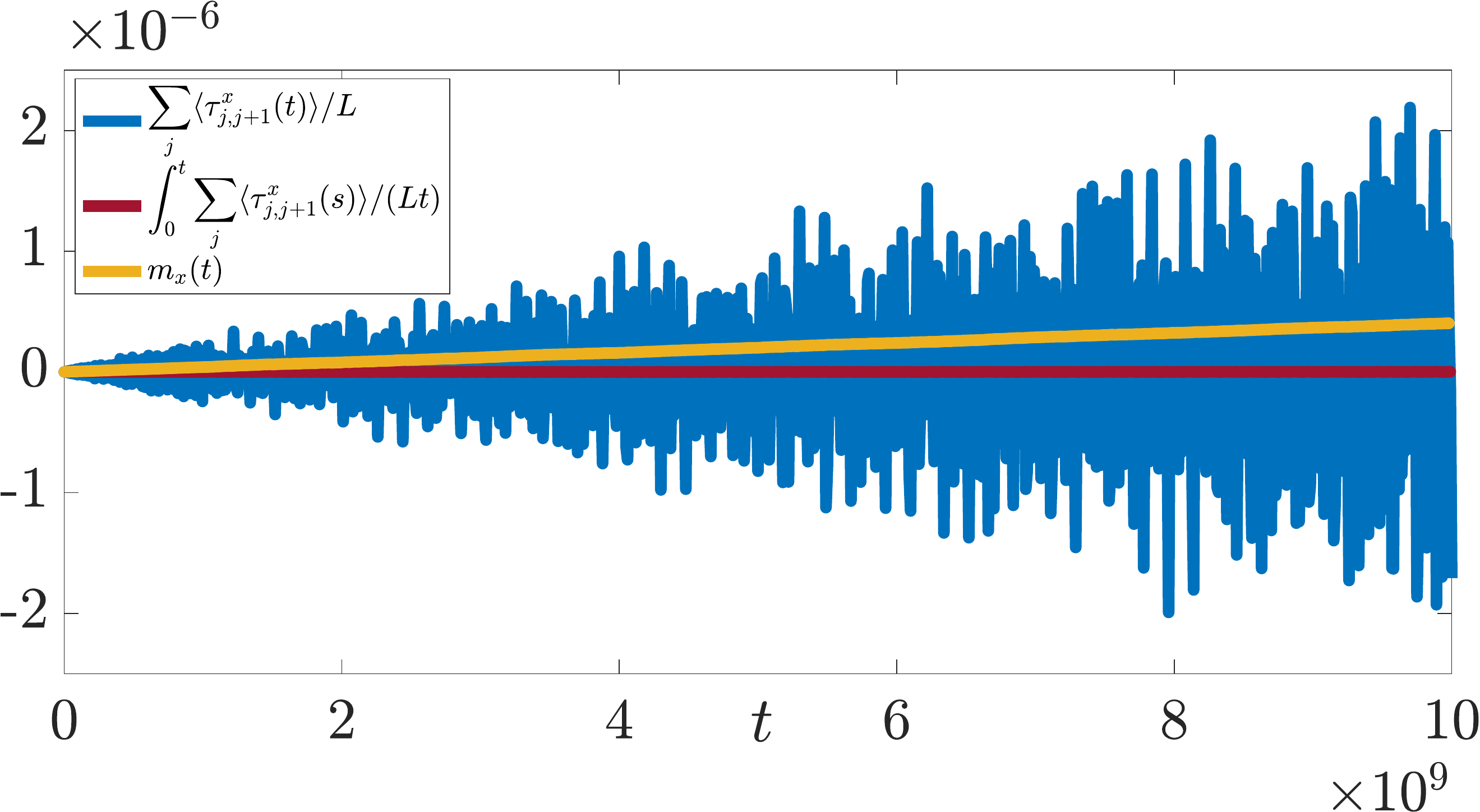}}\\
	\hspace{-.25 cm}
	\caption{(Color online). The ideal dynamics ($\lambda=V=0$) of the spatially averaged electric field's raw value (blue), its temporal average (red), and $m_x(t)$, which is the temporal average of the absolute value (yellow). From the plot, it is straightforward to see that $m_x(t)$ not only offers a \textit{cleaner} signal than the raw data, but it also captures dynamics~--~such as fluctuations~--~that cannot be captured by the mean of the raw data.} 
	\label{fig:clarification} 
\end{figure}

\section{Specific choice of observables}
We opt for temporal averages of our signals to reduce oscillation due to the small system sizes we are limited to in exact diagonalization. This does not alter the conclusions of our work, but merely presents them in a more eye-friendly manner. As an example, let us look at the suppression of the spatially (but not temporally) averaged gauge-invariance violation $\sum_j \langle G_j(t)\rangle/L$ in the $\mathrm{Z}_2$ gauge theory, as shown in Fig.~\ref{fig:UnaveragedViolation}. The qualitative picture remains unchanged when comparing $\sum_j \langle G_j(t)\rangle/L$ to its temporal average in Eq.~\eqref{eq:error}, shown as dashed lines in Fig.~\ref{fig:UnaveragedViolation}; cf.~Fig.~\ref{fig:Fig3}(a).

Another choice we make is to take the spatiotemporal average of the \emph{absolute value} of the observable. The reason behind this is twofold. Firstly, this provides a way to also look at the fluctuations in the observable. For example, if we consider Fig.~\ref{fig:Fig2}(b), we see a peculiar behavior in the ideal dynamics of $m_x(t)$ defined in Eq.~\eqref{eq:mag}. This picture is clarified when considering Fig.~\ref{fig:clarification}. The oscillations in the expectation value of the electric field, $\langle\tau_{j,j+1}^x(t)\rangle$ increases in amplitude with time, albeit its average remains zero. As such, $m_x(t)$, in capturing these fluctuations, offers a more stringent comparison when it comes to nonideal dynamics due to breaking of gauge invariance. Secondly, when gauge invariance is violated, even if slightly, this may lead to phase shifts that will never vanish even after adding a strong protection term that controllably suppresses the violation. By taking the spatiotemporal average of the absolute value of a quantity, such ``superficial'' differences are neglected.

\section{Specifications with regards to the numerics}
For our exact diagonalization simulations we use, in addition to our in-house code, the QuTiP \cite{Johansson2012-S,Johansson2013-S} and QuSpin \cite{Weinberg2017-S,Weinberg2019-S} toolkits in order to cross-check our results. For time evolution, we use our own exact exponentiation routine, because this does not require an iterative calculation that may hinder our ability to reach the long times we achieve. The solvers in QuTiP and QuSpin usually rely on solving an ordinary differential equation, which requires a small enough time-step in order to converge. This in turn demands a large computational time in order to achieve the long times we consider here.

Moreover, as can be seen in our results, the maximum size used for the $\mathrm{U}(1)$ gauge theory is $L=6$ matter sites (plus $6$ links), while for the $\mathrm{Z}_2$ gauge theory the largest chain is $L=8$ matter sites (plus $8$ links). The reason is because in the $\mathrm{Z}_2$ gauge theory, even after breaking gauge invariance according to Eq.~\eqref{eq:H1}, the particle number is conserved, thereby leading to a reduction in the effective Hilbert system in which the dynamics lives. In the case of the $\mathrm{U}(1)$ gauge theory, no such conservation laws are present, which therefore limits the largest $L$ we can reach.

\section{Mathematically rigorous bounds on emergent gauge symmetry at $V>0$}
In this and the following section, we corroborate our numerical results by analytic arguments. Assume $V$ is sufficiently strong to open an energy gap $\Delta\propto V$ from the gauge-invariant sector $G_j\left|\psi\right\rangle=0$, $\forall j$, to gauge-violating sectors. By adapting the results of Ref.~\cite{Chubb2017-S}, we can then mathematically rigorously show that the gauge-invariant sector is protected by an emergent symmetry that is close to the original one.  

Following Ref.~\cite{Chubb2017-S}, we define a \emph{ground symmetry} as an operator $\tilde U$ that commutes 
with the ground space projector ${\Pi}$, 
$[ {\tilde U},{\Pi}]=0$, 
and acts unitarily on the ground space $\Pi \tilde U^\dag \tilde U\Pi=\Pi \tilde U\tilde U^\dag \Pi=\Pi$. 
In our case, $\Pi$ is the projector onto the sector $G_j\left|\psi\right\rangle=0$, $\forall j$. 
Further, a unitary $U$ is called an \emph{$\epsilon$-approximate symmetry} if it approximately commutes with the Hamiltonian with respect to a given unitarily invariant norm
$ || {{U},{H}}||\leq \epsilon$. 
In our scenario of a gauge theory, $U=\exp(\mathrm{i} \sum_j \nu_j G_j)$ with $\nu_j\in\mathbb{R}$, i.e., the relevant unitary is a product of $\mathcal{O}(L)$ local symmetries. 
The relevant non-commutativity is governed by the gauge-violating error strength, so $\epsilon\propto\lambda$.

Using Lemma 2.3 from Ref.~\cite{Chubb2017-S}, we immediately get that there is a unitary $\tilde U$ that is an exact ground symmetry and that is perturbatively close in $(\lambda/V)^2$ to $U$, 
$||{\Pi\left(U-\tilde{U}\right)\Pi}||\leq 1-\sqrt{1-\left(\epsilon/\Delta\right)^2}=\epsilon^2/(2\Delta^2)+\mathcal{O}(\epsilon/\Delta)^4$.  
Thus, the protection term generates a deformed gauge invariance that is exactly retained even in the presence of gauge-violating errors. 

In a generic many-body system, $H_1$ is extensive and consists of few-body operators, so its non-commutativity with $U$ will give $\epsilon=\mathcal{O}(\lambda L)$. Moreover, typically the bandwidth of the ground manifold increases with $L$, which reduces the gap to the other symmetry sectors. Thus, from the above we may expect $||{\Pi\left(U-\tilde{U}\right)\Pi}||$ to deteriorate as a power (larger than 2) of $L$. 
As our numerics indicates, however, this requirement~--~though mathematically rigorous~--~is in practice too stringent, and a scaling with system size that is much more benign can be achieved.

\section{Perturbation Theory}
\noindent In this section, we further explain our numerical results through analytic arguments based on perturbation theory, first for $V=0$ and then for $V$ being the dominant energy scale. 

\subsection{Time-dependent perturbation theory for $V=0$}
In the case of $V=0$, our Hamiltonian is given by
\begin{align}\label{eq:H_unprotected}
H=H_0+\lambda H_1,
\end{align}
where $[H_0,G_j]=0$ and $[H_1,G_j]\neq0$, $\forall j$, with $G_j$ being a local gauge generator at position $j$. Consequently, the gauge invariance encapsulated in $H_0$ is broken up to a strength $\lambda$ by the gauge-noninvariant term $H_1$. We wish to see how the resulting violation grows in time using time-dependent perturbation theory in $\lambda$. 

The total Hilbert space contains gauge-invariant \textit{sectors}, where in each such sector Gauss's law takes on a unique value locally. If our initial state is in a given sector, $H_0$ propagates its dynamics only within that sector, while $H_1$ includes processes that drive the system into other sectors within the total Hilbert space of the system. 

Now since $[H_0,G_j]=0$ and $[G_j,G_x]=0$, we can find a common eigenbasis $\{\ket{\alpha,j}\}$ for $H_0$ and all $G_j$, where $\alpha=(\alpha_1,\alpha_2,\ldots)$ denotes the gauge-invariant sector defined by the unique set of local values $\alpha$, and $q$ stands for all remaining good quantum numbers. Thus, we have $H_0\ket{\alpha,q}=E_{\alpha,q}\ket{\alpha,q}$ and $G_j\ket{\alpha,q}=\alpha_j\ket{\alpha,q}$. Our initial state at $t=0$ is $\ket{\psi_0}$, which resides in the sector $0$, i.e., $G_j\ket{\psi_0}=0$ $\forall j$. As such, our initial state can be written as

\begin{align}
\ket{\psi_0}=\sum_{\alpha}\sum_q\ket{\alpha,q}\bra{\alpha,q}\ket{\psi_0}=\sum_q\ket{0,q}\bra{0,q}\ket{\psi_0},
\end{align}
where in the second equality we have indicated that $\ket{\psi_0}$ is prepared in the gauge-invariant sector $0$. Let us now consider a gauge-invariant observable $O$. This means that $[O,G_j]=0$ $\forall j$, but not necessarily that $[O,H_0]=0$. Thus, $OG_j\ket{\alpha,q}=\alpha_j O\ket{\alpha,q}=G_jO\ket{\alpha,q}$, meaning that $O\ket{\alpha,q}$ is also an eigenstate of $G_j$ in the same sector $\alpha$ as $\ket{\alpha,q}$. Consequently, the observable $O$ can be written as 
\begin{align}\label{eq:O}
O=\sum_{\alpha,\beta}\sum_{q,l}\ket{\alpha,q}\bra{\alpha,q}O\ket{\beta,l}\bra{\beta,l}=\sum_{\alpha}\sum_{q,l}\bra{\alpha,q}O\ket{\alpha,l}\ket{\alpha,q}\bra{\alpha,l},
\end{align}
where in the second equality we have utilized the gauge invariance of $O$. Another useful quantity to calculate, and which we will encounter later, is 
\begin{align}\label{eq:O_0}
O(t)=\mathrm{e}^{\mathrm{i}H_0t}O\mathrm{e}^{-\mathrm{i}H_0t}=\sum_{\alpha}\sum_{q,l}\mathrm{e}^{\mathrm{i}(E_{\alpha,q}-E_{\alpha,l})t}\bra{\alpha,q}O\ket{\alpha,l}\ket{\alpha,q}\bra{\alpha,l}.
\end{align}
We proceed by writing
\begin{align}\label{eq:H11}
H_1(t)=\sum_{\alpha,\beta}\sum_{q,l}\bra{\alpha,q}\mathrm{e}^{\mathrm{i}H_0t}H_1\mathrm{e}^{-\mathrm{i}H_0t}\ket{\beta,l}\ket{\alpha,q}\bra{\beta,l}=\sum_{\alpha,\beta}\sum_{q,l}\mathrm{e}^{\mathrm{i}(E_{\alpha,q}-E_{\beta,l})t}\bra{\alpha,q}H_1\ket{\beta,l}\ket{\alpha,q}\bra{\beta,l}.
\end{align}
As they will become handy later, we now write out the terms of the commutator of Eqs.~\eqref{eq:O_0} and~\eqref{eq:H11}:

\begin{align}\label{eq:OH}
O(t)H_1(\tau)&=\sum_{\alpha,\beta}\sum_{q,m,n}\mathrm{e}^{\mathrm{i}(E_{\alpha,q}-E_{\alpha,m})t}\mathrm{e}^{\mathrm{i}(E_{\alpha,m}-E_{\beta,n})\tau}\bra{\alpha,q}O\ket{\alpha,m}\bra{\alpha,m}H_1\ket{\beta,n}\ket{\alpha,q}\bra{\beta,n},\\[1em]\label{eq:HO}
H_1(\tau)O(t)&=\sum_{\alpha,\beta}\sum_{q,m,n}\mathrm{e}^{\mathrm{i}(E_{\alpha,q}-E_{\beta,m})\tau}\mathrm{e}^{\mathrm{i}(E_{\beta,m}-E_{\beta,n})t}\bra{\alpha,q}H_1\ket{\beta,m}\bra{\beta,m}O\ket{\beta,n}\ket{\alpha,q}\bra{\beta,n}.
\end{align}
The time-evolution operator can be written as
\begin{align}\nonumber
U(t)&=\mathrm{e}^{-\mathrm{i}(H_0+\lambda H_1)t}=\mathrm{e}^{-\mathrm{i}H_0t}\mathcal{T}\big\{\mathrm{e}^{-\mathrm{i}\lambda\int_0^t \d\tau H_1(\tau)}\big\}\\\label{eq:U}
&=\mathrm{e}^{-\mathrm{i}H_0t}\bigg\{1-\mathrm{i}\lambda\int_0^t\d t_1H_1(t_1)-\lambda^2\int_0^t\d t_2\int_0^{t_2}\d t_1H_1(t_2)H_1(t_1)+\mathcal{O}(\lambda^3)\bigg\}.
\end{align}
Consequently, we can now derive, up to second order in perturbation theory,
\begin{align}\nonumber
\bra{\psi_0}U^\dagger(t)OU(t)\ket{\psi_0}=&\,\bra{\psi_0}\bigg\{1+\mathrm{i}\lambda\int_0^t\d t_1H_1(t_1)-\lambda^2\int_0^t\d t_2\int_0^{t_2}\d t_1H_1(t_1)H_1(t_2)+\mathcal{O}(\lambda^3)\bigg\}\\\label{eq:O_PT}
&\times O(t)\bigg\{1-\mathrm{i}\lambda\int_0^t\d \tau_1H_1(\tau_1)-\lambda^2\int_0^t\d \tau_2\int_0^{\tau_2}\d \tau_1H_1(\tau_2)H_1(\tau_1)+\mathcal{O}(\lambda^3)\bigg\}\ket{\psi_0}.
\end{align}
The zeroth-order contribution from Eq.~\eqref{eq:O_PT} is
\begin{align}\label{eq:ZerothOrderPT}
\bra{\psi_0}O(t)\ket{\psi_0}=\sum_{q,l}\bra{\psi_0}\ket{0,q}\bra{0,l}\ket{\psi_0}\bra{0,q}O\ket{0,l}\mathrm{e}^{\mathrm{i}(E_{0,q}-E_{0,l})t}.
\end{align}
The first-order contribution is
\begin{align}\nonumber
&\mathrm{i}\lambda\int_0^t\d t_1\bra{\psi_0}[H_1(t_1),O(t)]\ket{\psi_0}\\\nonumber
=&\,\lambda\sum_{q,m,n}\frac{\mathrm{e}^{\mathrm{i}(E_{0,n}-E_{0,m})t}-1}{E_{0,n}-E_{0,m}}\bra{0,n}H_1\ket{0,m}\\\label{eq:FirstOrderPT}
&\times\big[\mathrm{e}^{\mathrm{i}(E_{0,m}-E_{0,q})t}\bra{0,m}O\ket{0,q}\bra{\psi_0}\ket{0,n}\bra{0,q}\ket{\psi_0}-\mathrm{e}^{\mathrm{i}(E_{0,q}-E_{0,n})t}\bra{0,q}O\ket{0,n}\bra{\psi_0}\ket{0,q}\bra{0,m}\ket{\psi_0}\big].
\end{align}
The second-order contribution is of two components, with the first reading
\begin{align}\nonumber
&\lambda^2\int_0^t\d t_1\int_0^t\d \tau_1\bra{\psi_0}H_1(t_1)O(t)H_1(\tau_1)\ket{\psi_0}\\\nonumber
=&\,-\lambda^2\sum_{\beta}\sum_{q,l,m,n}\mathrm{e}^{\mathrm{i}(E_{\beta,l}-E_{\beta,m})t}\frac{\mathrm{e}^{\mathrm{i}(E_{0,q}-E_{\beta,l})t}-1}{E_{0,q}-E_{\beta,l}}\frac{\mathrm{e}^{\mathrm{i}(E_{\beta,m}-E_{0,n})t}-1}{E_{\beta,m}-E_{0,n}}\bra{\psi_0}\ket{0,q}\bra{0,n}\ket{\psi_0}\\\label{eq:SecondOrderPTa}
&\times\bra{0,q}H_1\ket{\beta,l}\bra{\beta,l}O\ket{\beta,m}\bra{\beta,m}H_1\ket{0,n},
\end{align}
and the second taking the form
\begin{align}\nonumber
&-\lambda^2\int_0^t\d t_2\int_0^{t_2}\d t_1\bra{\psi_0}H_1(t_1)H_1(t_2)O(t)+O(t)H_1(t_2)H_1(t_1)\ket{\psi_0}\\\nonumber
=&\,2\lambda^2\sum_{\beta}\sum_{q,l,m,n}\real\bigg\{\bra{\psi_0}\ket{0,q}\bra{0,n}\ket{\psi_0}\bra{0,q}H_1\ket{\beta,l}\bra{\beta,l}H_1\ket{0,m}\bra{0,m}O\ket{0,n}\\\label{eq:SecondOrderPTb}
&\times\frac{1}{E_{0,q}-E_{\beta,l}}\left[\frac{\mathrm{e}^{\mathrm{i}(E_{0,q}-E_{0,n})t}-\mathrm{e}^{\mathrm{i}(E_{0,m}-E_{0,n})t}}{E_{0,q}-E_{0,m}}-\frac{\mathrm{e}^{\mathrm{i}(E_{\beta,l}-E_{0,n})t}-\mathrm{e}^{\mathrm{i}(E_{0,m}-E_{0,n})t}}{E_{\beta,l}-E_{0,m}}\right]\bigg\}\,.
\end{align}

The spatiotemporally averaged gauge-invariance violation in Eq.~\eqref{eq:error} at short times scales as $\varepsilon(t)\sim(\lambda t)^2$. This behavior can easily be understood by setting $O=\sum_j G_j^p$~--~recall that $p=1$ for the $\mathrm{Z}_2$ gauge theory and $p=2$ for the $\mathrm{U}(1)$ gauge theory~--~in Eqs.~\eqref{eq:ZerothOrderPT} and~\eqref{eq:FirstOrderPT}, where now the zeroth- and first-order contributions from perturbation theory reduce to zero since $\bra{0,m}\sum_jG_j^p\ket{0,n}=0$, $\forall m,n$. Consequently, first-order contributions from perturbation theory can never appear for $\varepsilon(t)$. On the other hand, looking at the second-order contributions of Eqs.~\eqref{eq:SecondOrderPTa} and~\eqref{eq:SecondOrderPTb} from perturbation theory, we see that when $O=\sum_j G_j^p$, Eq.~\eqref{eq:SecondOrderPTb} vanishes since $\langle 0,m|\sum_j G_j^p|0,n\rangle=0$ $\forall m,n$, while Eq.~\eqref{eq:SecondOrderPTa} does not in general. At short times, we can thus approximate Eq.~\eqref{eq:SecondOrderPTa} as

\begin{align}\nonumber
&\lambda^2\int_0^t\d t_1\int_0^t\d \tau_1\bra{\psi_0}H_1(t_1)\sum_j G_j^p(t)H_1(\tau_1)\ket{\psi_0}\\\nonumber
=&\,-\lambda^2\sum_\beta\sum_{j=1}^L\beta_j^p\sum_{q,m,n}\frac{\mathrm{e}^{\mathrm{i}(E_{0,q}-E_{\beta,m})t}-1}{E_{0,q}-E_{\beta,m}}\frac{\mathrm{e}^{\mathrm{i}(E_{\beta,m}-E_{0,n})t}-1}{E_{\beta,m}-E_{0,n}}\bra{\psi_0}\ket{0,q}\bra{0,n}\ket{\psi_0}\bra{0,q}H_1\ket{\beta,m}\bra{\beta,m}H_1\ket{0,n}\\\nonumber
\approx&\,\lambda^2 t^2\sum_\beta\sum_{j=1}^L\beta_j^p\sum_{q,m,n}\bra{\psi_0}\ket{0,q}\bra{0,n}\ket{\psi_0}\bra{0,q}H_1\ket{\beta,m}\bra{\beta,m}H_1\ket{0,n}\\
\sim&\,(\lambda t)^2,
\end{align}
thereby showing how the gauge-invariance violation always scales $\sim(\lambda t)^2$. 

On the other hand, gauge-invariant observables that do not commute with $H_0$ do not necessarily scale as $\sim(\lambda t)^2$ at short times. For example, this can be seen in the $\mathrm{Z}_2$ gauge theory where the deviation from ideal gauge-invariant dynamics of the spatiotemporally averaged staggered magnetization of Eq.~\eqref{eq:nstag} scales at short times as $\Delta N_\text{stag}\sim\lambda t^2$; cf.~Fig.~\ref{fig:Fig2}(c). This can be understood by noting that the zeroth-order contribution in Eq.~\eqref{eq:ZerothOrderPT} from perturbation theory vanishes in $\Delta N_\text{stag}$, whereas in Eq.~\eqref{eq:FirstOrderPT} the term $\bra{0,m}\sum_j(-1)^ja_j^\dagger a_j\ket{0,n}$ does not necessarily vanish because $[H_0,\sum_j(-1)^j a_j^\dagger a_j]\neq0$. As such, we can look at what happens to Eq.~\eqref{eq:FirstOrderPT} at short times for $O=\mathcal{N}=\sum_j(-1)^ja_j^\dagger a_j$ by approximating it as
\begin{align}\nonumber
&\mathrm{i}\lambda\int_0^t\d t_1\bra{\psi_0}[H_1(t_1),\mathcal{N}(t)]\ket{\psi_0}\\\nonumber
\approx&\,\lambda\sum_{q,m,n}\mathrm{i}t\bra{0,n}H_1\ket{0,m}\big\{[1+\mathrm{i}(E_{0,m}-E_{0,q})t]\bra{0,m}\mathcal{N}\ket{0,q}\bra{\psi_0}\ket{0,n}\bra{0,q}\ket{\psi_0}\\\nonumber
&-[1+\mathrm{i}(E_{0,q}-E_{0,n})t]\bra{0,q}\mathcal{N}\ket{0,n}\bra{\psi_0}\ket{0,q}\bra{0,m}\ket{\psi_0}\big\}\\
=&\,\lambda \big(\mathcal{A}t^2+\mathcal{B}t),
\end{align}
with

\begin{align}\nonumber
\mathcal{A}=&\,\sum_{q,m,n}\real\big\{\bra{0,n}H_1\ket{0,m}[(E_{0,q}-E_{0,m})\bra{0,m}\mathcal{N}\ket{0,q}\bra{\psi_0}\ket{0,n}\bra{0,q}\ket{\psi_0}\\
&+(E_{0,q}-E_{0,n})\bra{0,q}\mathcal{N}\ket{0,n}\bra{\psi_0}\ket{0,q}\bra{0,m}\ket{\psi_0}]\big\},\\[1em]
\mathcal{B}=&\,\sum_{q,m,n}\imaginary\big\{\bra{0,n}H_1\ket{0,m}[\bra{0,q}\mathcal{N}\ket{0,n}\bra{\psi_0}\ket{0,q}\bra{0,m}\ket{\psi_0}-\bra{0,m}\mathcal{N}\ket{0,q}\bra{\psi_0}\ket{0,n}\bra{0,q}\ket{\psi_0}]\big\}.
\end{align}
We note that the only nonvanishing contribution from $\bra{0,n}H_1\ket{0,m}$ is due to the terms whose coefficients are $c_1$ and $c_2$ in Eq.~\eqref{eq:H1}, otherwise $\bra{0,n}H_1\ket{0,m}=0$, and consequently Eq.~\eqref{eq:FirstOrderPT} also vanishes. One can easily check numerically that $\mathcal{B}=0$ while $\mathcal{A}\neq0$, further reducing Eq.~\eqref{eq:FirstOrderPT} to

\begin{align}\nonumber
\mathrm{i}\lambda\int_0^t\d t_1\bra{\psi_0}[H_1(t_1),\mathcal{N}(t)]\ket{\psi_0}\approx\lambda t^2\mathcal{A}\sim\lambda t^2,
\end{align}
thereby explaining the short-time scaling $\sim\lambda t^2$ of the deviation in the staggered boson number from ideal dynamics in Fig.~\ref{fig:Fig2}(c).

As a summary, in case $H_1$ contains only terms that bring us completely out of the initial gauge-invariant sector, all gauge-invariant observables scale as $\lambda^2$. For $\sum_jG_j^p$ and other gauge-invariant observables that commute with $H_0$, this scaling remains true even if $H_1$ contains also terms that generate dynamics within the initial gauge-invariant sector, but not necessarily for other gauge-invariant observables.

\subsection{Effective perturbative Hamiltonian for $V\gg 1$.}
Let us now turn our attention to the case of $V\gg 1$ at small $\lambda\lesssim\mathcal{O}(1)$. The analysis thus far has been valid only for short times. As described in the main text, the initial state $\ket{\psi_0}$ has staggered order on both matter and gauge sites in that even (odd) matter sites have one (zero) boson and even (odd) gauge sites have spin-$1/2$ particles polarized in the positive (negative) $x$-direction. Even though this renders $\ket{\psi_0}$ gauge-invariant, it is not an eigenstate of $H_0$ despite that $[H_0,G_j]=0$ $\forall j$. This is important since it means nontrivial dynamics will take place while always remaining in the same gauge-invariant sector when $\lambda=0$. 

Nevertheless, when $\lambda>0$, we get gauge-noninvariant processes that drive the system into gauge-invariant sectors different from the sector $\ket{\psi_0}$ started in. In order to protect gauge invariance, we add to Eq.~\eqref{eq:H_unprotected} the protection term $VH_G$ with $V\gg\lambda$, which energetically penalizes processes that want to drive the system away from the initial gauge-invariant sector. This results in the effective Hamiltonian \cite{cohen1992atom-S,Lewenstein_review-S}
\begin{align}\label{eq:Heff}
H_\text{eff}=P_\mathrm{0}H_0P_\mathrm{0}-\underbrace{P_\mathrm{0}\lambda H_1(1-P_\mathrm{0})(H_0+VH_G)^{-1}\lambda H_1P_\mathrm{0}}_{\propto\lambda^2/V\,\,\,\text{for}\,\,\,V\gg1}+\mathcal{O}(1/V^2).
\end{align}
where we have made use of the fact that $H_GP_\mathrm{0}=0$ and $P_\mathrm{0}H_1P_\mathrm{0}=0$. This indicates that for small $\lambda$, and $V\gg1$, the gauge invariance-violating error terms are perturbatively small with strength $\lambda^2/V$. This is why at $\lambda>0$ and $V\gg1$ (controlled-violation regime) $m_x(t)$ and $N_\text{stag}(t)$ deviate from their ideal gauge-invariant dynamics at $t\approx V/\lambda^2$.
Nevertheless, this deviation can be compensated for, as it is just due to a modification of the gauge-invariant Hamiltonian.


\begin{thebibliography}{47}%
	\makeatletter
	\providecommand \@ifxundefined [1]{%
		\@ifx{#1\undefined}
	}%
	\providecommand \@ifnum [1]{%
		\ifnum #1\expandafter \@firstoftwo
		\else \expandafter \@secondoftwo
		\fi
	}%
	\providecommand \@ifx [1]{%
		\ifx #1\expandafter \@firstoftwo
		\else \expandafter \@secondoftwo
		\fi
	}%
	\providecommand \natexlab [1]{#1}%
	\providecommand \enquote  [1]{``#1''}%
	\providecommand \bibnamefont  [1]{#1}%
	\providecommand \bibfnamefont [1]{#1}%
	\providecommand \citenamefont [1]{#1}%
	\providecommand \href@noop [0]{\@secondoftwo}%
	\providecommand \href [0]{\begingroup \@sanitize@url \@href}%
	\providecommand \@href[1]{\@@startlink{#1}\@@href}%
	\providecommand \@@href[1]{\endgroup#1\@@endlink}%
	\providecommand \@sanitize@url [0]{\catcode `\\12\catcode `\$12\catcode
		`\&12\catcode `\#12\catcode `\^12\catcode `\_12\catcode `\%12\relax}%
	\providecommand \@@startlink[1]{}%
	\providecommand \@@endlink[0]{}%
	\providecommand \url  [0]{\begingroup\@sanitize@url \@url }%
	\providecommand \@url [1]{\endgroup\@href {#1}{\urlprefix }}%
	\providecommand \urlprefix  [0]{URL }%
	\providecommand \Eprint [0]{\href }%
	\providecommand \doibase [0]{http://dx.doi.org/}%
	\providecommand \selectlanguage [0]{\@gobble}%
	\providecommand \bibinfo  [0]{\@secondoftwo}%
	\providecommand \bibfield  [0]{\@secondoftwo}%
	\providecommand \translation [1]{[#1]}%
	\providecommand \BibitemOpen [0]{}%
	\providecommand \bibitemStop [0]{}%
	\providecommand \bibitemNoStop [0]{.\EOS\space}%
	\providecommand \EOS [0]{\spacefactor3000\relax}%
	\providecommand \BibitemShut  [1]{\csname bibitem#1\endcsname}%
	\let\auto@bib@innerbib\@empty
	\bibitem [{\citenamefont {Cheng}\ and\ \citenamefont {Li}(1984)}]{Cheng_book}%
	\BibitemOpen
	\bibfield  {author} {\bibinfo {author} {\bibfnamefont {T.}~\bibnamefont
			{Cheng}}\ and\ \bibinfo {author} {\bibfnamefont {L.}~\bibnamefont {Li}},\
	}\href {https://books.google.it/books?id=lk8GEzVNb10C} {\emph {\bibinfo
			{title} {Gauge Theory of Elementary Particle Physics}}},\ Oxford science
	publications\ (\bibinfo  {publisher} {Clarendon Press},\ \bibinfo {year}
	{1984})\BibitemShut {NoStop}%
	\bibitem [{\citenamefont {Balents}(2010)}]{Balents_NatureReview}%
	\BibitemOpen
	\bibfield  {author} {\bibinfo {author} {\bibfnamefont {L.}~\bibnamefont
			{Balents}},\ }\href {\doibase 10.1038/nature08917} {\bibfield  {journal}
		{\bibinfo  {journal} {Nature}\ }\textbf {\bibinfo {volume} {464}},\ \bibinfo
		{pages} {199} (\bibinfo {year} {2010})}\BibitemShut {NoStop}%
	\bibitem [{\citenamefont {Savary}\ and\ \citenamefont
		{Balents}(2016)}]{Savary2016}%
	\BibitemOpen
	\bibfield  {author} {\bibinfo {author} {\bibfnamefont {L.}~\bibnamefont
			{Savary}}\ and\ \bibinfo {author} {\bibfnamefont {L.}~\bibnamefont
			{Balents}},\ }\href {\doibase 10.1088/0034-4885/80/1/016502} {\bibfield
		{journal} {\bibinfo  {journal} {Reports on Progress in Physics}\ }\textbf
		{\bibinfo {volume} {80}},\ \bibinfo {pages} {016502} (\bibinfo {year}
		{2016})}\BibitemShut {NoStop}%
	\bibitem [{\citenamefont {Gattringer}\ and\ \citenamefont
		{Lang}(2009)}]{Gattringer_book}%
	\BibitemOpen
	\bibfield  {author} {\bibinfo {author} {\bibfnamefont {C.}~\bibnamefont
			{Gattringer}}\ and\ \bibinfo {author} {\bibfnamefont {C.}~\bibnamefont
			{Lang}},\ }\href {https://books.google.de/books?id=l2hZKnlYDxoC} {\emph
		{\bibinfo {title} {Quantum Chromodynamics on the Lattice: An Introductory
				Presentation}}},\ Lecture Notes in Physics\ (\bibinfo  {publisher} {Springer
		Berlin Heidelberg},\ \bibinfo {year} {2009})\BibitemShut {NoStop}%
	\bibitem [{\citenamefont {Calzetta}\ and\ \citenamefont
		{Hu}(2008)}]{Calzetta_book}%
	\BibitemOpen
	\bibfield  {author} {\bibinfo {author} {\bibfnamefont {E.}~\bibnamefont
			{Calzetta}}\ and\ \bibinfo {author} {\bibfnamefont {B.}~\bibnamefont {Hu}},\
	}\href {https://books.google.de/books?id=BRJ7ryt2l1IC} {\emph {\bibinfo
			{title} {Nonequilibrium Quantum Field Theory}}},\ Cambridge Monographs on
	Mathematical Physics\ (\bibinfo  {publisher} {Cambridge University Press},\
	\bibinfo {year} {2008})\BibitemShut {NoStop}%
	\bibitem [{\citenamefont {Wiese}(2013)}]{Wiese2013}%
	\BibitemOpen
	\bibfield  {author} {\bibinfo {author} {\bibfnamefont {U.-J.}\ \bibnamefont
			{Wiese}},\ }\href {\doibase 10.1002/andp.201300104} {\bibfield  {journal}
		{\bibinfo  {journal} {Annalen der Physik}\ }\textbf {\bibinfo {volume}
			{525}},\ \bibinfo {pages} {777} (\bibinfo {year} {2013})}\BibitemShut
	{NoStop}%
	\bibitem [{\citenamefont {Zohar}\ \emph {et~al.}(2015)\citenamefont {Zohar},
		\citenamefont {Cirac},\ and\ \citenamefont {Reznik}}]{Zohar2015}%
	\BibitemOpen
	\bibfield  {author} {\bibinfo {author} {\bibfnamefont {E.}~\bibnamefont
			{Zohar}}, \bibinfo {author} {\bibfnamefont {J.~I.}\ \bibnamefont {Cirac}}, \
		and\ \bibinfo {author} {\bibfnamefont {B.}~\bibnamefont {Reznik}},\ }\href
	{\doibase 10.1088/0034-4885/79/1/014401} {\bibfield  {journal} {\bibinfo
			{journal} {Reports on Progress in Physics}\ }\textbf {\bibinfo {volume}
			{79}},\ \bibinfo {pages} {014401} (\bibinfo {year} {2015})}\BibitemShut
	{NoStop}%
	\bibitem [{\citenamefont {Dalmonte}\ and\ \citenamefont
		{Montangero}(2016)}]{Dalmonte2016}%
	\BibitemOpen
	\bibfield  {author} {\bibinfo {author} {\bibfnamefont {M.}~\bibnamefont
			{Dalmonte}}\ and\ \bibinfo {author} {\bibfnamefont {S.}~\bibnamefont
			{Montangero}},\ }\href {\doibase 10.1080/00107514.2016.1151199} {\bibfield
		{journal} {\bibinfo  {journal} {Contemporary Physics}\ }\textbf {\bibinfo
			{volume} {57}},\ \bibinfo {pages} {388} (\bibinfo {year} {2016})},\ \Eprint
	{http://arxiv.org/abs/https://doi.org/10.1080/00107514.2016.1151199}
	{https://doi.org/10.1080/00107514.2016.1151199} \BibitemShut {NoStop}%
	\bibitem [{\citenamefont {{Bañuls}}\ \emph {et~al.}(2019)\citenamefont
		{{Bañuls}}, \citenamefont {{Blatt}}, \citenamefont {{Catani}}, \citenamefont
		{{Celi}}, \citenamefont {{Cirac}}, \citenamefont {{Dalmonte}}, \citenamefont
		{{Fallani}}, \citenamefont {{Jansen}}, \citenamefont {{Lewenstein}},
		\citenamefont {{Montangero}}, \citenamefont {{Muschik}}, \citenamefont
		{{Reznik}}, \citenamefont {{Rico}}, \citenamefont {{Tagliacozzo}},
		\citenamefont {{Van Acoleyen}}, \citenamefont {{Verstraete}}, \citenamefont
		{{Wiese}}, \citenamefont {{Wingate}}, \citenamefont {{Zakrzewski}},\ and\
		\citenamefont {{Zoller}}}]{MariCarmen2019}%
	\BibitemOpen
	\bibfield  {author} {\bibinfo {author} {\bibfnamefont {M.~C.}\ \bibnamefont
			{{Bañuls}}}, \bibinfo {author} {\bibfnamefont {R.}~\bibnamefont {{Blatt}}},
		\bibinfo {author} {\bibfnamefont {J.}~\bibnamefont {{Catani}}}, \bibinfo
		{author} {\bibfnamefont {A.}~\bibnamefont {{Celi}}}, \bibinfo {author}
		{\bibfnamefont {J.~I.}\ \bibnamefont {{Cirac}}}, \bibinfo {author}
		{\bibfnamefont {M.}~\bibnamefont {{Dalmonte}}}, \bibinfo {author}
		{\bibfnamefont {L.}~\bibnamefont {{Fallani}}}, \bibinfo {author}
		{\bibfnamefont {K.}~\bibnamefont {{Jansen}}}, \bibinfo {author}
		{\bibfnamefont {M.}~\bibnamefont {{Lewenstein}}}, \bibinfo {author}
		{\bibfnamefont {S.}~\bibnamefont {{Montangero}}}, \bibinfo {author}
		{\bibfnamefont {C.~A.}\ \bibnamefont {{Muschik}}}, \bibinfo {author}
		{\bibfnamefont {B.}~\bibnamefont {{Reznik}}}, \bibinfo {author}
		{\bibfnamefont {E.}~\bibnamefont {{Rico}}}, \bibinfo {author} {\bibfnamefont
			{L.}~\bibnamefont {{Tagliacozzo}}}, \bibinfo {author} {\bibfnamefont
			{K.}~\bibnamefont {{Van Acoleyen}}}, \bibinfo {author} {\bibfnamefont
			{F.}~\bibnamefont {{Verstraete}}}, \bibinfo {author} {\bibfnamefont {U.-J.}\
			\bibnamefont {{Wiese}}}, \bibinfo {author} {\bibfnamefont {M.}~\bibnamefont
			{{Wingate}}}, \bibinfo {author} {\bibfnamefont {J.}~\bibnamefont
			{{Zakrzewski}}}, \ and\ \bibinfo {author} {\bibfnamefont {P.}~\bibnamefont
			{{Zoller}}},\ }\href {https://arxiv.org/abs/1911.00003} {\bibfield  {journal}
		{\bibinfo  {journal} {ArXiv e-prints}\ } (\bibinfo {year} {2019})},\ \Eprint
	{http://arxiv.org/abs/1911.00003} {arXiv:1911.00003 [quant-ph]} \BibitemShut
	{NoStop}%
	\bibitem [{\citenamefont {Martinez}\ \emph {et~al.}(2016)\citenamefont
		{Martinez}, \citenamefont {Muschik}, \citenamefont {Schindler}, \citenamefont
		{Nigg}, \citenamefont {Erhard}, \citenamefont {Heyl}, \citenamefont {Hauke},
		\citenamefont {Dalmonte}, \citenamefont {Monz}, \citenamefont {Zoller},\ and\
		\citenamefont {Blatt}}]{Martinez2016}%
	\BibitemOpen
	\bibfield  {author} {\bibinfo {author} {\bibfnamefont {E.~A.}\ \bibnamefont
			{Martinez}}, \bibinfo {author} {\bibfnamefont {C.~A.}\ \bibnamefont
			{Muschik}}, \bibinfo {author} {\bibfnamefont {P.}~\bibnamefont {Schindler}},
		\bibinfo {author} {\bibfnamefont {D.}~\bibnamefont {Nigg}}, \bibinfo {author}
		{\bibfnamefont {A.}~\bibnamefont {Erhard}}, \bibinfo {author} {\bibfnamefont
			{M.}~\bibnamefont {Heyl}}, \bibinfo {author} {\bibfnamefont {P.}~\bibnamefont
			{Hauke}}, \bibinfo {author} {\bibfnamefont {M.}~\bibnamefont {Dalmonte}},
		\bibinfo {author} {\bibfnamefont {T.}~\bibnamefont {Monz}}, \bibinfo {author}
		{\bibfnamefont {P.}~\bibnamefont {Zoller}}, \ and\ \bibinfo {author}
		{\bibfnamefont {R.}~\bibnamefont {Blatt}},\ }\href {\doibase
		10.1038/nature18318} {\bibfield  {journal} {\bibinfo  {journal} {Nature}\
		}\textbf {\bibinfo {volume} {534}},\ \bibinfo {pages} {516} (\bibinfo {year}
		{2016})}\BibitemShut {NoStop}%
	\bibitem [{\citenamefont {Bernien}\ \emph {et~al.}(2017)\citenamefont
		{Bernien}, \citenamefont {Schwartz}, \citenamefont {Keesling}, \citenamefont
		{Levine}, \citenamefont {Omran}, \citenamefont {Pichler}, \citenamefont
		{Choi}, \citenamefont {Zibrov}, \citenamefont {Endres}, \citenamefont
		{Greiner}, \citenamefont {Vuleti{\'c}},\ and\ \citenamefont
		{Lukin}}]{Bernien2017}%
	\BibitemOpen
	\bibfield  {author} {\bibinfo {author} {\bibfnamefont {H.}~\bibnamefont
			{Bernien}}, \bibinfo {author} {\bibfnamefont {S.}~\bibnamefont {Schwartz}},
		\bibinfo {author} {\bibfnamefont {A.}~\bibnamefont {Keesling}}, \bibinfo
		{author} {\bibfnamefont {H.}~\bibnamefont {Levine}}, \bibinfo {author}
		{\bibfnamefont {A.}~\bibnamefont {Omran}}, \bibinfo {author} {\bibfnamefont
			{H.}~\bibnamefont {Pichler}}, \bibinfo {author} {\bibfnamefont
			{S.}~\bibnamefont {Choi}}, \bibinfo {author} {\bibfnamefont {A.~S.}\
			\bibnamefont {Zibrov}}, \bibinfo {author} {\bibfnamefont {M.}~\bibnamefont
			{Endres}}, \bibinfo {author} {\bibfnamefont {M.}~\bibnamefont {Greiner}},
		\bibinfo {author} {\bibfnamefont {V.}~\bibnamefont {Vuleti{\'c}}}, \ and\
		\bibinfo {author} {\bibfnamefont {M.~D.}\ \bibnamefont {Lukin}},\ }\href
	{\doibase 10.1038/nature24622} {\bibfield  {journal} {\bibinfo  {journal}
			{Nature}\ }\textbf {\bibinfo {volume} {551}},\ \bibinfo {pages} {579}
		(\bibinfo {year} {2017})}\BibitemShut {NoStop}%
	\bibitem [{\citenamefont {Dai}\ \emph {et~al.}(2017)\citenamefont {Dai},
		\citenamefont {Yang}, \citenamefont {Reingruber}, \citenamefont {Sun},
		\citenamefont {Xu}, \citenamefont {Chen}, \citenamefont {Yuan},\ and\
		\citenamefont {Pan}}]{Dai2017}%
	\BibitemOpen
	\bibfield  {author} {\bibinfo {author} {\bibfnamefont {H.-N.}\ \bibnamefont
			{Dai}}, \bibinfo {author} {\bibfnamefont {B.}~\bibnamefont {Yang}}, \bibinfo
		{author} {\bibfnamefont {A.}~\bibnamefont {Reingruber}}, \bibinfo {author}
		{\bibfnamefont {H.}~\bibnamefont {Sun}}, \bibinfo {author} {\bibfnamefont
			{X.-F.}\ \bibnamefont {Xu}}, \bibinfo {author} {\bibfnamefont {Y.-A.}\
			\bibnamefont {Chen}}, \bibinfo {author} {\bibfnamefont {Z.-S.}\ \bibnamefont
			{Yuan}}, \ and\ \bibinfo {author} {\bibfnamefont {J.-W.}\ \bibnamefont
			{Pan}},\ }\href {\doibase 10.1038/nphys4243} {\bibfield  {journal} {\bibinfo
			{journal} {Nature Physics}\ }\textbf {\bibinfo {volume} {13}},\ \bibinfo
		{pages} {1195} (\bibinfo {year} {2017})}\BibitemShut {NoStop}%
	\bibitem [{\citenamefont {Klco}\ \emph {et~al.}(2018)\citenamefont {Klco},
		\citenamefont {Dumitrescu}, \citenamefont {McCaskey}, \citenamefont {Morris},
		\citenamefont {Pooser}, \citenamefont {Sanz}, \citenamefont {Solano},
		\citenamefont {Lougovski},\ and\ \citenamefont {Savage}}]{Klco2018}%
	\BibitemOpen
	\bibfield  {author} {\bibinfo {author} {\bibfnamefont {N.}~\bibnamefont
			{Klco}}, \bibinfo {author} {\bibfnamefont {E.~F.}\ \bibnamefont
			{Dumitrescu}}, \bibinfo {author} {\bibfnamefont {A.~J.}\ \bibnamefont
			{McCaskey}}, \bibinfo {author} {\bibfnamefont {T.~D.}\ \bibnamefont
			{Morris}}, \bibinfo {author} {\bibfnamefont {R.~C.}\ \bibnamefont {Pooser}},
		\bibinfo {author} {\bibfnamefont {M.}~\bibnamefont {Sanz}}, \bibinfo {author}
		{\bibfnamefont {E.}~\bibnamefont {Solano}}, \bibinfo {author} {\bibfnamefont
			{P.}~\bibnamefont {Lougovski}}, \ and\ \bibinfo {author} {\bibfnamefont
			{M.~J.}\ \bibnamefont {Savage}},\ }\href {\doibase
		10.1103/PhysRevA.98.032331} {\bibfield  {journal} {\bibinfo  {journal} {Phys.
				Rev. A}\ }\textbf {\bibinfo {volume} {98}},\ \bibinfo {pages} {032331}
		(\bibinfo {year} {2018})}\BibitemShut {NoStop}%
	\bibitem [{\citenamefont {G{\"o}rg}\ \emph {et~al.}(2019)\citenamefont
		{G{\"o}rg}, \citenamefont {Sandholzer}, \citenamefont {Minguzzi},
		\citenamefont {Desbuquois}, \citenamefont {Messer},\ and\ \citenamefont
		{Esslinger}}]{Goerg2019}%
	\BibitemOpen
	\bibfield  {author} {\bibinfo {author} {\bibfnamefont {F.}~\bibnamefont
			{G{\"o}rg}}, \bibinfo {author} {\bibfnamefont {K.}~\bibnamefont
			{Sandholzer}}, \bibinfo {author} {\bibfnamefont {J.}~\bibnamefont
			{Minguzzi}}, \bibinfo {author} {\bibfnamefont {R.}~\bibnamefont
			{Desbuquois}}, \bibinfo {author} {\bibfnamefont {M.}~\bibnamefont {Messer}},
		\ and\ \bibinfo {author} {\bibfnamefont {T.}~\bibnamefont {Esslinger}},\
	}\href {\doibase 10.1038/s41567-019-0615-4} {\bibfield  {journal} {\bibinfo
			{journal} {Nature Physics}\ }\textbf {\bibinfo {volume} {15}},\ \bibinfo
		{pages} {1161} (\bibinfo {year} {2019})}\BibitemShut {NoStop}%
	\bibitem [{\citenamefont {Kokail}\ \emph {et~al.}(2019)\citenamefont {Kokail},
		\citenamefont {Maier}, \citenamefont {van Bijnen}, \citenamefont {Brydges},
		\citenamefont {Joshi}, \citenamefont {Jurcevic}, \citenamefont {Muschik},
		\citenamefont {Silvi}, \citenamefont {Blatt}, \citenamefont {Roos},\ and\
		\citenamefont {Zoller}}]{Kokail2019}%
	\BibitemOpen
	\bibfield  {author} {\bibinfo {author} {\bibfnamefont {C.}~\bibnamefont
			{Kokail}}, \bibinfo {author} {\bibfnamefont {C.}~\bibnamefont {Maier}},
		\bibinfo {author} {\bibfnamefont {R.}~\bibnamefont {van Bijnen}}, \bibinfo
		{author} {\bibfnamefont {T.}~\bibnamefont {Brydges}}, \bibinfo {author}
		{\bibfnamefont {M.~K.}\ \bibnamefont {Joshi}}, \bibinfo {author}
		{\bibfnamefont {P.}~\bibnamefont {Jurcevic}}, \bibinfo {author}
		{\bibfnamefont {C.~A.}\ \bibnamefont {Muschik}}, \bibinfo {author}
		{\bibfnamefont {P.}~\bibnamefont {Silvi}}, \bibinfo {author} {\bibfnamefont
			{R.}~\bibnamefont {Blatt}}, \bibinfo {author} {\bibfnamefont {C.~F.}\
			\bibnamefont {Roos}}, \ and\ \bibinfo {author} {\bibfnamefont
			{P.}~\bibnamefont {Zoller}},\ }\href {\doibase 10.1038/s41586-019-1177-4}
	{\bibfield  {journal} {\bibinfo  {journal} {Nature}\ }\textbf {\bibinfo
			{volume} {569}},\ \bibinfo {pages} {355} (\bibinfo {year}
		{2019})}\BibitemShut {NoStop}%
	\bibitem [{\citenamefont {Schweizer}\ \emph {et~al.}(2019)\citenamefont
		{Schweizer}, \citenamefont {Grusdt}, \citenamefont {Berngruber},
		\citenamefont {Barbiero}, \citenamefont {Demler}, \citenamefont {Goldman},
		\citenamefont {Bloch},\ and\ \citenamefont {Aidelsburger}}]{Schweizer2019}%
	\BibitemOpen
	\bibfield  {author} {\bibinfo {author} {\bibfnamefont {C.}~\bibnamefont
			{Schweizer}}, \bibinfo {author} {\bibfnamefont {F.}~\bibnamefont {Grusdt}},
		\bibinfo {author} {\bibfnamefont {M.}~\bibnamefont {Berngruber}}, \bibinfo
		{author} {\bibfnamefont {L.}~\bibnamefont {Barbiero}}, \bibinfo {author}
		{\bibfnamefont {E.}~\bibnamefont {Demler}}, \bibinfo {author} {\bibfnamefont
			{N.}~\bibnamefont {Goldman}}, \bibinfo {author} {\bibfnamefont
			{I.}~\bibnamefont {Bloch}}, \ and\ \bibinfo {author} {\bibfnamefont
			{M.}~\bibnamefont {Aidelsburger}},\ }\href {\doibase
		10.1038/s41567-019-0649-7} {\bibfield  {journal} {\bibinfo  {journal} {Nature
				Physics}\ } (\bibinfo {year} {2019}),\ 10.1038/s41567-019-0649-7}\BibitemShut
	{NoStop}%
	\bibitem [{\citenamefont {{Mil}}\ \emph {et~al.}(2019)\citenamefont {{Mil}},
		\citenamefont {{Zache}}, \citenamefont {{Hegde}}, \citenamefont {{Xia}},
		\citenamefont {{Bhatt}}, \citenamefont {{Oberthaler}}, \citenamefont
		{{Hauke}}, \citenamefont {{Berges}},\ and\ \citenamefont
		{{Jendrzejewski}}}]{Mil2019}%
	\BibitemOpen
	\bibfield  {author} {\bibinfo {author} {\bibfnamefont {A.}~\bibnamefont
			{{Mil}}}, \bibinfo {author} {\bibfnamefont {T.~V.}\ \bibnamefont {{Zache}}},
		\bibinfo {author} {\bibfnamefont {A.}~\bibnamefont {{Hegde}}}, \bibinfo
		{author} {\bibfnamefont {A.}~\bibnamefont {{Xia}}}, \bibinfo {author}
		{\bibfnamefont {R.~P.}\ \bibnamefont {{Bhatt}}}, \bibinfo {author}
		{\bibfnamefont {M.~K.}\ \bibnamefont {{Oberthaler}}}, \bibinfo {author}
		{\bibfnamefont {P.}~\bibnamefont {{Hauke}}}, \bibinfo {author} {\bibfnamefont
			{J.}~\bibnamefont {{Berges}}}, \ and\ \bibinfo {author} {\bibfnamefont
			{F.}~\bibnamefont {{Jendrzejewski}}},\ }\href
	{https://arxiv.org/abs/1909.07641} {\bibfield  {journal} {\bibinfo  {journal}
			{ArXiv e-prints}\ } (\bibinfo {year} {2019})},\ \Eprint
	{http://arxiv.org/abs/1909.07641} {arXiv:1909.07641 [cond-mat.quant-gas]}
	\BibitemShut {NoStop}%
	\bibitem [{Yan()}]{Yang2019}%
	\BibitemOpen
	\href@noop {} {}\bibinfo {howpublished} {B.~Yang, H.~Sun, R.~Ott, H.-Y.~Wang,
		T.~V.~Zache, J.~C.~Halimeh, Z.-S.~Yuan, P.~Hauke, and J.-W.~Pan; (in
		preparation, 2019).}\BibitemShut {Stop}%
	\bibitem [{\citenamefont {Hauke}\ \emph {et~al.}(2012)\citenamefont {Hauke},
		\citenamefont {Cucchietti}, \citenamefont {Tagliacozzo}, \citenamefont
		{Deutsch},\ and\ \citenamefont {Lewenstein}}]{Hauke2012}%
	\BibitemOpen
	\bibfield  {author} {\bibinfo {author} {\bibfnamefont {P.}~\bibnamefont
			{Hauke}}, \bibinfo {author} {\bibfnamefont {F.~M.}\ \bibnamefont
			{Cucchietti}}, \bibinfo {author} {\bibfnamefont {L.}~\bibnamefont
			{Tagliacozzo}}, \bibinfo {author} {\bibfnamefont {I.}~\bibnamefont
			{Deutsch}}, \ and\ \bibinfo {author} {\bibfnamefont {M.}~\bibnamefont
			{Lewenstein}},\ }\href {\doibase 10.1088/0034-4885/75/8/082401} {\bibfield
		{journal} {\bibinfo  {journal} {Reports on Progress in Physics}\ }\textbf
		{\bibinfo {volume} {75}},\ \bibinfo {pages} {082401} (\bibinfo {year}
		{2012})}\BibitemShut {NoStop}%
	\bibitem [{\citenamefont {Berges}(2019)}]{Berges2019}%
	\BibitemOpen
	\bibfield  {author} {\bibinfo {author} {\bibfnamefont {J.}~\bibnamefont
			{Berges}},\ }\href {\doibase 10.1038/d41586-019-01483-1} {\bibfield
		{journal} {\bibinfo  {journal} {Nature}\ }\textbf {\bibinfo {volume} {569}},\
		\bibinfo {pages} {339} (\bibinfo {year} {2019})}\BibitemShut {NoStop}%
	\bibitem [{SM5()}]{SM5}%
	\BibitemOpen
	\href@noop {} {}\bibinfo {howpublished} {See Supplemental Material for a
		derivation of the emergent deformed gauge symmetry due to the violation of
		the original one.}\BibitemShut {Stop}%
	\bibitem [{\citenamefont {Chubb}\ and\ \citenamefont
		{Flammia}(2017)}]{Chubb2017}%
	\BibitemOpen
	\bibfield  {author} {\bibinfo {author} {\bibfnamefont {C.~T.}\ \bibnamefont
			{Chubb}}\ and\ \bibinfo {author} {\bibfnamefont {S.~T.}\ \bibnamefont
			{Flammia}},\ }\href {\doibase 10.1063/1.4998921} {\bibfield  {journal}
		{\bibinfo  {journal} {Journal of Mathematical Physics}\ }\textbf {\bibinfo
			{volume} {58}},\ \bibinfo {pages} {082202} (\bibinfo {year} {2017})},\
	\Eprint {http://arxiv.org/abs/https://doi.org/10.1063/1.4998921}
	{https://doi.org/10.1063/1.4998921} \BibitemShut {NoStop}%
	\bibitem [{\citenamefont {Banerjee}\ \emph {et~al.}(2012)\citenamefont
		{Banerjee}, \citenamefont {Dalmonte}, \citenamefont {M\"uller}, \citenamefont
		{Rico}, \citenamefont {Stebler}, \citenamefont {Wiese},\ and\ \citenamefont
		{Zoller}}]{Banerjee2012}%
	\BibitemOpen
	\bibfield  {author} {\bibinfo {author} {\bibfnamefont {D.}~\bibnamefont
			{Banerjee}}, \bibinfo {author} {\bibfnamefont {M.}~\bibnamefont {Dalmonte}},
		\bibinfo {author} {\bibfnamefont {M.}~\bibnamefont {M\"uller}}, \bibinfo
		{author} {\bibfnamefont {E.}~\bibnamefont {Rico}}, \bibinfo {author}
		{\bibfnamefont {P.}~\bibnamefont {Stebler}}, \bibinfo {author} {\bibfnamefont
			{U.-J.}\ \bibnamefont {Wiese}}, \ and\ \bibinfo {author} {\bibfnamefont
			{P.}~\bibnamefont {Zoller}},\ }\href {\doibase
		10.1103/PhysRevLett.109.175302} {\bibfield  {journal} {\bibinfo  {journal}
			{Phys. Rev. Lett.}\ }\textbf {\bibinfo {volume} {109}},\ \bibinfo {pages}
		{175302} (\bibinfo {year} {2012})}\BibitemShut {NoStop}%
	\bibitem [{\citenamefont {Hauke}\ \emph {et~al.}(2013)\citenamefont {Hauke},
		\citenamefont {Marcos}, \citenamefont {Dalmonte},\ and\ \citenamefont
		{Zoller}}]{Hauke2013}%
	\BibitemOpen
	\bibfield  {author} {\bibinfo {author} {\bibfnamefont {P.}~\bibnamefont
			{Hauke}}, \bibinfo {author} {\bibfnamefont {D.}~\bibnamefont {Marcos}},
		\bibinfo {author} {\bibfnamefont {M.}~\bibnamefont {Dalmonte}}, \ and\
		\bibinfo {author} {\bibfnamefont {P.}~\bibnamefont {Zoller}},\ }\href
	{\doibase 10.1103/PhysRevX.3.041018} {\bibfield  {journal} {\bibinfo
			{journal} {Phys. Rev. X}\ }\textbf {\bibinfo {volume} {3}},\ \bibinfo {pages}
		{041018} (\bibinfo {year} {2013})}\BibitemShut {NoStop}%
	\bibitem [{\citenamefont {K\"uhn}\ \emph {et~al.}(2014)\citenamefont {K\"uhn},
		\citenamefont {Cirac},\ and\ \citenamefont {Ba\~nuls}}]{Kuehn2014}%
	\BibitemOpen
	\bibfield  {author} {\bibinfo {author} {\bibfnamefont {S.}~\bibnamefont
			{K\"uhn}}, \bibinfo {author} {\bibfnamefont {J.~I.}\ \bibnamefont {Cirac}}, \
		and\ \bibinfo {author} {\bibfnamefont {M.-C.}\ \bibnamefont {Ba\~nuls}},\
	}\href {\doibase 10.1103/PhysRevA.90.042305} {\bibfield  {journal} {\bibinfo
			{journal} {Phys. Rev. A}\ }\textbf {\bibinfo {volume} {90}},\ \bibinfo
		{pages} {042305} (\bibinfo {year} {2014})}\BibitemShut {NoStop}%
	\bibitem [{\citenamefont {Dehkharghani}\ \emph {et~al.}(2017)\citenamefont
		{Dehkharghani}, \citenamefont {Rico}, \citenamefont {Zinner},\ and\
		\citenamefont {Negretti}}]{Negretti2017}%
	\BibitemOpen
	\bibfield  {author} {\bibinfo {author} {\bibfnamefont {A.~S.}\ \bibnamefont
			{Dehkharghani}}, \bibinfo {author} {\bibfnamefont {E.}~\bibnamefont {Rico}},
		\bibinfo {author} {\bibfnamefont {N.~T.}\ \bibnamefont {Zinner}}, \ and\
		\bibinfo {author} {\bibfnamefont {A.}~\bibnamefont {Negretti}},\ }\href
	{\doibase 10.1103/PhysRevA.96.043611} {\bibfield  {journal} {\bibinfo
			{journal} {Phys. Rev. A}\ }\textbf {\bibinfo {volume} {96}},\ \bibinfo
		{pages} {043611} (\bibinfo {year} {2017})}\BibitemShut {NoStop}%
	\bibitem [{\citenamefont {{Barros}}\ \emph {et~al.}(2019)\citenamefont
		{{Barros}}, \citenamefont {{Burrello}},\ and\ \citenamefont
		{{Trombettoni}}}]{Barros2019}%
	\BibitemOpen
	\bibfield  {author} {\bibinfo {author} {\bibfnamefont {J.~C.~P.}\
			\bibnamefont {{Barros}}}, \bibinfo {author} {\bibfnamefont {M.}~\bibnamefont
			{{Burrello}}}, \ and\ \bibinfo {author} {\bibfnamefont {A.}~\bibnamefont
			{{Trombettoni}}},\ }\href {https://arxiv.org/abs/1911.06022} {\bibfield
		{journal} {\bibinfo  {journal} {ArXiv e-prints}\ } (\bibinfo {year}
		{2019})},\ \Eprint {http://arxiv.org/abs/1911.06022} {arXiv:1911.06022
		[quant-ph]} \BibitemShut {NoStop}%
	\bibitem [{\citenamefont {Hastings}\ and\ \citenamefont
		{Wen}(2005)}]{Hastings2005}%
	\BibitemOpen
	\bibfield  {author} {\bibinfo {author} {\bibfnamefont {M.~B.}\ \bibnamefont
			{Hastings}}\ and\ \bibinfo {author} {\bibfnamefont {X.-G.}\ \bibnamefont
			{Wen}},\ }\href {\doibase 10.1103/PhysRevB.72.045141} {\bibfield  {journal}
		{\bibinfo  {journal} {Phys. Rev. B}\ }\textbf {\bibinfo {volume} {72}},\
		\bibinfo {pages} {045141} (\bibinfo {year} {2005})}\BibitemShut {NoStop}%
	\bibitem [{\citenamefont {Wetterich}(2017)}]{Wetterich2017}%
	\BibitemOpen
	\bibfield  {author} {\bibinfo {author} {\bibfnamefont {C.}~\bibnamefont
			{Wetterich}},\ }\href {\doibase
		https://doi.org/10.1016/j.nuclphysb.2016.12.008} {\bibfield  {journal}
		{\bibinfo  {journal} {Nuclear Physics B}\ }\textbf {\bibinfo {volume}
			{915}},\ \bibinfo {pages} {135 } (\bibinfo {year} {2017})}\BibitemShut
	{NoStop}%
	\bibitem [{\citenamefont {Foerster}\ \emph {et~al.}(1980)\citenamefont
		{Foerster}, \citenamefont {Nielsen},\ and\ \citenamefont
		{Ninomiya}}]{Foerster1980}%
	\BibitemOpen
	\bibfield  {author} {\bibinfo {author} {\bibfnamefont {D.}~\bibnamefont
			{Foerster}}, \bibinfo {author} {\bibfnamefont {H.}~\bibnamefont {Nielsen}}, \
		and\ \bibinfo {author} {\bibfnamefont {M.}~\bibnamefont {Ninomiya}},\ }\href
	{\doibase https://doi.org/10.1016/0370-2693(80)90842-4} {\bibfield  {journal}
		{\bibinfo  {journal} {Physics Letters B}\ }\textbf {\bibinfo {volume} {94}},\
		\bibinfo {pages} {135 } (\bibinfo {year} {1980})}\BibitemShut {NoStop}%
	\bibitem [{\citenamefont {Poppitz}\ and\ \citenamefont
		{Shang}(2008)}]{Poppitz2008}%
	\BibitemOpen
	\bibfield  {author} {\bibinfo {author} {\bibfnamefont {E.}~\bibnamefont
			{Poppitz}}\ and\ \bibinfo {author} {\bibfnamefont {Y.}~\bibnamefont
			{Shang}},\ }\href {\doibase 10.1142/S0217751X08041281} {\bibfield  {journal}
		{\bibinfo  {journal} {International Journal of Modern Physics A}\ }\textbf
		{\bibinfo {volume} {23}},\ \bibinfo {pages} {4545} (\bibinfo {year}
		{2008})},\ \Eprint
	{http://arxiv.org/abs/https://doi.org/10.1142/S0217751X08041281}
	{https://doi.org/10.1142/S0217751X08041281} \BibitemShut {NoStop}%
	\bibitem [{\citenamefont {Golterman}(2001)}]{Golterman2001}%
	\BibitemOpen
	\bibfield  {author} {\bibinfo {author} {\bibfnamefont {M.}~\bibnamefont
			{Golterman}},\ }\href {\doibase
		https://doi.org/10.1016/S0920-5632(01)00953-7} {\bibfield  {journal}
		{\bibinfo  {journal} {Nuclear Physics B - Proceedings Supplements}\ }\textbf
		{\bibinfo {volume} {94}},\ \bibinfo {pages} {189 } (\bibinfo {year}
		{2001})},\ \bibinfo {note} {proceedings of the XVIIIth International
		Symposium on Lattice Field Theory}\BibitemShut {NoStop}%
	\bibitem [{\citenamefont {Zohar}\ \emph {et~al.}(2017)\citenamefont {Zohar},
		\citenamefont {Farace}, \citenamefont {Reznik},\ and\ \citenamefont
		{Cirac}}]{Zohar2017}%
	\BibitemOpen
	\bibfield  {author} {\bibinfo {author} {\bibfnamefont {E.}~\bibnamefont
			{Zohar}}, \bibinfo {author} {\bibfnamefont {A.}~\bibnamefont {Farace}},
		\bibinfo {author} {\bibfnamefont {B.}~\bibnamefont {Reznik}}, \ and\ \bibinfo
		{author} {\bibfnamefont {J.~I.}\ \bibnamefont {Cirac}},\ }\href {\doibase
		10.1103/PhysRevLett.118.070501} {\bibfield  {journal} {\bibinfo  {journal}
			{Phys. Rev. Lett.}\ }\textbf {\bibinfo {volume} {118}},\ \bibinfo {pages}
		{070501} (\bibinfo {year} {2017})}\BibitemShut {NoStop}%
	\bibitem [{\citenamefont {Barbiero}\ \emph {et~al.}(2019)\citenamefont
		{Barbiero}, \citenamefont {Schweizer}, \citenamefont {Aidelsburger},
		\citenamefont {Demler}, \citenamefont {Goldman},\ and\ \citenamefont
		{Grusdt}}]{Barbiero2019}%
	\BibitemOpen
	\bibfield  {author} {\bibinfo {author} {\bibfnamefont {L.}~\bibnamefont
			{Barbiero}}, \bibinfo {author} {\bibfnamefont {C.}~\bibnamefont {Schweizer}},
		\bibinfo {author} {\bibfnamefont {M.}~\bibnamefont {Aidelsburger}}, \bibinfo
		{author} {\bibfnamefont {E.}~\bibnamefont {Demler}}, \bibinfo {author}
		{\bibfnamefont {N.}~\bibnamefont {Goldman}}, \ and\ \bibinfo {author}
		{\bibfnamefont {F.}~\bibnamefont {Grusdt}},\ }\href {\doibase
		10.1126/sciadv.aav7444} {\bibfield  {journal} {\bibinfo  {journal} {Science
				Advances}\ }\textbf {\bibinfo {volume} {5}} (\bibinfo {year} {2019}),\
		10.1126/sciadv.aav7444}\BibitemShut {NoStop}%
	\bibitem [{\citenamefont {{Frank}}\ \emph {et~al.}(2019)\citenamefont
		{{Frank}}, \citenamefont {{Huffman}},\ and\ \citenamefont
		{{Chandrasekharan}}}]{Frank2019}%
	\BibitemOpen
	\bibfield  {author} {\bibinfo {author} {\bibfnamefont {J.}~\bibnamefont
			{{Frank}}}, \bibinfo {author} {\bibfnamefont {E.}~\bibnamefont {{Huffman}}},
		\ and\ \bibinfo {author} {\bibfnamefont {S.}~\bibnamefont
			{{Chandrasekharan}}},\ }\href {https://arxiv.org/abs/1904.05414} {\bibfield
		{journal} {\bibinfo  {journal} {ArXiv e-prints}\ } (\bibinfo {year}
		{2019})},\ \Eprint {http://arxiv.org/abs/1904.05414} {arXiv:1904.05414
		[cond-mat.str-el]} \BibitemShut {NoStop}%
	\bibitem [{\citenamefont {{Borla}}\ \emph {et~al.}(2019)\citenamefont
		{{Borla}}, \citenamefont {{Verresen}}, \citenamefont {{Grusdt}},\ and\
		\citenamefont {{Moroz}}}]{Borla2019}%
	\BibitemOpen
	\bibfield  {author} {\bibinfo {author} {\bibfnamefont {U.}~\bibnamefont
			{{Borla}}}, \bibinfo {author} {\bibfnamefont {R.}~\bibnamefont {{Verresen}}},
		\bibinfo {author} {\bibfnamefont {F.}~\bibnamefont {{Grusdt}}}, \ and\
		\bibinfo {author} {\bibfnamefont {S.}~\bibnamefont {{Moroz}}},\ }\href
	{https://arxiv.org/abs/1909.07399} {\bibfield  {journal} {\bibinfo  {journal}
			{ArXiv e-prints}\ } (\bibinfo {year} {2019})},\ \Eprint
	{http://arxiv.org/abs/1909.07399} {arXiv:1909.07399 [cond-mat.str-el]}
	\BibitemShut {NoStop}%
	\bibitem [{SM0()}]{SM0}%
	\BibitemOpen
	\href@noop {} {}\bibinfo {howpublished} {See Supplemental Material for all
		the above results repeated for a $\mathrm{U}(1)$ gauge theory, where the main
		conclusions remain unchanged with respect to the $\mathrm{Z}_2$ gauge
		theory.}\BibitemShut {Stop}%
	\bibitem [{foo()}]{footnote}%
	\BibitemOpen
	\href@noop {} {}\bibinfo {howpublished} {Although Ref.~\cite{Schweizer2019}
		considers a minimal model, we are interested here in gauge invariance in
		extended systems.}\BibitemShut {Stop}%
	\bibitem [{SM3()}]{SM3}%
	\BibitemOpen
	\href@noop {} {}\bibinfo {howpublished} {See Supplemental Material for a
		derivation of the constants $c_l$, $\l\in\{1,2,3,4\}$.}\BibitemShut {Stop}%
	\bibitem [{SM4()}]{SM4}%
	\BibitemOpen
	\href@noop {} {}\bibinfo {howpublished} {See Supplemental Material for
		further supporting results for the $\mathrm{Z}_2$ gauge theory.}\BibitemShut
	{Stop}%
	\bibitem [{\citenamefont {Johansson}\ \emph {et~al.}(2012)\citenamefont
		{Johansson}, \citenamefont {Nation},\ and\ \citenamefont
		{Nori}}]{Johansson2012}%
	\BibitemOpen
	\bibfield  {author} {\bibinfo {author} {\bibfnamefont {J.}~\bibnamefont
			{Johansson}}, \bibinfo {author} {\bibfnamefont {P.}~\bibnamefont {Nation}}, \
		and\ \bibinfo {author} {\bibfnamefont {F.}~\bibnamefont {Nori}},\ }\href
	{\doibase https://doi.org/10.1016/j.cpc.2012.02.021} {\bibfield  {journal}
		{\bibinfo  {journal} {Computer Physics Communications}\ }\textbf {\bibinfo
			{volume} {183}},\ \bibinfo {pages} {1760 } (\bibinfo {year}
		{2012})}\BibitemShut {NoStop}%
	\bibitem [{\citenamefont {Johansson}\ \emph {et~al.}(2013)\citenamefont
		{Johansson}, \citenamefont {Nation},\ and\ \citenamefont
		{Nori}}]{Johansson2013}%
	\BibitemOpen
	\bibfield  {author} {\bibinfo {author} {\bibfnamefont {J.}~\bibnamefont
			{Johansson}}, \bibinfo {author} {\bibfnamefont {P.}~\bibnamefont {Nation}}, \
		and\ \bibinfo {author} {\bibfnamefont {F.}~\bibnamefont {Nori}},\ }\href
	{\doibase https://doi.org/10.1016/j.cpc.2012.11.019} {\bibfield  {journal}
		{\bibinfo  {journal} {Computer Physics Communications}\ }\textbf {\bibinfo
			{volume} {184}},\ \bibinfo {pages} {1234 } (\bibinfo {year}
		{2013})}\BibitemShut {NoStop}%
	\bibitem [{\citenamefont {Weinberg}\ and\ \citenamefont
		{Bukov}(2017)}]{Weinberg2017}%
	\BibitemOpen
	\bibfield  {author} {\bibinfo {author} {\bibfnamefont {P.}~\bibnamefont
			{Weinberg}}\ and\ \bibinfo {author} {\bibfnamefont {M.}~\bibnamefont
			{Bukov}},\ }\href {\doibase 10.21468/SciPostPhys.2.1.003} {\bibfield
		{journal} {\bibinfo  {journal} {SciPost Phys.}\ }\textbf {\bibinfo {volume}
			{2}},\ \bibinfo {pages} {003} (\bibinfo {year} {2017})}\BibitemShut {NoStop}%
	\bibitem [{\citenamefont {Weinberg}\ and\ \citenamefont
		{Bukov}(2019)}]{Weinberg2019}%
	\BibitemOpen
	\bibfield  {author} {\bibinfo {author} {\bibfnamefont {P.}~\bibnamefont
			{Weinberg}}\ and\ \bibinfo {author} {\bibfnamefont {M.}~\bibnamefont
			{Bukov}},\ }\href {\doibase 10.21468/SciPostPhys.7.2.020} {\bibfield
		{journal} {\bibinfo  {journal} {SciPost Phys.}\ }\textbf {\bibinfo {volume}
			{7}},\ \bibinfo {pages} {20} (\bibinfo {year} {2019})}\BibitemShut {NoStop}%
	\bibitem [{SM1()}]{SM1}%
	\BibitemOpen
	\href@noop {} {}\bibinfo {howpublished} {See Supplemental Material where we
		present the detailed derivation in perturbation theory of the short-time
		dynamics due to a gauge invariance-breaking term of strength
		$\lambda>0$.}\BibitemShut {Stop}%
	\bibitem [{\citenamefont {Jiang}\ and\ \citenamefont
		{Rieffel}(2017)}]{Jiang2015}%
	\BibitemOpen
	\bibfield  {author} {\bibinfo {author} {\bibfnamefont {Z.}~\bibnamefont
			{Jiang}}\ and\ \bibinfo {author} {\bibfnamefont {E.~G.}\ \bibnamefont
			{Rieffel}},\ }\href {\doibase https://doi.org/10.1007/s11128-017-1527-9}
	{\bibfield  {journal} {\bibinfo  {journal} {Quantum Inf Process}\ }\textbf
		{\bibinfo {volume} {16}},\ \bibinfo {pages} {89} (\bibinfo {year}
		{2017})}\BibitemShut {NoStop}%
	\bibitem [{SM2()}]{SM2}%
	\BibitemOpen
	\href@noop {} {}\bibinfo {howpublished} {See Supplemental Material where we
		present the detailed derivation in perturbation theory of the effective
		Hamiltonian due to introducing an energy-penalty term with strength $V$ to
		protect against the gauge invariance-breaking processes.}\BibitemShut {Stop}%
\end{thebibliography}

\begin{thebibliography}{13}
	\expandafter\ifx\csname natexlab\endcsname\relax\def\natexlab#1{#1}\fi
	\expandafter\ifx\csname bibnamefont\endcsname\relax
	\def\bibnamefont#1{#1}\fi
	\expandafter\ifx\csname bibfnamefont\endcsname\relax
	\def\bibfnamefont#1{#1}\fi
	\expandafter\ifx\csname citenamefont\endcsname\relax
	\def\citenamefont#1{#1}\fi
	\expandafter\ifx\csname url\endcsname\relax
	\def\url#1{\texttt{#1}}\fi
	\expandafter\ifx\csname urlprefix\endcsname\relax\def\urlprefix{URL }\fi
	\providecommand{\bibinfo}[2]{#2}
	\providecommand{\eprint}[2][]{\url{#2}}
	
	\bibitem[{\citenamefont{Schweizer et~al.}(2019)\citenamefont{Schweizer, Grusdt,
			Berngruber, Barbiero, Demler, Goldman, Bloch, and
			Aidelsburger}}]{Schweizer2019-S}
	\bibinfo{author}{\bibfnamefont{C.}~\bibnamefont{Schweizer}},
	\bibinfo{author}{\bibfnamefont{F.}~\bibnamefont{Grusdt}},
	\bibinfo{author}{\bibfnamefont{M.}~\bibnamefont{Berngruber}},
	\bibinfo{author}{\bibfnamefont{L.}~\bibnamefont{Barbiero}},
	\bibinfo{author}{\bibfnamefont{E.}~\bibnamefont{Demler}},
	\bibinfo{author}{\bibfnamefont{N.}~\bibnamefont{Goldman}},
	\bibinfo{author}{\bibfnamefont{I.}~\bibnamefont{Bloch}}, \bibnamefont{and}
	\bibinfo{author}{\bibfnamefont{M.}~\bibnamefont{Aidelsburger}},
	\bibinfo{journal}{Nature Physics}  (\bibinfo{year}{2019}),
	\urlprefix\url{https://doi.org/10.1038/s41567-019-0649-7}.
	
	\bibitem[{\citenamefont{Coleman}(1976)}]{Coleman1976-S}
	\bibinfo{author}{\bibfnamefont{S.}~\bibnamefont{Coleman}},
	\bibinfo{journal}{Annals of Physics} \textbf{\bibinfo{volume}{101}},
	\bibinfo{pages}{239 } (\bibinfo{year}{1976}), ISSN \bibinfo{issn}{0003-4916},
	\urlprefix\url{http://www.sciencedirect.com/science/article/pii/0003491676902803}.
	
	\bibitem[{\citenamefont{Huang et~al.}(2019)\citenamefont{Huang, Banerjee, and
			Heyl}}]{Huang2019-S}
	\bibinfo{author}{\bibfnamefont{Y.-P.} \bibnamefont{Huang}},
	\bibinfo{author}{\bibfnamefont{D.}~\bibnamefont{Banerjee}}, \bibnamefont{and}
	\bibinfo{author}{\bibfnamefont{M.}~\bibnamefont{Heyl}},
	\bibinfo{journal}{Phys. Rev. Lett.} \textbf{\bibinfo{volume}{122}},
	\bibinfo{pages}{250401} (\bibinfo{year}{2019}),
	\urlprefix\url{https://link.aps.org/doi/10.1103/PhysRevLett.122.250401}.
	
	\bibitem[{\citenamefont{Brenes et~al.}(2018)\citenamefont{Brenes, Dalmonte,
			Heyl, and Scardicchio}}]{Brenes2018-S}
	\bibinfo{author}{\bibfnamefont{M.}~\bibnamefont{Brenes}},
	\bibinfo{author}{\bibfnamefont{M.}~\bibnamefont{Dalmonte}},
	\bibinfo{author}{\bibfnamefont{M.}~\bibnamefont{Heyl}}, \bibnamefont{and}
	\bibinfo{author}{\bibfnamefont{A.}~\bibnamefont{Scardicchio}},
	\bibinfo{journal}{Phys. Rev. Lett.} \textbf{\bibinfo{volume}{120}},
	\bibinfo{pages}{030601} (\bibinfo{year}{2018}),
	\urlprefix\url{https://link.aps.org/doi/10.1103/PhysRevLett.120.030601}.
	
	\bibitem[{\citenamefont{Hauke et~al.}(2013)\citenamefont{Hauke, Marcos,
			Dalmonte, and Zoller}}]{Hauke2013-S}
	\bibinfo{author}{\bibfnamefont{P.}~\bibnamefont{Hauke}},
	\bibinfo{author}{\bibfnamefont{D.}~\bibnamefont{Marcos}},
	\bibinfo{author}{\bibfnamefont{M.}~\bibnamefont{Dalmonte}}, \bibnamefont{and}
	\bibinfo{author}{\bibfnamefont{P.}~\bibnamefont{Zoller}},
	\bibinfo{journal}{Phys. Rev. X} \textbf{\bibinfo{volume}{3}},
	\bibinfo{pages}{041018} (\bibinfo{year}{2013}),
	\urlprefix\url{https://link.aps.org/doi/10.1103/PhysRevX.3.041018}.
	
	\bibitem[{\citenamefont{Yang et~al.}(2016)\citenamefont{Yang, Giri, Johanning,
			Wunderlich, Zoller, and Hauke}}]{Yang2016-S}
	\bibinfo{author}{\bibfnamefont{D.}~\bibnamefont{Yang}},
	\bibinfo{author}{\bibfnamefont{G.~S.} \bibnamefont{Giri}},
	\bibinfo{author}{\bibfnamefont{M.}~\bibnamefont{Johanning}},
	\bibinfo{author}{\bibfnamefont{C.}~\bibnamefont{Wunderlich}},
	\bibinfo{author}{\bibfnamefont{P.}~\bibnamefont{Zoller}}, \bibnamefont{and}
	\bibinfo{author}{\bibfnamefont{P.}~\bibnamefont{Hauke}},
	\bibinfo{journal}{Phys. Rev. A} \textbf{\bibinfo{volume}{94}},
	\bibinfo{pages}{052321} (\bibinfo{year}{2016}),
	\urlprefix\url{https://link.aps.org/doi/10.1103/PhysRevA.94.052321}.
	
	\bibitem[{\citenamefont{Johansson et~al.}(2012)\citenamefont{Johansson, Nation,
			and Nori}}]{Johansson2012-S}
	\bibinfo{author}{\bibfnamefont{J.}~\bibnamefont{Johansson}},
	\bibinfo{author}{\bibfnamefont{P.}~\bibnamefont{Nation}}, \bibnamefont{and}
	\bibinfo{author}{\bibfnamefont{F.}~\bibnamefont{Nori}},
	\bibinfo{journal}{Computer Physics Communications}
	\textbf{\bibinfo{volume}{183}}, \bibinfo{pages}{1760 }
	(\bibinfo{year}{2012}), ISSN \bibinfo{issn}{0010-4655},
	\urlprefix\url{http://www.sciencedirect.com/science/article/pii/S0010465512000835}.
	
	\bibitem[{\citenamefont{Johansson et~al.}(2013)\citenamefont{Johansson, Nation,
			and Nori}}]{Johansson2013-S}
	\bibinfo{author}{\bibfnamefont{J.}~\bibnamefont{Johansson}},
	\bibinfo{author}{\bibfnamefont{P.}~\bibnamefont{Nation}}, \bibnamefont{and}
	\bibinfo{author}{\bibfnamefont{F.}~\bibnamefont{Nori}},
	\bibinfo{journal}{Computer Physics Communications}
	\textbf{\bibinfo{volume}{184}}, \bibinfo{pages}{1234 }
	(\bibinfo{year}{2013}), ISSN \bibinfo{issn}{0010-4655},
	\urlprefix\url{http://www.sciencedirect.com/science/article/pii/S0010465512003955}.
	
	\bibitem[{\citenamefont{Weinberg and Bukov}(2017)}]{Weinberg2017-S}
	\bibinfo{author}{\bibfnamefont{P.}~\bibnamefont{Weinberg}} \bibnamefont{and}
	\bibinfo{author}{\bibfnamefont{M.}~\bibnamefont{Bukov}},
	\bibinfo{journal}{SciPost Phys.} \textbf{\bibinfo{volume}{2}},
	\bibinfo{pages}{003} (\bibinfo{year}{2017}),
	\urlprefix\url{https://scipost.org/10.21468/SciPostPhys.2.1.003}.
	
	\bibitem[{\citenamefont{Weinberg and Bukov}(2019)}]{Weinberg2019-S}
	\bibinfo{author}{\bibfnamefont{P.}~\bibnamefont{Weinberg}} \bibnamefont{and}
	\bibinfo{author}{\bibfnamefont{M.}~\bibnamefont{Bukov}},
	\bibinfo{journal}{SciPost Phys.} \textbf{\bibinfo{volume}{7}},
	\bibinfo{pages}{20} (\bibinfo{year}{2019}),
	\urlprefix\url{https://scipost.org/10.21468/SciPostPhys.7.2.020}.
	
	\bibitem[{\citenamefont{Chubb and Flammia}(2017)}]{Chubb2017-S}
	\bibinfo{author}{\bibfnamefont{C.~T.} \bibnamefont{Chubb}} \bibnamefont{and}
	\bibinfo{author}{\bibfnamefont{S.~T.} \bibnamefont{Flammia}},
	\bibinfo{journal}{Journal of Mathematical Physics}
	\textbf{\bibinfo{volume}{58}}, \bibinfo{pages}{082202}
	(\bibinfo{year}{2017}), \eprint{https://doi.org/10.1063/1.4998921},
	\urlprefix\url{https://doi.org/10.1063/1.4998921}.
	
	\bibitem[{\citenamefont{Cohen-Tannoudji
			et~al.}(1992)\citenamefont{Cohen-Tannoudji, Dupont-Roc, and
			Grynberg}}]{cohen1992atom-S}
	\bibinfo{author}{\bibfnamefont{C.}~\bibnamefont{Cohen-Tannoudji}},
	\bibinfo{author}{\bibfnamefont{J.}~\bibnamefont{Dupont-Roc}},
	\bibnamefont{and} \bibinfo{author}{\bibfnamefont{G.}~\bibnamefont{Grynberg}},
	\emph{\bibinfo{title}{Atom-photon interactions: basic processes and
			applications}}, Wiley-Interscience publication (\bibinfo{publisher}{J.
		Wiley}, \bibinfo{year}{1992}), ISBN \bibinfo{isbn}{9780471293361},
	\urlprefix\url{https://books.google.de/books?id=m7gPAQAAMAAJ}.
	
	\bibitem[{\citenamefont{Lewenstein et~al.}(2007)\citenamefont{Lewenstein,
			Sanpera, Ahufinger, Damski, Sen(De), and Sen}}]{Lewenstein_review-S}
	\bibinfo{author}{\bibfnamefont{M.}~\bibnamefont{Lewenstein}},
	\bibinfo{author}{\bibfnamefont{A.}~\bibnamefont{Sanpera}},
	\bibinfo{author}{\bibfnamefont{V.}~\bibnamefont{Ahufinger}},
	\bibinfo{author}{\bibfnamefont{B.}~\bibnamefont{Damski}},
	\bibinfo{author}{\bibfnamefont{A.}~\bibnamefont{Sen(De)}}, \bibnamefont{and}
	\bibinfo{author}{\bibfnamefont{U.}~\bibnamefont{Sen}},
	\bibinfo{journal}{Advances in Physics} \textbf{\bibinfo{volume}{56}},
	\bibinfo{pages}{243} (\bibinfo{year}{2007}),
	\eprint{https://doi.org/10.1080/00018730701223200},
	\urlprefix\url{https://doi.org/10.1080/00018730701223200}.
	
\end{thebibliography}
\end{document}